\documentclass[11pt, onecolumn,draftclsnofoot, journal]{IEEEtran}
\usepackage[utf8]{inputenc}
\usepackage[mathscr]{eucal}
\usepackage[cmex10]{amsmath}
\usepackage{epsfig,epsf,psfrag}
\usepackage{amssymb,amsmath,amsthm,amsfonts,latexsym,bm}
\usepackage{amsmath,graphicx,bm,xcolor,url}
\usepackage[caption=false]{subfig} 
\usepackage{array}%array and tabular environments
\usepackage{verbatim}
\usepackage{bm}
\usepackage{cite}
\usepackage{algpseudocode}
\usepackage{algorithm}
\usepackage{verbatim}
\usepackage{textcomp}
\usepackage{mathrsfs}
\usepackage{multirow}
\usepackage{epstopdf}
\catcode`~=11 \def\UrlSpecials{\do\~{\kern -.15em\lower .7ex\hbox{~}\kern .04em}} \catcode`~=13 

\allowdisplaybreaks[1]
 
\newcommand{\nn}{\nonumber}

\newcommand{\calB}{\mathcal{B}}
\newcommand{\calH}{\mathcal{H}}

\newcommand{\calN}{\mathcal{N}}

\newcommand{\ba}{\boldsymbol{a}}

\newcommand{\bI}{\boldsymbol{I}}

\newcommand{\bP}{\boldsymbol{P}}

\newcommand{\bR}{\boldsymbol{R}}
\newcommand{\bu}{\boldsymbol{u}}
\newcommand{\bU}{\boldsymbol{U}}

\newcommand{\bV}{\boldsymbol{V}}

\newcommand{\bbC}{\mathbb{C}}

\newcommand{\bbN}{\mathbb{N}}

\newcommand{\bbS}{\mathbb{S}}

\DeclareMathAlphabet{\mathbsf}{OT1}{cmss}{bx}{n}
\DeclareMathAlphabet{\mathssf}{OT1}{cmss}{m}{sl}% slanted sans serif

\DeclareSymbolFont{bsfletters}{OT1}{cmss}{bx}{n}  
\DeclareSymbolFont{ssfletters}{OT1}{cmss}{m}{n}
\DeclareMathSymbol{\bsfGamma}{0}{bsfletters}{'000}
\DeclareMathSymbol{\ssfGamma}{0}{ssfletters}{'000}
\DeclareMathSymbol{\bsfDelta}{0}{bsfletters}{'001}
\DeclareMathSymbol{\ssfDelta}{0}{ssfletters}{'001}
\DeclareMathSymbol{\bsfTheta}{0}{bsfletters}{'002}
\DeclareMathSymbol{\ssfTheta}{0}{ssfletters}{'002}
\DeclareMathSymbol{\bsfLambda}{0}{bsfletters}{'003}
\DeclareMathSymbol{\ssfLambda}{0}{ssfletters}{'003}
\DeclareMathSymbol{\bsfXi}{0}{bsfletters}{'004}
\DeclareMathSymbol{\ssfXi}{0}{ssfletters}{'004}
\DeclareMathSymbol{\bsfPi}{0}{bsfletters}{'005}
\DeclareMathSymbol{\ssfPi}{0}{ssfletters}{'005}
\DeclareMathSymbol{\bsfSigma}{0}{bsfletters}{'006}
\DeclareMathSymbol{\ssfSigma}{0}{ssfletters}{'006}
\DeclareMathSymbol{\bsfUpsilon}{0}{bsfletters}{'007}
\DeclareMathSymbol{\ssfUpsilon}{0}{ssfletters}{'007}
\DeclareMathSymbol{\bsfPhi}{0}{bsfletters}{'010}
\DeclareMathSymbol{\ssfPhi}{0}{ssfletters}{'010}
\DeclareMathSymbol{\bsfPsi}{0}{bsfletters}{'011}
\DeclareMathSymbol{\ssfPsi}{0}{ssfletters}{'011}
\DeclareMathSymbol{\bsfOmega}{0}{bsfletters}{'012}
\DeclareMathSymbol{\ssfOmega}{0}{ssfletters}{'012}

\newcommand{\hatC}{\hat{C}}

\newcommand{\tilC}{\tilde{C}}

\newcommand{\hatQ}{\hat{Q}}

\newcommand{\tilX}{\tilde{X}}

\newcommand{\barG}{\bar{G}}

\newcommand{\bpi}{\bm{\pi}}

\newcommand{\bphi}{\bm{\phi}}

\newcommand{\bSigma	}{\bm{\Sigma}}

\DeclareMathOperator{\diag}{diag}

\DeclareMathOperator{\tr}{tr}

\DeclareMathOperator{\var}{\mathsf{Var}}

\newtheorem{theorem}{Theorem} 
\newtheorem{lemma}[theorem]{Lemma}
\newtheorem{claim}[theorem]{Claim}

\newtheorem{corollary}[theorem]{Corollary}
\newtheorem{definition}[theorem]{Definition}

\newtheorem{remark}[theorem]{Remark}

\newcommand{\qednew}{\nobreak \ifvmode \relax \else
      \ifdim\lastskip<1.5em \hskip-\lastskip
      \hskip1.5em plus0em minus0.5em \fi \nobreak
      \vrule height0.75em width0.5em depth0.25em\fi}

\usepackage{xspace}
\usepackage[colorlinks = true,
linkcolor = blue,
urlcolor  = blue,
citecolor = red,
anchorcolor = green,
]{hyperref}
\usepackage{url}
\newcommand{\blue}{\textcolor{black}}
\newcommand{\red}{\textcolor{black}}
\newcommand{\bW}{\boldsymbol{W}}
\newcommand{\bx}{\boldsymbol{x}}
\newcommand{\bX}{\boldsymbol{X}}
\newcommand{\by}{\boldsymbol{y}}
\newcommand{\bY}{\boldsymbol{Y}}
\newcommand{\bA}{\boldsymbol{A}}

\newcommand{\bS}{\boldsymbol{S}}
\newcommand{\bUpsilon}{\boldsymbol{\Upsilon}}
\newcommand{\bPhi}{\boldsymbol{\Phi}}
\newcommand{\bQ}{\boldsymbol{Q}}
\newcommand{\bT}{\boldsymbol{T}}

\newcommand{\calA}{\mathcal{A}}
\newcommand{\calC}{\mathcal{C}}
\newcommand{\calE}{\mathcal{E}}
\newcommand{\calX}{\mathcal{X}}
\newcommand{\calS}{\mathcal{S}}
\newcommand{\calF}{\mathcal{F}}
\newcommand{\calQ}{\mathcal{Q}}
\newcommand{\calM}{\mathcal{M}}
\newcommand{\calV}{\mathcal{V}}

\newcommand{\calG}{\mathcal{G}} % Free energy function (Markov source)
\newcommand{\tilG}{\tilde{G}} % Free energy function (hidden Markov source)
\newcommand{\tilQ}{\tilde{Q}} % Scaling matrix
\newcommand{\hatG}{\hat{G}} % Auxiliary free energy function 

\newcommand{\barQ}{\bar{Q}} % States in the Markov chain by covariance matrices

\newcommand{\bbR}{\mathbb{R}} % Real Numbers
\newcommand{\bbZ}{\mathbb{Z}} % Integer
\newcommand{\bbE}{\mathbb{E}} % Expectation
\newcommand{\bbP}{\mathbb{P}} % Probability of a set

\newcommand{\rvF}{\mathsf{F}}  % Closed set
\newcommand{\rvU}{\mathsf{U}}  % Open set

\newcommand{\normal}{\calN}
\newcommand{\prob}{\mathbb{P}}
\newcommand{\ups}{\upsilon}
\newcommand{\Ups}{\Upsilon}

\newcommand{\sfX}{{\sf X}}

\newcommand{\E}{\mathbb{E}}

\newcommand{\beq}{\begin{equation}}
\newcommand{\eeq}{\end{equation}}

\begin{document}
\title{Replica Analysis of the Linear Model with Markov or Hidden Markov Signal Priors}
	
\author{Lan~V.~Truong % \!   and \! Ramji~Venkataramanan
\thanks{L.~V~Truong is with the Department of Engineering, University of Cambridge, United Kingdom (e-mail: lt407@cam.ac.uk).}  \thanks{This paper was presented in part at the 2021 International Symposium on Information Theory (ISIT) and the 25th International Conference on Artificial Intelligence and Statistics (AISTATS).}}
		
\maketitle

	\begin{abstract}
		This paper estimates free energy, average mutual information, and minimum mean square error (MMSE) of a linear model under two assumptions: (1) the source is generated by a Markov chain, (2) the source is generated via a hidden Markov model. Our estimates are based on the replica method in statistical physics. We show that under the posterior mean estimator, the linear model with Markov sources or hidden Markov sources is decoupled into single-input AWGN channels with state information available at both encoder and decoder where the state distribution follows the left Perron-Frobenius eigenvector with unit Manhattan norm of the stochastic matrix of Markov chains. Numerical results show that the free energies and MSEs obtained via the replica method are closely approximate to their counterparts achieved by the Metropolis–Hastings algorithm or some well-known approximate message passing algorithms in the research literature.
\end{abstract}

\section{Introduction} \label{sec:intro}
%\RV{Section \ref{sec:intro} has not been edited. }
In the canonical compressed sensing problem, the primary goal is to reconstruct an $n$-dimensional vector $\bX=(X_1,X_2,\cdots,X_n)$ with independent and identical prior from an $m$-dimensional vector of noisy linear observations $\bY=(Y_1,Y_2,\cdots,Y_m)$ of the form $Y_k=\langle \bPhi_k, \bX\rangle+ W_k, k=1,2,\cdots,m$, where $\{\bPhi_k\}$ is a sequence of $n$-dimensional measurement vectors, $\{W_k\}$ is a sequence of standard Gaussian random variables, and $\langle \cdot, \cdot \rangle$ denotes the Euclidean inner product between vectors. In this paper, under the assumption that $\bX$ has a Markov or hidden Markov prior, we wish to estimate the asymptotic mutual information $\lim_{n\to \infty} \frac{1}{n}I(\bX;\bY)$ and the MMSE $\lim_{n \to \infty} \frac{1}{n}\bbE[\|\bX-\bbE[\bX|\bY,\bPhi]\|^2]$. Our estimates are based on the replica method which was developed originally to study mean field approximations in spin glasses \cite{Edwards1975a}. Although this method lacks of rigorous mathematical proof in some particular parts, it has been widely accepted as an analytic tool and utilized to investigate a variety of problems in applied mathematics, information processing, machine learning, and coding \cite{Bereyhi2019a}.
\subsection{Related Work}
The use of the replica method for studying multiuser estimators goes back to \cite{Tanaka2002a} where Tanaka determined the asymptotic bit error rate of Marginal-Posterior-Mode (MPM) estimators by employing the replica method. The study demonstrated interesting large-system properties of multiuser estimators. As a result, the statistical physics approach received more attention in the context of multiuser systems \cite{GuoVerdu2005, Mller2003ChannelCA} with a subsequent work focusing on the compressed sensing directly \cite{Kabashima2009TypicalRL,Guo2009ASC,RFG2012,Reeves2012TheSR,Reeves2012CompressedSP, Tulino2013}.  Guo and Verd\'u \cite{GuoVerdu2005} studied the same CDMA detection problem as \cite{Tanaka2002a} but under more general (arbitrary) input distributions. They assumed that a generic posterior mean estimator is applied before single-user decoding. The generic detector can be particularized to the matched filter, decorrelator, linear minimum mean-square error (MMSE) detector, the jointly or the individual optimal detector, and others. It is found that the detection output for each user, although in general asymptotically non-Gaussian conditioned on the transmitted symbol, converges as the number of users go to infinity to a deterministic function of a ``hidden" Gaussian statistic independent of the interferers. Thus, the multi-user channel can be decoupled.  

The results of replica method have been rigorously in a number of settings in compressed sensing. One example is given by message passing on matrices with special structure, such as sparsity \cite{Guo2006AsymptoticMO,Montanari2006AnalysisOB,Baron2010BayesianCS,Korada2010a,Barbier2020a} or spatial coupling \cite{Kudekar2010,Krzakala2012,Donoho2011OC}. In \cite{RFG2012}, Rangan et al. studied the asymptotic performance of a class of Maximize-A-Posterior (MAP) estimators. Using standard large deviation techniques, the authors represented the MAP estimator as the limit of an indexed MMSE estimator's sequence. Consequently, they determined the estimator's asymptotics employing the results from \cite{GuoVerdu2005} and justified the decoupling property of MAP estimators under Replica Symmetry (RS) assumption for an i.i.d. measurement matrix $\bPhi$. The asymptotic performance for the MAP estimator where the RS assumption does not hold but satisfies some looser symmetric assumptions, called Replica Symmetry Breaking (RSB) is considered in \cite{Bereyhi2019a}. Under the RSB assumption with $b$ steps of breaking (bRSB), the  equivalent noisy single-user channel is given in form of an input term added by an impairment term. The impairment term, moreover, is expressed as a sum of an independent Gaussian random variable and $b$ correlated non-Gaussian interference terms. 

Recently, there have been some works which aim to close the gap between mathematically rigorous proof and results from the replica method. Reeves and Pfister considered the fundamental limit of compressed sensing for i.i.d. signal distributions and i.i.d. Gaussian measurement matrices \cite{Reeves2019}. Under some mild technical conditions, their results show that the limiting mutual information and Minimum Mean Square Error (MMSE) are equal to the values predicted by the replica method. Their proof techniques are based on establishing relationships between mutual information  and MMSE at finite $n,m$ and $n\sim m$ such as \cite{GuoShamaiVerdu2005}, and extending obtained results in large system limits. In \cite{Barbier2018OptimalEA}, Barbier et al. showed that the results for Generalized Linear Models (GLM) and i.i.d. sources stemming from the replica method are indeed correct and imply the optimal value of both estimation and generalization error. The proof is based on the adaptive interpolation method \cite{Barbier2017TheAI} which is an extension of interpolation method developed by Guerra and Toninelli \cite{Guerra2002TheTL} in the context of spin glasses, with an adaptive interpolation path. More specifically, this scheme interpolates between the original problem and the solution via replica method in small steps, each step involving its own set of trial parameters and Gaussian mean-fields in the spirit of Guerra and Toninelli. We are then able to choose the set of trial parameters in various ways so that the upper and lower bounds are eventually matched.  By a generalization of the adaptive interpolation method, Truong \cite{Truong2021FundamentalLA} has recently established exact asymptotic expressions for the normalized mutual information and MMSE of sparse linear regression in the sub-linear sparsity regime, i.e., $m=n^{\alpha}$ for some $\alpha \in (0,1)$.  This work shows that the traditional linear assumption between the signal dimension and number of observations in the replica and adaptive interpolation methods is not necessary for sparse signals.

In all above research literature, the authors assume that the source is independently and identically distributed (i.i.d.). In many practical applications, samples of input data may be dependent on each other, e.g., Markov chains or hidden Markov models. There are a few non-rigorous literatures handling Markov chains using the replica method 
\cite{NIPS1998_655ea4bd,Takeda_2011,Takeda2010}. 
However, to the best of our knowledge, there exists no rigorously analytic result which was developed based on replica-related methods for these models. 
Some recent works considered the linear model with random generative priors where the signal is the output of a Bayesian neural network with specific structures with the input being an  i.i.d. sequence \cite{Aubin2020ExactAF,Tramel2016InferringSC,Aubin2021TheSM}. Although these papers are to recover the structured signal, however, the signal structure is different from Markov or hidden Markov. For example, if we use a classifier (one layer neural network) with ReLU activation function, i.e., $\bx=\sigma(\ba^T \bu)$ where $\ba$ is Gaussian as the assumptions in these papers and $\bu$ is an i.i.d. vector, then $\bx$ is not Markov or Hidden Markov. The adaptive interpolation method looks hard to apply for the linear model with Markov sources or hidden Markov sources since it requires that $X_1,X_2,\ldots,X_n$ are i.i.d. (or at least i.i.d. block-by-block) to guarantee a fixed interpolating free energy at the final $(k,t)$-interpolation model for each finite value of $n$ \cite{Barbier2017TheAI}. There were also some existing works related to Mean Square Errors (MSE) achieved by Approximate Message Passing algorithms (AMP) for the linear model with Markov or hidden Markov sources \cite{Philip,MaRushBaron2019,Berthier2017StateEF}. Approximate message passing (AMP) refers to a class of efficient algorithms for statistical estimation in high-dimensional problems such as compressed sensing and low-rank matrix estimation. AMP is initially proposed for sparse signal recovery and compressed sensing \cite{Donoho06,Candes06,Metzler16a}. AMP algorithms have been proved to be effective in reconstructing sparse signals from a small number of incoherent linear measurements. Their dynamics are accurately tracked by a simple one-dimensional iteration termed \emph{state evolution} \cite{BayatiMonta2011}. The state evolution is redefined in non-asymptotic sense for the sparse linear regression with sublinear sparsity in \cite{Truong2021FundamentalLA}. AMP algorithms achieve state-of-the-art performance for several high-dimensional statistical estimation problems, including compressed sensing \cite{Donoho2009a,BayatiMonta2011,Krzakala2012} and low-rank matrix estimation \cite{BayatiMonta2011, MontanariRamji2020}.
\subsection{Main Contributions}
In this paper, based on the same replica assumptions as \cite{GuoVerdu2005}, we establish free energy, mutual information, and MMSE for the linear model with Markov or hidden Markov sources. When limiting to the linear model with i.i.d. sources as case, we recover Guo and Verd\'u's results \cite{GuoVerdu2005}, which extends Tanaka work \cite{Tanaka2002a} to more general alphabets. More specially, our main contributions are as follows:
\begin{itemize}
	\item Using the replica method, we estimate the free energy, the normalized mutual information in the large system limit for two models: linear model with Markov sources and linear model with hidden Markov sources (cf. Claim \ref{claim:maincontri1} and Claim \ref{claim:hiddenMarkov}).
	\item Using the replica method, we characterize MMSEs in the large system limit for two estimation problems (cf. Claim \ref{claim:jointmoments} and Claim \ref{claim:hiddenMarkov}). We show that under the posterior mean estimator, the linear model with Markov sources or hidden Markov sources is decoupled into single-input AWGN channels with state information available at both encoder and decoder where the state distribution follows the left Perron-Frobenius eigenvector with unit Manhattan norm of the stochastic matrix of Markov chains\footnote{For any irreducible Markov process $\{Z_n\}_{n=1}^{\infty}$, the left Perron-Frobenius eigenvector with unit Manhattan norm is the stationary distribution of this Markov process, and the Perron-Frobenius eigenvalue is equal to $1$ \cite{Lancaster1985b}.}.
	\item We show that the free energies and MSEs obtained via the replica method are closely approximate to their counterparts achieved by the Metropolis–Hastings algorithm or some well-known approximate message passing algorithms in the research literature (cf. Section \ref{sec:numerical}). 
	%Especially, the replica prediction (MMSE) for the linear model with hidden Markov sources is very close to the MSE of the Turbo AMP algorithm in \cite{Philip} for some simulation cases.
\end{itemize}
Essentially, our results show that in the large system limit, we can convert the estimation in high-dimensional space for the linear model with Markov or hidden Markov signal prior to the estimation problems in one-dimensional spaces. 
Compared with the linear model with i.i.d. sources \cite{GuoVerdu2005}, we need to deal with some new technical challenges related to the estimation of the derivative of Perron-Frobenius eigenvalue of  non-negative matrices. For example, in the following Lemma \ref{lem18mod}, we develop a new technique to estimate this derivative in the large system limit.  

MMSE and free energy are very important fundamental limits, which are benchmarks to check if a coding scheme or a learning algorithm for the linear model is optimal. In this work, we aim to characterize these fundamental limits by using replica method. Our simulation results (cf. Section \ref{sec:numerical}) show that some existing MCMC algorithms (for example, Metropolis–Hastings algorithm) and AMP (for example, Turbo AMP \cite{Philip}) are (potentially) optimal for the linear model with Markov or hidden Markov signal prior. Before our work, whether these interesting algorithms are optimal or not is an open question.
\subsection{Paper Organization}
	
The problem setting is placed in Section \ref{sec:setting}, where we introduce the system model, posterior mean estimation, free energy and replica method in statistical physics.  We also introduce some new concepts such as single-symbol Posterior Mean Estimation (PME) channel with state information, free energy functions, and other related notations in this section. Our main results are stated and proved in Section \ref{sec:main}. We apply our main results to estimate free energy, mutual information, and MMSE for some specific Markov chains or hidden Markov models in Section \ref{sec:numerical}, where we also compare our obtained MMSEs with achievable MSEs by the classical Metropolis–Hastings algorithm and some well-known AMP algorithms in research literature. In Appendix \ref{sec:largedev}, we introduce some related results on large deviations and develop new ones for specific applications in this paper. Appendix \ref{sec:perronfrobenius} begins with some results on Perron-Frobenius eigenvalues, and ends with an estimation of the derivative of the Perron-Frobenius eigenvalue for a Markov chain formed by covariance matrices. The proof of joint moments in Section  \ref{sec:main} is placed in Appendix \ref{proof:jointmoment} since the proof technique is similar to the proof of another theorem in Section \ref{sec:main}. Appendix \ref{extend} provides extensions to Markov chains on a general Polish space in $\bbR$.
\subsection{Notation} \label{sub:nota}

Use $[n]$ to denote the set $\{1, \ldots, n \}$. Random vectors and matrices are in bold letters. Expectations with respect to ``quenched" random variables (i.e., the variables that are fixed by the realization of the problem) are denoted by $\bbE$ and those with respect to ``annealed" random variables (i.e., dynamical variables) are denoted by Gibbs bracket $\langle - \rangle$ possibly with appropriate subscripts. This choice follows the stardards of statistical physics. 

As standard literature, we define $x^n=(x_1,x_2,\cdots, x_n)^T$ to denote a vector of length $n$. However, if the dimension of a vector $x$ is clear from context, we omit it for simplicity. Define two loss functions $l_1: \bbR \times \bbR \to \bbR$ and $l_2: \bbR \times \bbR \to \bbR$ as $l_1(x,y)=|x-y|$ and $l_2(x,y)=(x-y)^2$. Let $\log x:=\log_2 x$ and $\ln x$ be the natural logarithm of $x$ for all $x \in \bbR^+$. Manhattan and Euclidean norms of a vector $x \in \bbR^n$ are defined as
\begin{align}
\|x\|_1&:=\sum_{i=1}^n |x_i|,\\
\|x\|_2&:=\sqrt{\sum_{i=1}^n |x_i|^2},
\end{align} respectively. In addition, $\rm{vec}(\cdot)$ denotes the vectorization operator. Besides, for any $A \in \bbR^{p\times q}$ and $B \in (\bbR^{n\times n})^{p\times q}$, we define $Ao_{\mathrm{tr}}B:=\sum_{i,j} A_{ij} \odot B_{ij}$, where $A\odot B$ is the Hadamard product between $A$ and $B$.

The moment generating function of a random vector $\bX \in \bbR^n$ is defined as $\calM(\lambda):=\bbE[\exp(\lambda^T \bX)]$ for all $\lambda \in \bbR^n$. Let $\calM(\tilQ):=\bbE[\exp(\mathrm{tr}(\tilQ \bQ))]$ be 
the moment generating function of a random matrix $\bQ \in \bbR^{n \times n}$ for all matrix $\tilQ \in \bbR^{n\times n}$.

Denote by
\begin{align}
\calQ:=\bigg\{s x x^T \enspace \mbox{for some} \enspace x \in \calS \times \calX^{\nu+1} \bigg\} \label{defspaceQ}.
\end{align}
For simplicity of presentation, we enumerate all matrices in $\calQ$ as $\barQ_0,\barQ_1,\cdots,\barQ_M$ where $M:=|\calQ|-1$.  

\section{Problem Setting}\label{sec:setting}

We consider the linear model 
\begin{align}
	\bY= \bPhi \bX+\bW= \bA \bS^{1/2} \bX +\bW \label{eqn:key}.
	\end{align} Here $\bY \in \bbR^m$ is a vector of observations, $\bX \in \bbR^n$ is the signal vector, $\bA \in \bbR^{m\times n}$ is a measurement matrix, $\bS$ is diagonal  matrix of positive scale factors:
	\begin{align}
	\bS=\diag(S_1,S_2,\cdots,S_n), \quad S_j\in \bbR^{+},
	\end{align} 
and $\bW \in \bbR^m$ is a noise vector. We consider a sequence of problems indexed by $n$, and make the following assumptions on the model. These assumptions are identical to those in earlier works \cite{GuoVerdu2005,RFG2012} except for the signal prior, which we allow to be Markov or hidden Markov in contrast to the i.i.d. priors considered in earlier works. 

\begin{enumerate}

\item We assume that the number of measurements $m$ scales linearly with $n$, and 
$\lim_{n \to \infty} \frac{n}{m} = \beta$, for some $\beta >0$.

\item The elements $\{ A_{ij} \}_{i \in [m], j \in [n]}$ of the matrix $\bA$ are i.i.d. and distributed as $A_{ij} \stackrel{\text{d}}{=}  \frac{1}{\sqrt{m}}  A$, where $A$ is a random variable with zero mean, unit variance and all moments finite.

\item The scale factors $(S_1, \ldots, S_n)$ are i.i.d. according to $P_S$, which is supported on a set $\calS \subset \bbR^+$. The scale factors $(S_1, \ldots, S_n)$ are independent of $\bA, \bX$, and $\bW$.

\item The noise vector $\bW$ is standard normal, i.e.,  $W_j \sim_{\text{i.i.d.}} \normal(0, 1)$ for $j \in [m]$.
 
\item \emph{Signal prior}: We assume that the components of $\bX$ take values \blue{on a Polish space on $\bbR$}, and are distributed according to either a Markov or a hidden Markov prior.
\begin{itemize}
\item \emph{Markov chain prior}: This model assumes that 
\beq 
\prob(\bX = (x_1, \ldots, x_n)) = p(x_1)\pi(x_1,x_2)\cdots \pi(x_{n-1},x_n) \label{defmc1}
\eeq
for some initial probability distribution $p(\cdot)$  on $\calX$, where $\pi(\cdot,\cdot)$ is the transition probability of a time-homogeneous, irreducible Markov chain on $\calX$. 

\item \emph{Hidden Markov prior}: The second model assumes that $\{X_n\}_{n=1}^{\infty}$ are generated by a Hidden Markov Model (HMM), with hidden states $\{\Upsilon_n\}_{n=1}^{\infty}$ take values \blue{on a Polish space on 
$\calH_{\Upsilon}$}.  That is, $\prob( \bUpsilon = (\upsilon_1, \ldots, \upsilon_n))=p_{\Upsilon}(\upsilon_1)\pi_{\Upsilon}(\upsilon_1,\upsilon_2)\cdots \pi_{\Upsilon}(\upsilon_{n-1},\upsilon_n)$ for some initial probability distribution $p_{\Upsilon}(\cdot)$  on $\calH_{\Upsilon}$,  where $\pi_{\Upsilon}(\cdot,\cdot)$ is the transition probability 
of a time-homogeneous, irreducible Markov chain on $\calH_{\Upsilon}$. Then,
\[ 
\prob(X_i = x_i \mid \Ups_1 = \ups_1, \ldots, \Ups_i= \ups_i ) = p_{X|\Upsilon}(x_i \mid \ups_i), \quad i \in [n],
\]
for some stationary emission probability $p_{X|\Upsilon}(\cdot|\cdot)$ on $\calS_{\Upsilon} \times \calX$. 
\end{itemize}
\end{enumerate}

\blue{For simplicity of presentation, we assume that Markov chains $\{X_n\}_{n=1}^{\infty}, \{\Upsilon_n\}_{n=1}^{\infty}$ have finite state spaces and $\calS$ has a finite number of elements in some proofs. However, it is not hard to extend these proofs to Markov chains on Polish spaces in $\bbR$ with an infinite set $\calS$\footnote{In the Appendix \ref{extend}, we show how to extend our analysis to Markov chain on a general Polish spaces in $\bbR$.} by referring to a more general definition of Markov chain in \cite{Tuominen1979} and noting that the Varadhan's large deviation theorem holds for Markov chains on the general Polish space. An \emph{irreducible} and \emph{recurrent} Markov chain on an infinite state-space is called a Harris chain \cite{Tuominen1979}, which owns many similar properties to the finite state-space version such as the existence of an unique stationary distribution. For both models, we denote the joint probability mass distribution (pmf) of the signal by $p(x_1, \ldots, x_n)$. For general proofs, we use Radon–Nikodym derivatives with respect to corresponding measures \cite{Royden}.}

%\RV{since $\calX$ is now assumed to be finite, the condition that for each $q \in \reals$, there is only a countable number of pairs $(s,x) \in \calS \times \calX$ such that $sx^2=q$ is automatically satisfied.}

%\RV{remark about integrals vs sums }

%%%%

\subsection{Posterior Mean Estimation}\label{sec:PME}

The problem setting described above induces a posterior distribution $p_{\bX| \bY, \bPhi}$, given by
\beq
p_{\bX| \bY, \bPhi}(\bx \mid \by, \bphi) = \frac{ p_{\bY \mid \bX,\bPhi}( \by \mid \bx,\bphi)p_{\bX}(\bx)}{p_{\bY | \bPhi}(\by \mid \bphi)},
\label{eq:true_post}
\eeq
where
\begin{align}
p_{\bY|\bX,\bPhi}(\by \mid \bx,\Phi)=(2\pi)^{-m/2} \exp\bigg[-\frac{\| \by -\bphi \bx\|^2}{2}\bigg]
\label{channelstat},
\end{align} 
and 
\[ p_{\bY | \bPhi}(\by \mid \bphi) = \E_p[ p_{\bY|\bX,\bPhi}(\by \mid \bX,\bphi) ] = \sum_{\bx} p_{\bY|\bX,\bPhi}(\by \mid \bx,\bphi) p_{\bX}(\bx). \]

%\RV{remark on using integrals instead of sums for discrete valued random variables.}

The (canonical) posterior mean estimator (PME), which computes the mean value of the posterior distribution $p_{\bX|\bY,\bPhi}$ is given by, %hereafter denoted by angle brackets $\langle \cdot \rangle$. 
\begin{align}
[\bX] =\bbE_{p}\big[\bX|\bY,\bPhi] \label{MMSEEstimator}.
\end{align}
This estimator achieves MMSE between the estimated and the original signal.

As in Guo and Verd{\'u} \cite{GuoVerdu2005}, we  consider a more general class of posterior mean estimators, based on a \emph{postulated} posterior distribution $q_{\bX | \bY, \bPhi}$, to model that scenario that the true posterior mean may be infeasible to compute or the estimator may not know the exact prior and the noise variance. The postulated posterior distribution is defined via a postulated prior and a postulated noise variance. The postulated prior $q_{\bX}(\bx)$ is of the form
\begin{align}
q_{\bX}( \bx )=q(x_1)\tilde{\pi}(x_1,x_2)\cdots \tilde{\pi}(x_{n-1},x_n) \label{priorq}
\end{align}
for some initial distribution $q(\cdot)$ on $\calX$, and $\tilde{\pi}(\cdot,\cdot)$ is the transition probability of an irreducible Markov chain on $\calX$. The postulated likelihood is Gaussian with variance $\sigma^2$, which may not be equal to the true noise variance $1$:
\begin{align}
q_{\bY|\bX,\bPhi}( \by \mid \bx,\bphi)=(2\pi)^{-m/2} \exp\bigg[-\frac{\|\by - \bphi \bx\|^2}{2\sigma^2}\bigg]
\label{ps1}
\end{align}
The postulated prior and noise variance induce the posterior distribution $q_{\bX|\bY, \bPhi}$ given by
\begin{align}
q_{\bX|\bY,\bPhi}(\bx\mid \by,\bphi)=\frac{q_{\bY|\bX,\bPhi}(\by\mid \bx,\bphi) \, q_{\bX}(\bx)}{q_{\bY|\bPhi}(\by|\bphi)} \label{ch:gePME},
\end{align}  where
\begin{align}
q_{\bY|\bPhi}(\by|\bphi)&=\bbE_q[q_{\bY|\bX,\bPhi}(\by\mid \bx,\bphi) \mid \bPhi = \bphi]
=\sum_{\bx} q_{\bX}(\bx) \, q_{\bY|\bX,\bPhi}(\by\mid \bx,\bphi).
\end{align} 
The posterior mean estimator computed using $\eqref{ch:gePME}$, which we call  the `generalized PME', is denoted by 
	\begin{align}
	[\bX]_q=\bbE_{q}\big[\bX|\bY,\bPhi] \label{def1}.
	\end{align} 
As described in \cite{GuoVerdu2005}, with suitable choices of the postulated distribution,  the generalized PME can  recover various commonly used sub-optimal estimators such as the linear MMSE estimator and the matched filter.  The postulated prior can also be used to model estimators that ignore the memory in the signal $\bX$, e.g., estimator based on an i.i.d. prior. 

In the remainder of the paper, we will use the subscript $p$ to denote expectations computed using the true prior/posterior, and $q$ to denote expectations using the postulated prior/posterior.

\subsection{Free Energy and Replica Method}\label{sec:RepMethod}
Let
\begin{align}
Z(\bY,\bPhi):=q_{\bY|\bPhi}(\bY|\bPhi). \label{defyphi}
\end{align}
The free energy of the model in \eqref{eqn:key} is defined as
	\begin{align}
	\calF_n:= -\frac{1}{n}\log Z(\bY,\bPhi) \label{freeenergy}.
	\end{align}
	The expectation of the free energy (with respect to $q_{\bY|\bPhi}(\bY|\bPhi)$)  is equal to the conditional entropy of the observation $\frac{1}{n}H_{q}(\bY|\bPhi)$ as well as (up to an additive constant) to the mutual information density between the signal and the observations $\frac{1}{n}I_{q}(\bX,\bY)$.
	
	The asymptotic free energy is the limit of the sequence $\{F_n\}_{n=1}^{\infty}$, i.e.,
	\begin{align}
	\calF_q:=\lim_{n\to \infty} \calF_n \label{freelimit}.
	\end{align}
	In general, it is very challenging to prove the existence and estimate the limit in \eqref{freelimit}. Replica method, originally developed in statistical physics, is usually used to evaluate this limit \cite{Tanaka2002a,GuoVerdu2005} because the linear model is similar to the thermodynamic system. For this model, replica method is based on the following assumptions (A) and facts (F):
\begin{itemize}
\item (A1) The free energy $\calF_n$ has the self-averaging property as $n\to \infty$. This means that
\begin{align}
\calF:=\lim_{n\to \infty} \bbE[\calF_n] \label{eq:A1}.
\end{align}
The self-averaging property essentially assumes that the variations of $Z(\bY,\bPhi)$ due to the randomness of the measurement matrix $\Phi$ vanish in the limit $n\to \infty$. Although a large number of statistical physics quantities exhibit such self-averaging, the self-averaging of the relevant quantities for the general PME (PMMSE) and Postulated MAP (PMAP) analyses has not been rigorously established \cite{RFG2012}. For the purpose of estimating the average mutual information of the Markov model only, we don't need to make use of this assumption. 
	\item (F1) The following identity holds:
	\begin{align}
	\bbE[\log Z(\bY,\bPhi)]=\lim_{\nu \to 0}\frac{\partial}{\partial \nu}\log \bbE[Z^{\nu}(\bY,\bPhi)] \label{eq:F1}.
	\end{align}
	\item (A2) Estimation of $\bbE[Z(\bY,\bPhi)^{\nu}]$ for a positive real number $\nu$ in the neighbourhood of $0$ can be done by two steps: (1) Estimate $\bbE[Z^{\nu}(\bY,\bPhi)]$ for a general positive integer $\nu$ (2) Take the limit of the obtained result as $\nu \to 0$. This is called ``replica trick" in statistical physics.
	\item (F2) For any positive integer $\nu$ and a realization $(\by,\Phi)$ of $(\bY,\bPhi)$, the quantity $Z^{\nu}(\by,\bPhi)$ can be written as
	\begin{align}
	Z^{\nu}(\by,\Phi)&=\bigg\{q_{\bY|\bPhi}(\by|\Phi)\bigg\}^{\nu}\\
	&=\bigg\{\bbE_{q_{\bX}} \bigg[q_{\bY|\bX,\bPhi}(\by|\bX,\Phi)\bigg]\bigg\}^{\nu}\\
	&=\bbE_{q_{\bX}}\bigg\{\prod_{a=1}^{\nu} q_{\bY|\bX,\bPhi}(\by|\bX^{(a)},\Phi)\bigg\}.
		\end{align} where the last expectation is taken over relicated vectors $\bX^{(a)}, a=1,2,\cdots,\nu$ which are independent copies of a random vector with postulated distribution $q_{\bX}$.
	\item (A3) The order of limit $n \to \infty$ and $\nu \to 0$ can be interchanged. Mathematically, under some conditions such as Theorem Moore-Osgood \cite{Stewart}, the interchange between limits work. This theorem is used in \cite{BarbierALT2016} for a similar purpose. 
	\item (A4) Usually, the free energy can be expressed an optimal value of an optimization problem over the space of covariance matrices of replica samples, say $\calQ$. This optimization is general difficult to perform. To overcome this, the replica method also makes an additional assumption that the optimizer $Q^*$ is symmetric with respect to permutations of $\nu$ replica indices. This assumption is called Replica Symmetry (RS) in statistical physics. See Definition \ref{defrs} for our assumption about RS in this paper.
	\end{itemize} 
	
\section{Main results}\label{sec:main}

  \subsection{Results for Markov Priors} \label{freEnergyx0}

  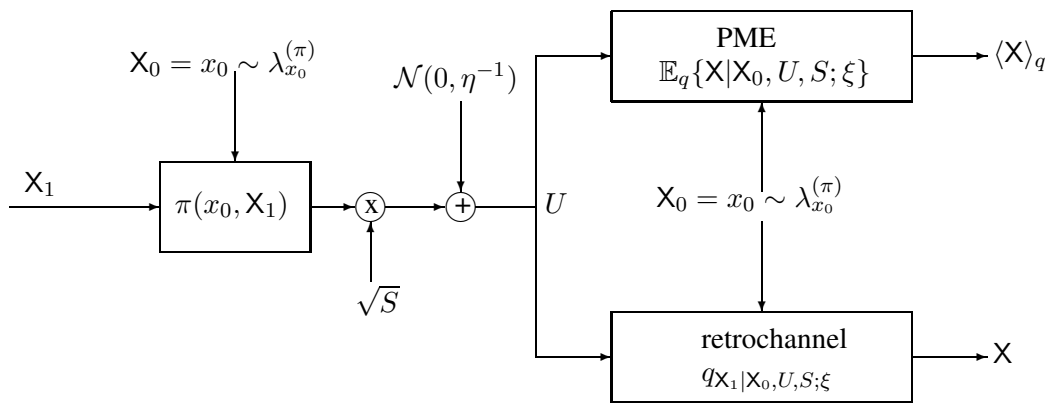
\begin{figure}[ht]
	\centering
	\setlength{\unitlength}{.4mm}
	\begin{picture}(350,150)
	\put(50,70){\line(1,0){50}}
	\put(50,100){\line(1,0){50}}
	\put(50,70){\line(0,1){30}}
	\put(100,70){\line(0,1){30}}
	\put(00,85){\vector(1,0){50}}
	\put(120,85){\circle{10}}  	
	\put(118,83){$\mbox{x}$}
	\put(100,85){\vector(1,0){14.6}}
	\put(125,85){\vector(1,0){19.6}}
	\put(120,60){\vector(0,1){20}}
	\put(115,50){$\sqrt{S}$}
	\put(5,90){$\sfX_1$}
	\put(55,83){$\pi(x_0,\sfX_1)$}
	\put(40,130){$\sfX_0=x_0 \sim \lambda_{x_0}^{(\pi)}$}
	\put(75,130){\vector(0,-1){30}}
	\put(150,85){\circle{10}}  
	\put(148,83){$\mbox{+}$}
	\put(128,125){$\calN(0,\eta^{-1})$}
	\put(150,120){\vector(0,-1){30}}

	\put(200,20){\line(1,0){100}}
	\put(200,50){\line(1,0){100}}
	\put(200,20){\line(0,1){30}}
	\put(300,20){\line(0,1){30}}
	\put(230,38){$\mbox{retrochannel}$}
	\put(230,27){$q_{\sfX_1|\sfX_0,U,S;\xi}$}

	\put(200,120){\line(1,0){100}}
	\put(200,150){\line(1,0){100}}
	\put(200,120){\line(0,1){30}}
	\put(300,120){\line(0,1){30}}
	\put(175,85){\line(0,1){50}}
	\put(178,83){$U$}
	\put(155,85){\line(1,0){20}}
	\put(175,135){\vector(1,0){25}}
	\put(175,85){\line(0,-1){50}}
	\put(175,35){\vector(1,0){25}}
	\put(300,35){\vector(1,0){25}}
	\put(327,33){$\sfX$}

	\put(235,138){$\mbox{PME}$}
	\put(216,127){$\bbE_q\{\sfX|\sfX_0,U,S;\xi\}$}
	\put(215,85){$\sfX_0=x_0 \sim \lambda_{x_0}^{(\pi)}$}
	\put(250,80){\vector(0,-1){30}}
	\put(250,90){\vector(0,1){30}}
	\put(300,135){\vector(1,0){25}}
	\put(327,133){$\langle \sfX|\sfX_0 \rangle_q$}
	
	\end{picture}
	\caption{The equivalent single-symbol Gaussian channel with state available at both encoder and decoder, PME, and retrochannel.}
	\label{fig:PME}
\end{figure}
  
Our results on the free energy and  MMSE will be stated in terms of a \emph{single-symbol} channel, similar to the equivalent single-user Gaussian channel which is obtained via decoupling as in \cite[Section D]{GuoVerdu2005}. \blue{Let $\lambda^{(\pi)}$ be the left Perron-Frobenius eigenvector with unit Manhattan norm\footnote{Since there exists a unique left Perron-Frobenius eigenvector up to a positive scaling factor \cite{Lancaster1985b}, $\lambda^{(\pi)}$ exists uniquely, which is the stationary distribution of the Markov chain.} of 
$ P_{\pi}=\{\pi(x,y)\}_{x \in \calX, y \in \calX}$
which is the stochastic matrix of the Markov chain $\{X_n\}_{n=1}^{\infty}$, and let $\lambda_{x_0}^{(\pi)}$ be the component of $\lambda^{(\pi)}$ associated with the $x_0$-th row of $P_{\pi}$. Let us consider the composition of a Gaussian channel with one state $\sfX_0$ available at both encoder and decoder such that $\sfX_0 =x_0 \sim \lambda_{x_0}^{(\pi)}$, a one-state PME, and a companion retrochannel in the single-symbol setting depicted in Fig. \ref{fig:PME}.} Given the state information $\sfX_0 =x_0 \sim \lambda_{x_0}^{(\pi)}$, the input-output relationship of this single-symbol channel is given by
  \begin{align}
  U=\sqrt{S} \ \sfX_1+\frac{1}{\sqrt{\eta}}W \label{singleCPME},
  \end{align} where the input $\sfX_1 \sim p_{\sfX_1|\sfX_0}(\cdot|x_0):=\pi(x_0,\cdot)$, $S \sim P_S$ which is independent $\sfX_0$ and $\sfX_1$, $W \sim \calN(0,1)$ the noise independent of $\sfX_0$ and $\sfX_1$, and $\eta>0$ the inverse noise variance. 
  % Let us consider the composition of a Gaussian channel with state $\sfX_0=x_0$ available at both encoder and decoder, a PME, and a companion retrochannel in the single-symbol setting depicted in Fig. \ref{fig:PME}. 
  The conditional distribution associated with the channel is
  \begin{align}
  p_{U|\sfX_0,\sfX_1,S;\eta}(u\mid x_0,x_1,s;\eta)=\sqrt{\frac{\eta}{2\pi}} \exp\bigg[-\frac{\eta}{2}(u-\sqrt{s}x_1)^2\bigg] \label{pertain}.
  \end{align}
  Let $q_{U|\sfX_0,\sfX_1,S;\xi}$ represent Gaussian channel with state $\sfX_0$ available at both encoder and decoder akin to \eqref{singleCPME}, the only difference being that the inverse noise variance is $\xi$ instead of $\eta$
   \begin{align}
  q_{U|\sfX_0,\sfX_1,S;\xi}(u\mid x_0,x_1,s;\xi)=\sqrt{\frac{\xi}{2\pi}} \exp\bigg[-\frac{\xi}{2}(u-\sqrt{s}x_1)^2\bigg].
  \end{align}
  Similar to that in the vector channel setting, by postulating the input distribution to be $q_{\sfX_1|\sfX_0}(\cdot|x_0)=\tilde{\pi}(x_0,\cdot)$, a posterior probability distribution $q_{\sfX_1|\sfX_0,U,S;\xi}$ is induced by $q_{\sfX_1|\sfX_0}$ and $q_{U|\sfX_0,\sfX_1,S;\xi}$ using the Bayes rule, i.e.,
  \begin{align}
  q_{\sfX_1|\sfX_0,S,U;\xi}(x\mid x_0,s,u;\xi)=\frac{q_{\sfX_1|\sfX_0}(x\mid x_0)q_{U|\sfX_0,\sfX_1,S;\xi}(u\mid x_0,x_1,s;\xi)}{q_{U|\sfX_0,S;\xi}(u\mid x_0,s;\xi)} \label{retrodist}.
  \end{align}
  This induces a single-use retrochannel with random transformation $q_{\sfX_1|\sfX_0,U,S;\xi}$, which outputs a random variable $\sfX$ given the channel output $U$ and the channel state $\sfX_0$ (Fig. \ref{fig:PME}). A (generalized) single-symbol PME with state available $\sfX_0=x_0$ is defined naturally as (cf. \eqref{def1})
  \begin{align}
  \langle \sfX \big|\sfX_0=x_0 \rangle_q=\bbE_q\big[ \sfX|\sfX_0=x_0,U,S;\xi\big]\label{singPME},
  \end{align} where the expectation is taken over the (conditionally) postulated distribution in \eqref{retrodist}.
   
  The single-symbol PME \eqref{singPME} is merely a decision function applied to the Gaussian channel output with state $\sfX_0=x_0$ available at both encoder and decoder (or input and output), which can expressed explicitly as
  \begin{align}
  \bbE_q\big[\sfX|U, \sfX_0=x_0,S;\xi\big]=\frac{q_1(U,x_0,S;\xi)}{q_0(U,x_0,S;\xi)} \label{estxq},
  \end{align} where
  \begin{align}
  q_0(u,x_0,S;\xi)&:=q_{U|\sfX_0,S;\xi}(u\mid x_0,s;\xi)=\bbE\bigg[q_{U|\sfX_0,\sfX_1,S;\xi}(u\mid x_0,\sfX_1,S;\xi)\bigg|S\bigg] \label{estq0},\\
  q_1(z,x_0,S;\xi)&=\bbE\bigg[\sfX \, q_{U|\sfX_0,\sfX_1,S;\xi}(z\mid x_0,\sfX_1,S;\xi)\bigg|S\bigg] \label{estq1}.
  \end{align}

  The probability law of the (composite) single-symbol channel depicted by Fig. \ref{fig:PME} is determined by $S$ and two parameters $\eta$ and $\xi$ given state $\sfX_0$. We define the conditional mean-square error of the PME as
  \begin{align}
  \calE(S;\eta,\xi|x_0)=\bbE[(\sfX_1-\langle \sfX \, \big| \, \sfX_0\rangle_q)^2 \mid \sfX_0=x_0, S;\eta,\xi] \label{condME}
  \end{align}
  and also define the conditional variance of the retrochannel as
  \begin{align}
  \calV(S;\eta,\xi|x_0)=\bbE\big[(\sfX-\langle \sfX|\sfX_0 \rangle_q)^2 \mid \sfX_0=x_0,S;\eta,\xi\big]. \label{condMV}
  \end{align}

Define
   \begin{align}
   \calG:=\sum_{x_0 \in \calX} \lambda_{x_0}^{(\pi)}\calG(x_0) \label{defFx0},
   \end{align}
  where
  \begin{align}
  \calG(x_0)&:=-\bbE\bigg\{\int p_{U|\sfX_0,\sfX_1,S;\eta}(u\mid x_0,\sfX_1,S;\eta)\log q_{U|\sfX_0,\sfX_1,S;\xi}(u\mid x_0,S;\xi)du\bigg\}\nn\\
  &\qquad + \frac{1}{2\beta}\bigg[(\xi-1)\log e -\log \xi\bigg]-\frac{1}{2}\log \frac{2\pi}{\xi}-\frac{\xi}{2\eta}\log e\nn\\
  &\qquad + \frac{\sigma^2\xi(\eta-\xi)}{2\beta \eta}\log e+\frac{1}{2\beta}\log(2\pi)+\frac{\xi}{2\beta \eta}\log e,
  \end{align} and $\eta$ and $\xi$ is the solution of the following equation system
  \begin{align}
  \eta^{-1}&=1+\beta\sum_{x_0 \in \calX} \lambda_{x_0}^{(\pi)} \bbE[S \calE(S;\eta,\xi\mid x_0)], \label{eqeta} \\
  \xi^{-1}&=\sigma^2+ \beta \sum_{x_0 \in \calX} \lambda_{x_0}^{(\pi)} \bbE[S \calV(S;\eta,\xi\mid x_0)]
  \end{align} 
  such that they minimize $\calG$. Observe that for the case $\sfX_0,\sfX_1,\cdots, \sfX_n$ are i.i.d., $\calG(x_0)$ does not depend on $x_0$ and is defined in \cite[Eq. (22)]{GuoVerdu2005}. 
  
  \begin{claim}\label{claim:maincontri1} The free energy of the linear model with Markov sources in Section \ref{sec:setting} satisfies
  	\begin{align}
  		\calF_q=\calG, 
  	\end{align} where $\calG$ is defined in \eqref{defFx0}. In addition, the average mutual information of this model satisfies:
  	\begin{align}
  		C=\lim_{n \to \infty}\frac{1}{n}I\big(\bX^n;\bY^m|\bPhi\big)=\calF_q\bigg|_{\sigma=1}-\frac{1}{2\beta}.
  	\end{align}
  \end{claim}

 \begin{claim}\label{claim:jointmoments}
	Recall the definition of $\{\lambda_{x_0}^{(\pi)}\}_{x_0 \in \calX}$ in Section \ref{freEnergyx0}. Assume that the generalized PME defined in \eqref{def1} is used for estimation. Then, for all $k \in [n]$ and $(i_0,j_0,l_0) \in \bbZ_+ \times \bbZ_+ \times \bbZ_+$, the joint moments satisfy:
	\begin{align}
		\lim_{n\to \infty}\bbE\big[X_k^{i_0} \tilX_k^{j_0}  [X_k]_q^{l_0} \big]=\sum_{x_0 \in \calX} \lambda_{x_0}^{(\pi)} \bbE\big[\sfX_1^{i_0} \sfX^{j_0} \langle \sfX\big|\sfX_0 \rangle_q^{l_0}\big|\sfX_0=x_0\big],
	\end{align} where $(\sfX_1,\sfX_0, \sfX,\langle \sfX|\sfX_0\rangle_q)$ is the input, channel state, and outputs defined in the (composite) single-symbol PME channel in Fig. \ref{fig:PME}, and $(X_k, \tilX_k,  [X_k]_q)$ is the $k$-th symbol in the vector $\bX \in \calX^n$, the $k$-th output of the vector retrochanel defined in \eqref{ch:gePME}, and its corresponding estimated symbol by using the PME estimate in \eqref{def1}.
	
	In addition, the average MMSE satisfies:
	\begin{align}
		\frac{1}{n}\bbE\big[\|\bX-[\bX]\|_2^2\big] = \bbE\big[\sfX_1^2\big]-\sum_{x_0 \in \calX}\lambda_{x_0}^{(\pi)} \bbE\big[\langle \sfX_1|\sfX_0=x_0 \rangle^2\big],
	\end{align} where $\sfX_1 \sim \sum_{x_0 \in \calX} \pi(x_0,\cdot) \lambda_{x_0}^{(\pi)}$.
\end{claim}

  \subsection{Results for Hidden Markov Priors} \label{freEnergyx0_hmm}
  
  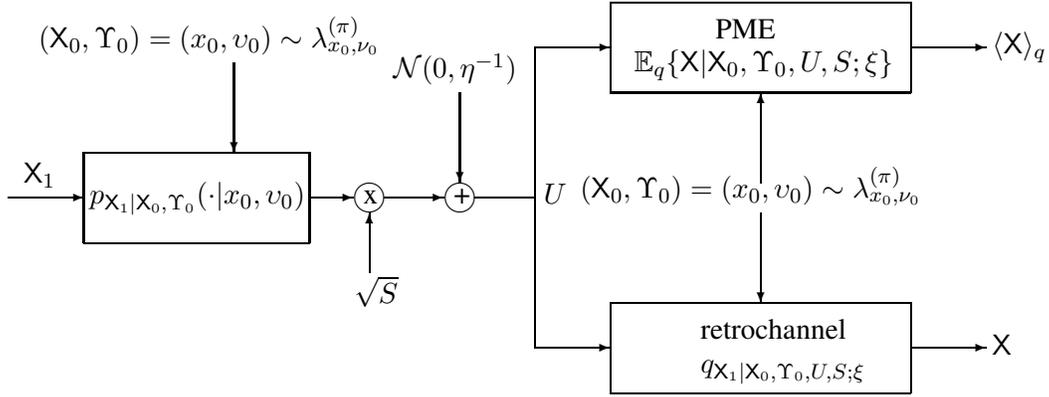
\begin{figure}[ht]
  	\centering
  	\setlength{\unitlength}{.4mm}
  	\begin{picture}(350,150)
  	\put(25,70){\line(1,0){75}}
  	\put(25,100){\line(1,0){75}}
  	\put(25,70){\line(0,1){30}}
  	\put(100,70){\line(0,1){30}}
  	\put(00,85){\vector(1,0){25}}
  	\put(120,85){\circle{10}}  	
  	\put(118,83){$\mbox{x}$}
  	\put(100,85){\vector(1,0){14.6}}
  	\put(125,85){\vector(1,0){19.6}}
  	\put(120,60){\vector(0,1){20}}
  	\put(115,50){$\sqrt{S}$}
  	\put(5,90){$\sfX_1$}
  	\put(27,83){$p_{\sfX_1|\sfX_0,\Upsilon_0}(\cdot|x_0,\upsilon_0)$}
  	\put(10,135){$(\sfX_0,\Upsilon_0)=(x_0, \upsilon_0) \sim \lambda_{x_0,\nu_0}^{(\pi)}$}
  	\put(75,130){\vector(0,-1){30}}
  	\put(150,85){\circle{10}}  
  	\put(148,83){$\mbox{+}$}
  	\put(128,125){$\calN(0,\eta^{-1})$}
  	\put(150,120){\vector(0,-1){30}}

  	\put(200,20){\line(1,0){100}}
  	\put(200,50){\line(1,0){100}}
  	\put(200,20){\line(0,1){30}}
  	\put(300,20){\line(0,1){30}}
  	\put(230,38){$\mbox{retrochannel}$}
  	\put(230,27){$q_{\sfX_1|\sfX_0,\Upsilon_0, U,S;\xi}$}

  	\put(200,120){\line(1,0){100}}
  	\put(200,150){\line(1,0){100}}
  	\put(200,120){\line(0,1){30}}
  	\put(300,120){\line(0,1){30}}
  	\put(175,85){\line(0,1){50}}
  	\put(178,83){$U$}
  	\put(155,85){\line(1,0){20}}
  	\put(175,135){\vector(1,0){25}}
  	\put(175,85){\line(0,-1){50}}
  	\put(175,35){\vector(1,0){25}}
  	\put(300,35){\vector(1,0){25}}
  	\put(327,33){$\sfX $}

  	\put(235,138){$\mbox{PME}$}
  	\put(208,127){$\bbE_q\{\sfX|\sfX_0,\Upsilon_0, U,S;\xi\}$}
  	\put(190,85){$(\sfX_0,\Upsilon_0)=(x_0, \upsilon_0) \sim \lambda_{x_0,\nu_0}^{(\pi)}$}
  	\put(250,80){\vector(0,-1){30}}
  	\put(250,90){\vector(0,1){30}}
  	\put(300,135){\vector(1,0){25}}
  	\put(327,133){$\langle \sfX|\sfX_0, \Upsilon_0 \rangle_q$}
  	
  	\end{picture}
  	\caption{The equivalent single-symbol Gaussian channel with two states available at both encoder and decoder, PME, and retrochannel.}
  	\label{fig:PMEHM}
  \end{figure}
    
  As the previous section, for the case that $\{X_n\}_{n=1}^{\infty}$ are hidden states of a Markov chain $\{\Upsilon_n\}_{n=1}^{\infty}$ on the space $\calS_{\Upsilon}$, we define a new single-symbol channel with state which is similar to the conditional PME channel defined in Section \ref{freEnergyx0}. \blue{Let $\lambda^{(\pi_{\Upsilon})}$ be the left Perron-Frobenius eigenvector with unit Manhattan norm\footnote{Since there exists a unique left Perron-Frobenius eigenvector unique up to a positive scaling factor \cite{Lancaster1985b}, so $\lambda^{(\pi_{\Upsilon})}$ exists uniquely.} of 
  \begin{align*} P_{\pi,\calX}&=\bigg\{P_{\sfX_1,\Upsilon_1| \sfX_0,\Upsilon_0}(x_1,\upsilon_1\mid x_0,\upsilon_0)\bigg\}_{(x_0,\upsilon_0) \in \calX \times \calS_{\Upsilon}, (x_1,\upsilon_1) \in \calX \times \calS_{\Upsilon}},\\
  &=\bigg\{\pi_{\Upsilon}(\upsilon_0,\upsilon_1)P_{\sfX|\Upsilon}(x_1\mid \upsilon_1)\bigg\}_{(x_0,\upsilon_0) \in \calX \times \calS_{\Upsilon}, (x_1,\upsilon_1) \in \calX \times \calS_{\Upsilon}},
  \end{align*}
  which is the stochastic matrix of the Markov chain $\{(X_n,\Upsilon_n)\}_{n=1}^{\infty}$\footnote{The fact that $\{(X_n,\Upsilon_n)\}_{n=1}^{\infty}$ forms a Markov chain will be proved in Claim \ref{HMMFreeEnergy}.}, and let $\lambda_{x_0,\upsilon_0}^{(\pi)}$ be the component of $\lambda^{(\pi)}$ associated with the $(x_0,\upsilon_0)$-th row of $P_{\pi,\calX}$. Let us consider the composition of a Gaussian channel with two states $(\sfX_0,\Upsilon_0)$ available at both encoder and decoder such that $(\sfX_0,\Upsilon_0)=(x_0,\upsilon_0)\sim \lambda_{x_0,\upsilon_0}^{(\pi)}$, a two-state PME, and a companion retrochannel in the single-symbol setting depicted in Fig. \ref{fig:PMEHM}. Given the state information $(\sfX_0,\Upsilon_0)=(x_0,\upsilon_0)$, the input-output relationship of this single-symbol channel is given by}
  \begin{align}
  U=\sqrt{S}\sfX_1+\frac{1}{\sqrt{\eta}}W \label{singleCPMEHM},
  \end{align} where the input $\sfX_1 \sim p_{\sfX_1|\sfX_0,\Upsilon_0}(\cdot|x_0,\upsilon_0)$ such that
  \begin{align}
  p_{\sfX_1|\sfX_0,\Upsilon_0}(x_1\mid x_0,\upsilon_0)&=\sum_{\upsilon_1 \in \calS_{\Upsilon}} p_{\sfX_1,\Upsilon_1|\sfX_0,\Upsilon_0}(x_1,\upsilon_1\mid x_0,\upsilon_0)\\
  &=\sum_{\upsilon_1 \in \calS_{\Upsilon}}\pi_{\Upsilon}(\upsilon_0,\upsilon_1)p_{\sfX|\Upsilon}(x_1\mid\upsilon_1),
  \end{align}
  $S \sim P_S$ which is independent $\sfX_0, \Upsilon_0$ and $\sfX_1, \Upsilon_1$, $W \sim \calN(0,1)$ the noise independent of $\sfX_0,\Upsilon_0$ and $\sfX_1,\Upsilon_1$, and $\eta>0$ the inverse noise variance. The conditional distribution associated with the channel is
  \begin{align}
  p_{U|\sfX_0,\Upsilon_0, \sfX_1,S;\eta}(u\mid x_0,\upsilon_0, x_1,s;\eta)=\sqrt{\frac{\eta}{2\pi}} \exp\bigg[-\frac{\eta}{2}(u-\sqrt{s}x_1)^2\bigg] \label{pertainHM}.
  \end{align}
  Let $q_{U|\sfX_0,\Upsilon_0, \sfX_1,S;\xi}$ represent Gaussian channel with two states $\sfX_0$ and $\Upsilon_0$ available at both encoder and decoder akin to \eqref{singleCPMEHM}, the only difference being that the inverse noise variance is $\xi$ instead of $\eta$
  \begin{align}
  q_{U|\sfX_0,\Upsilon_0, \sfX_1,S;\xi}(u\mid x_0,\upsilon_0, x_1,s;\eta,x_0)=\sqrt{\frac{\xi}{2\pi}} \exp\bigg[-\frac{\xi}{2}(u-\sqrt{s}x_1)^2\bigg].
  \end{align}
  Similar to that in the vector channel setting, by postulating the input distribution to be $q_{\Upsilon_1|\Upsilon_0}(\cdot|\upsilon_0)=\tilde{\pi}_{\Upsilon}(\upsilon_0,\cdot)$, a posterior probability distribution $q_{\sfX_1|\sfX_0,\Upsilon_0, U,S;\xi}$ is induced by $q_{\sfX_1|\sfX_0,\Upsilon_0}$ and $q_{U|\sfX_0,\Upsilon_0, \sfX_1,S;\xi}$ using the Bayes rule, i.e.,
  \begin{align}
  q_{\sfX_1|\sfX_0,\Upsilon_0, S,U;\xi}(x\mid x_0,\upsilon_0, s,u;\xi)=\frac{q_{\sfX_1|\sfX_0,\Upsilon_0}(x\mid x_0,\upsilon_0)q_{U|\sfX_0,\Upsilon_0, \sfX_1,S;\xi}(u\mid x_0,\upsilon_0, x_1,s;\xi)}{q_{U|\sfX_0,\Upsilon_0, S;\xi}(u\mid x_0,\upsilon_0,s;\xi)} \label{retrodistHM}.
  \end{align}
  This induces a single-use retrochannel with random transformation $q_{\sfX_1|\sfX_0,\Upsilon_0, U,S;\xi}$, which outputs a random variable $\sfX$ given the channel output $U$ and the channel states $\sfX_0, \Upsilon_0$ (Fig. \ref{fig:PMEHM}). A (generalized) single-symbol PME with two available  states $\sfX_0=x_0$ and $\Upsilon_0=\upsilon_0$ is defined naturally as (cf. \eqref{singPME})
  \begin{align}
  \langle \sfX \big|\sfX_0=x_0,\Upsilon=\upsilon_0 \rangle_q=\bbE_q\big[\sfX|\sfX_0=x_0,\Upsilon_0=\upsilon_0, U,S;\xi\big]\label{singPMEHM},
  \end{align} where the expectation is taken over the (conditionally) postulated distribution in \eqref{retrodist}.
  
  The single-symbol PME \eqref{singPME} is merely a decision function applied to the Gaussian channel output with two states $\sfX_0=x_0$ and $\Upsilon_0=\upsilon_0$ available at both encoder and decoder (or input and output), which can expressed explicitly as
  \begin{align}
  \bbE_q\big[\sfX|U,\sfX_0=x_0,\Upsilon_0=\upsilon_0, S;\xi\big]=\frac{q_1(U,x_0,\upsilon_0, S;\xi)}{q_0(U,x_0,\upsilon_0, S;\xi)} \label{estxqHM},
  \end{align} where
  \begin{align}
  q_0(u,x_0,\upsilon_0, S;\xi)&:=q_{U|\sfX_0,\Upsilon_0, S;\xi}(u\mid x_0,\upsilon_0, S;\xi)=\bbE\big[q_{U|\sfX_0,\Upsilon_0, \sfX_1,S;\xi}(u\mid x_0,\upsilon_0, \sfX_1,S;\xi)\big|S\big] \label{estq0HM},\\
  q_1(u,x_0,\upsilon_0, S;\xi)&=\bbE\big[\sfX q_{U|\sfX_0,\Upsilon_0, \sfX_1,S;\xi}(u\mid x_0,\upsilon_0, \sfX_1,S;\xi)\big|S\big] \label{estq1HM}.
  \end{align}
  
  The probability law of the (composite) single-symbol channel depicted by Fig. \ref{fig:PMEHM} is determined by $S$ and two parameters $\eta$ and $\xi$ given states $\sfX_0$ and $\Upsilon_0$. We define the conditional mean-square error of the PME as
  \begin{align}
  \calE(S;\eta,\xi|x_0,\upsilon_0)=\bbE[(\sfX_1-\langle \sfX\big|\sfX_0,\Upsilon_0 \rangle_q)^2|\sfX_0=x_0,\Upsilon_0=\upsilon_0, S;\eta,\xi] \label{condMEHM}
  \end{align}
  and also define the conditional variance of the retrochannel as
  \begin{align}
  \calV(S;\eta,\xi|x_0,\upsilon_0)=\bbE\big[(\sfX-\langle \sfX|\sfX_0, \Upsilon_0\rangle_q)^2|\sfX_0=x_0,\Upsilon_0=\upsilon_0, S;\eta,\xi\bigg]. \label{condMVHM}
  \end{align}
  
 Define
  \begin{align}
  \tilG:=\sum_{(x_0,\upsilon_0) \in \calX \times \calS_{\Upsilon}}\lambda_{x_0,\upsilon_0}^{(\pi_{\Upsilon})}\tilG(x_0,\upsilon_0),\label{defFx0HM}
  \end{align} where
  \begin{align}
  \tilG(x_0,\upsilon_0)&:=-\bbE\bigg\{\int p_{U|\sfX_0,\Upsilon_0, S;\eta}(u\mid x_0,\upsilon_0, S;\eta)\log q_{U|\sfX_0,\Upsilon_0, S;\xi}(u\mid x_0,\upsilon_0, S;\xi)du\bigg\}\nn\\
  &\qquad + \frac{1}{2\beta}\bigg[(\xi-1)\log e -\log \xi\bigg]-\frac{1}{2}\log \frac{2\pi}{\xi}-\frac{\xi}{2\eta}\log e\nn\\
  &\qquad + \frac{\sigma^2\xi(\eta-\xi)}{2\beta \eta}\log e+\frac{1}{2\beta}\log(2\pi)+\frac{\xi}{2\beta \eta}\log e \label{deftilG},
  \end{align} and $\eta$ and $\xi$ is the solution of the following equation system
  \begin{align}
  \eta^{-1}&=1+\beta\sum_{(x_0,\upsilon_0) \in \calX \times \calS_{\Upsilon}}\lambda_{x_0,\upsilon_0}^{(\pi_{\Upsilon})}\bbE[S \calE(S;\eta,\xi\mid x_0,\upsilon_0)], \label{eqetaHM} \\
  \xi^{-1}&=\sigma^2+ \beta \sum_{(x_0,\upsilon_0) \in \calX \times \calS_{\Upsilon}}\lambda_{x_0,\upsilon_0}^{(\pi_{\Upsilon})}\bbE[S \calV(S;\eta,\xi \mid x_0,\upsilon_0)]
  \end{align} such that they minimize $\tilG$. 
    
 \begin{claim} \label{claim:hiddenMarkov} Assume that $\{X_n\}_{n=1}^{\infty}$ is the hidden states (outputs) of a hidden Markov model generated by a Markov chain $\{\Upsilon_n\}_{n=1}^{\infty}$ with transition probability (function) $\pi_{\Upsilon}(\cdot,\cdot)$ on some Polish space $\calS_{\Upsilon}$, i.e.,
	\begin{itemize}
		\item $\Upsilon_n$ is a Markov process and is not directly observable.
		\item $\bbP(X_n \in \calA|\Upsilon_1=\upsilon_1, \Upsilon_2=\upsilon_2,\cdots, \Upsilon_n=\upsilon_n)=\bbP(X_n\in \calA|\Upsilon_n=\upsilon_n)=P_{\sfX|\Upsilon}(\calA|\upsilon_n)$,
	\end{itemize} for every $n\geq 1$, $\upsilon_1,\upsilon_2,\cdots, \upsilon_n$, and an arbitrary measurable set $\calA$, where $P_{\sfX|\Upsilon}(\cdot|\cdot)$ is some probability measure called emission probability. Then, the following holds:
	\begin{itemize}
		\item $\{X_n,\Upsilon_n\}_{n=1}^{\infty}$ forms a Markov chain on $\calX \times \calS_{\Upsilon}$ with transition probability $P_{\sfX_1,\Upsilon_1|\sfX_0,\Upsilon_0}(x_1,\upsilon_1|x_0,\upsilon_0)=P_{\sfX|\Upsilon}(x_1|\upsilon_1) \pi_{\Upsilon}(\upsilon_0,\upsilon_1)$.
		\item Recall the definitions of  $\{\lambda_{x_0,\upsilon_0}^{(\pi_{\Upsilon})}\}_{(x_0,\upsilon_0) \in \calX \times \calS_{\Upsilon}}$ and  $\tilG$ in \eqref{deftilG}. Then, the free energy, mutual information, joint moments, the average MMSE of the linear model with hidden Markov sources in \ref{sec:setting} satisfy:
		\begin{align}
		\calF_q&= \tilG, \label{eq246HMmod} \\
		&C=\calF_q\bigg|_{\sigma=1}- \frac{1}{2\beta}, \label{eq247HMmod} \\ 
		&\lim_{n\to \infty}\bbE\big[X_k^{i_0} \tilX_k^{j_0} [X_k]_q^{l_0} \big] = \sum_{x_0, \upsilon_0 \in \calX \times \calS_{\Upsilon}}\lambda_{x_0,\upsilon_0}^{(\pi_{\Upsilon})} \bbE\big[\sfX_1^{i_0} \sfX^{j_0} \langle \sfX\big|\sfX_0,\Upsilon_0 \rangle_q^{l_0}\big|\sfX_0=x_0,\Upsilon_0=\upsilon_0\big],\\
		&\qquad \qquad \qquad \qquad \qquad \qquad   \forall i_0,j_0,l_0 \in \bbZ_+ \label{jointmomentHMmod},\\
		&\lim_{n \to \infty} \frac{1}{n}\bbE[\|\bX-[\bX]\|_2^2]= \bbE[\sfX_1^2]-\sum_{x_0,\upsilon_0 \in \calX_1 \times \calS_{\gamma}} \lambda_{x_0,\upsilon_0}^{(\pi_{\Upsilon})}\bbE[\langle \sfX|\sfX_0=x_0,\Upsilon_0=\upsilon_0\rangle^2 ] \label{uplowMMSEHMmod}, 
		\end{align}  where $(\sfX_1,\sfX,\langle \sfX|\sfX_0=x_0,\Upsilon_0=\upsilon_0\rangle_q)$ is the input and outputs defined in the (composite) single-symbol PME channel in Fig. \ref{fig:PMEHM}, and $(X_k, \tilX_k, [X_k]_q)$ is the $k$-th symbol in the vector $\bX \in \calX^n$, the $k$-th output of the vector retrochanel defined in \eqref{ch:gePME}, and its corresponding estimated symbol by using the generalized PME estimate in \eqref{def1}. In addition, in \eqref{uplowMMSEHMmod}, $\sfX_1 \sim \sum_{\upsilon \in \calS_{\Upsilon}} P_{\sfX|\Upsilon}(\cdot|\upsilon)\pi_{\Upsilon}(\upsilon_0,\upsilon)$, where $P_{\sfX|\Upsilon}$ is the stationary emission probability of the hidden Markov process.
	\end{itemize}
\end{claim}

  \section{Numerical examples and comparison with algorithmic performance}\label{sec:numerical}
  
  \subsection{Binary-valued Markov Prior}\label{sub:BVMP}
Assume that $\bX$ is a homogeneous Markov chain on the alphabet $\calX=\{-1,1\}$ with the stochastic matrix as follows:
\begin{align}
P_{\pi}=\begin{bmatrix} \pi(-1,-1)& \pi(-1,1)\\ \pi(1,-1)& \pi(1,1)\end{bmatrix}&=\begin{bmatrix} 1-\alpha & \alpha \\ \delta& 1-\delta \end{bmatrix} \label{exmarkov} 
\end{align} for some $\alpha$ and $\delta$ in $(0,1)$. 
\subsubsection{Free Energy and Average Mutual Information} \label{sec:spec1}
It is easy to see that the left Perron-Frobenius eigenvector $\lambda^{(\pi)}$ of $P_{\pi}$ defined in Subsection \ref{freEnergyx0} is
\begin{align}
\lambda^{(\pi)}=\bigg(\frac{\delta}{\alpha+\delta},\frac{\alpha}{\alpha+\delta}\bigg)^T \label{PF1example}. 
\end{align}

First, we estimate $\calG(-1)$ as a function of $\alpha$. We assume that all postulated distributions are the same as their true ones for simplicity. We also assume that $S=1$ with probability $1$. Now from \eqref{estq0}, we have
\begin{align}
q_0(u,-1,1;\eta)&=\bbE_{\pi(-1,\cdot)}\bigg[q_{U|\sfX_0,\sfX_1,S;\xi}(u\mid-1,\sfX,1;\xi)|S=1\bigg]\\
&=\bbE_{\pi(-1,\cdot)}\bigg[\sqrt{\frac{\eta}{2\pi}} \exp\bigg[-\frac{\eta}{2}(u-\sfX)^2\bigg]\bigg]\\
&=\sum_{x \in \calX}\sqrt{\frac{\eta}{2\pi}} \exp\bigg[-\frac{\eta}{2}(u-x)^2\bigg]\pi(-1,x) \\
&=(1-\alpha) \sqrt{\frac{\eta}{2\pi}} \exp\bigg[-\frac{\eta}{2}(u+1)^2\bigg]+\alpha \sqrt{\frac{\eta}{2\pi}} \exp\bigg[-\frac{\eta}{2}(u-1)^2\bigg] \label{estq0est}.
\end{align}
Similarly, from \eqref{estq1}, we also have
\begin{align}
q_1(u,-1,1;\eta)&=\bbE_{\pi(-1,\cdot)}\bigg[Xq_{U|\sfX_0,\sfX_1,S;\xi}(u\mid -1,\sfX,1;\xi)|S=1\bigg]\\
&=\bbE_{\pi(-1,\cdot)}\bigg[\sfX\sqrt{\frac{\eta}{2\pi}} \exp\bigg[-\frac{\eta}{2}(u-\sfX)^2\bigg]\bigg]\\
&=\sum_{x \in \calX}x\sqrt{\frac{\eta}{2\pi}} \exp\bigg[-\frac{\eta}{2}(u-x)^2\bigg]\pi(-1,x) \\
&=-(1-\alpha) \sqrt{\frac{\eta}{2\pi}} \exp\bigg[-\frac{\eta}{2}(u+1)^2\bigg]+\alpha \sqrt{\frac{\eta}{2\pi}} \exp\bigg[-\frac{\eta}{2}(u-1)^2\bigg]  \label{estq1est}.
\end{align}
Therefore, from \eqref{singPME}, \eqref{estxq}, \eqref{estq0est}, and \eqref{estq1est}, we have
\begin{align}
\langle \sfX|\sfX_0=-1\rangle_q&=\bbE_{q}\bigg[\sfX\big|\sfX_0=-1,U,1;\eta\bigg]\\
&=\frac{q_1(U,-1,1;\eta)}{q_0(U,-1,1;\eta)}\\
&=\frac{1-\big(\frac{1-\alpha}{\alpha}\big)\exp(-2\eta U)}{1+\big(\frac{1-\alpha}{\alpha}\big)\exp(-2\eta U)}.
\end{align} 
It follows from \eqref{condME} and \eqref{condMV} that
\begin{align}
\calV(1;\eta,\eta|-1)&=\calE(1;\eta,\eta|-1)\\
&=\bbE\bigg[\big(\sfX_1-\langle \sfX |\sfX_0=-1 \rangle_q\big)^2|\sfX_0=-1, 1;\eta,\eta\bigg]\\
&=\bbE_U\bigg[\bbE\bigg[\sfX_1^2\big|\sfX_0=-1, 1;\eta,\eta\bigg]-\langle \sfX |\sfX_0=-1 \rangle_q^2\bigg] \\
&=\sum_{x \in \calX} x^2 \pi(-1,x)-\bbE_U\bigg[\bigg(\frac{1-\big(\frac{1-\alpha}{\alpha}\big)\exp(-2\eta U)}{1+\big(\frac{1-\alpha}{\alpha}\big)\exp(-2\eta U)}\bigg)^2\bigg]\\
&=1-\bbE_U\bigg[\bigg(\frac{1-\big(\frac{1-\alpha}{\alpha}\big)\exp(-2\eta U)}{1+\big(\frac{1-\alpha}{\alpha}\big)\exp(-2\eta U)}\bigg)^2\bigg] \label{eq242eq}.
\end{align}
Similarly, we have
\begin{align}
\calV(1;\eta,\eta|1)&=\calE(1;\eta,\eta|1)\\
&=1-\bbE_U\bigg[\bigg(\frac{1-\big(\frac{\delta}{1-\delta}\big)\exp(-2\eta U)}{1+\big(\frac{\delta}{1-\delta}\big)\exp(-2\eta U)}\bigg)^2\bigg] \label{eq242eqb}.
\end{align}
Now, by \eqref{singPME}, the single-symbol PME for this special case is
\begin{align}
U=\sfX_1+\frac{1}{\sqrt{\eta}}W. 
\end{align} 
Now, observe that
\begin{itemize}
	\item Under the condition that $\sfX_1 \sim \pi(-1,\cdot)$, we have
	\begin{align}
	F_U(u)&=\bbP(U \leq u)\\
	&=\bbP(U \leq u|\sfX_1=-1)\pi(-1,-1)+\bbP(U \leq u|\sfX_1=1)\pi(-1,1)\\
	&=(1-\alpha) \bbP(W \leq (u+1)\sqrt{\eta} )+\alpha \bbP(W \leq (u-1)\sqrt{\eta})\\
	&=(1-\alpha)\Phi((u+1) \sqrt{\eta})+ \alpha \Phi((u-1) \sqrt{\eta}) \label{cupet1}
	\end{align}  where $\Phi(x):=\frac{1}{\sqrt{2\pi}}\int_{-\infty}^x \exp(-t^2/2)dt$. From \eqref{cupet1}, we obtain\footnote{We can derive it use the convolution since $W$ and $X_1$ are independent random variables.}
	\begin{align}
U=u \sim 	f_U^{(\rm{MS},1)}(u)&=\frac{1}{\sqrt{2\pi}}(1-\alpha)\sqrt{\eta} \exp\bigg(-\frac{(u+1)^2 \eta}{2}\bigg)+ \frac{1}{\sqrt{2\pi}}\alpha \sqrt{\eta} \exp\bigg(-\frac{(u-1)^2 \eta}{2}\bigg) \label{deffz}.
	\end{align}
	\item Similarly, under the condition $\sfX_1 \sim \pi(1,\cdot)$, we have
	\begin{align}
U=u \sim 	f_U^{(\rm{MS},2)}(u)=\frac{1}{\sqrt{2\pi}}\delta \sqrt{\eta} \exp\bigg(-\frac{(u+1)^2 \eta}{2}\bigg)+ \frac{1}{\sqrt{2\pi}}(1-\delta) \sqrt{\eta} \exp\bigg(-\frac{(u-1)^2 \eta}{2}\bigg) \label{deffzcMS}.
	\end{align}
\end{itemize}
Hence, Eq.~\eqref{eqeta} in Claim \ref{claim:maincontri1}, $\eta$ is a solution of the following equation
\begin{align}
\eta^{-1}&=1+\beta\bigg(\frac{\delta}{\delta+\alpha}\bbE[1 \calE(1;\eta,\eta|-1)]+\frac{\alpha}{\delta+\alpha}\bbE[1 \calE(1;\eta,\eta|1)]\bigg),\\
&=1+ \beta \bigg(\frac{\delta}{\delta+\alpha}\bbE_{f_U^{(\rm{MS},1)}}\bigg[1-\bigg(\frac{1-\big(\frac{1-\alpha}{\alpha}\big)\exp(-2\eta U)}{1+\big(\frac{1-\alpha}{\alpha}\big)\exp(-2\eta U)}\bigg)^2\bigg]\nn\\
&\qquad + \frac{\alpha}{\delta+\alpha}\bbE_{f_U^{(\rm{MS},2)}}\bigg[1-\bigg(\frac{1-\big(\frac{\delta}{1-\delta}\big)\exp(-2\eta U)}{1+\big(\frac{\delta}{1-\delta}\big)\exp(-2\eta U)}\bigg)^2\bigg] \bigg)\label{drawbeta}.
\end{align}

Now, since $S=1, \xi=\eta, \sigma=1, \tilde{\pi}=\pi$, from \eqref{defFx0}, we have
\begin{align}
\calG(-1)\bigg|_{\sigma=1,\tilde{\pi}=\pi}&=-\int p_{U|\sfX_0,S;\eta}(u\mid -1,1;\eta)\log p_{U|\sfX_0,S;\eta}(u\mid -1,1;\eta)du \nn\\
&\qquad + \frac{1}{2\beta}\bigg[(\eta-1)\log e -\log \eta \bigg]-\frac{1}{2}\log \frac{2\pi}{\eta}-\frac{1}{2}\log e\nn\\
&\qquad +\frac{1}{2\beta}\log(2\pi)+\frac{1}{2\beta}\log e.
\end{align}
On the other hand, 
\begin{align}
p_{U|\sfX_0,1;\eta}(u\mid -1,1;\eta)&=\sum_{x \in \calX}p_{U|\sfX_0,\sfX_1, 1;\eta}(u\mid -1,x,1;\eta)\pi(-1,x) \\
&=\sum_{x \in \calX}\sqrt{\frac{\eta}{2\pi}} \exp\bigg[-\frac{\eta}{2}(u-x)^2\bigg]\pi(-1,x)\\
&=(1-\alpha)\sqrt{\frac{\eta}{2\pi}}\exp\bigg[-\frac{\eta}{2}(u+1)^2\bigg]+ \alpha \sqrt{\frac{\eta}{2\pi}}\exp\bigg[-\frac{\eta}{2}(u-1)^2\bigg]. 
\end{align}
It follows that
\begin{align}
\calG(-1)\bigg|_{S=1,\sigma=1,\tilde{\pi}=\pi}= \barG(-1,\eta,\alpha) \label{optim1},
\end{align} where
\begin{align}
\barG(-1,\eta,\alpha)&:=-\int_{-\infty}^{\infty}\bigg((1-\alpha)\sqrt{\frac{\eta}{2\pi}}\exp\bigg[-\frac{\eta}{2}(u+1)^2\bigg]+ \alpha \sqrt{\frac{\eta}{2\pi}}\exp\bigg[-\frac{\eta}{2}(u-1)^2\bigg]\bigg)\nn\\
&\qquad \times  \log\bigg((1-\alpha)\sqrt{\frac{\eta}{2\pi}}\exp\bigg[-\frac{\eta}{2}(u+1)^2\bigg]+ \alpha \sqrt{\frac{\eta}{2\pi}}\exp\bigg[-\frac{\eta}{2}(u-1)^2\bigg]\bigg)du\nn\\
&\qquad + \frac{1}{2\beta}\bigg[(\eta-1)\log e -\log \eta \bigg]-\frac{1}{2}\log \frac{2\pi}{\eta}-\frac{1}{2}\log e\nn\\
&\qquad +\frac{1}{2\beta}\log(2\pi)+\frac{1}{2\beta}\log e.
\end{align} 

\begin{figure}[ht]
	\centering
	\includegraphics[width=0.7\linewidth]{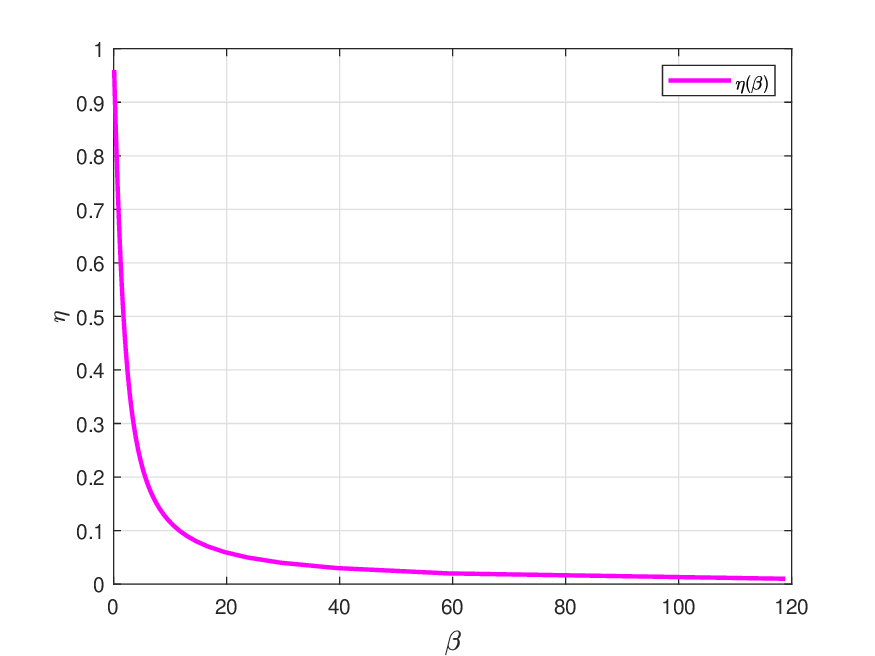}
	\caption{$\eta$ as a decreasing function in $\beta$ for the symmetric case $\alpha=\delta=0.3$.}
	\label{fig:DMK1}
\end{figure}

By the symmetry, it is not hard to see that
\begin{align}
\calG(1)\bigg|_{S=1,\sigma=1,\tilde{\pi}=\pi}=\barG(1,\eta,\delta) \label{optim2},
\end{align} where
\begin{align}
\barG(1,\eta,\delta)&:=-\int_{-\infty}^{\infty}\bigg(\delta \sqrt{\frac{\eta}{2\pi}}\exp\bigg[-\frac{\eta}{2}(u+1)^2\bigg]+ (1-\delta) \sqrt{\frac{\eta}{2\pi}}\exp\bigg[-\frac{\eta}{2}(u-1)^2\bigg]\bigg)\nn\\
&\qquad \times  \log\bigg(\delta \sqrt{\frac{\eta}{2\pi}}\exp\bigg[-\frac{\eta}{2}(u+1)^2\bigg]+ (1-\delta) \sqrt{\frac{\eta}{2\pi}}\exp\bigg[-\frac{\eta}{2}(u-1)^2\bigg]\bigg)du\nn\\
&\qquad + \frac{1}{2\beta}\bigg[(\eta-1)\log e -\log \eta \bigg]-\frac{1}{2}\log \frac{2\pi}{\eta}-\frac{1}{2}\log e\nn\\
&\qquad +\frac{1}{2\beta}\log(2\pi)+\frac{1}{2\beta}\log e.
\end{align} 

Now, let $\calC_{\beta}(\alpha,\delta)$ is the set of all solutions $\eta$ of the equation \eqref{drawbeta} given $\beta$ and $\alpha$ and $\delta$. Then, by Claim \ref{claim:maincontri1} and \eqref{PF1example}, the free energy can be expressed as
\begin{align}
\calF\bigg|_{S=1,\sigma=1,\tilde{\pi}=\pi}= \min_{\eta \in \calC_{\beta}(\alpha,\delta)}\bigg[\frac{\delta}{\alpha+\delta} \calG(-1) + \frac{\alpha}{\alpha+\delta} \calG(1)\bigg] \label{FEoptm},
\end{align} where $\calG(-1)$ and $\calG(1)$ are given in \eqref{optim1} and \eqref{optim2}, respectively.

\begin{figure}[ht]
	\centering
	\includegraphics[width=0.7\linewidth]{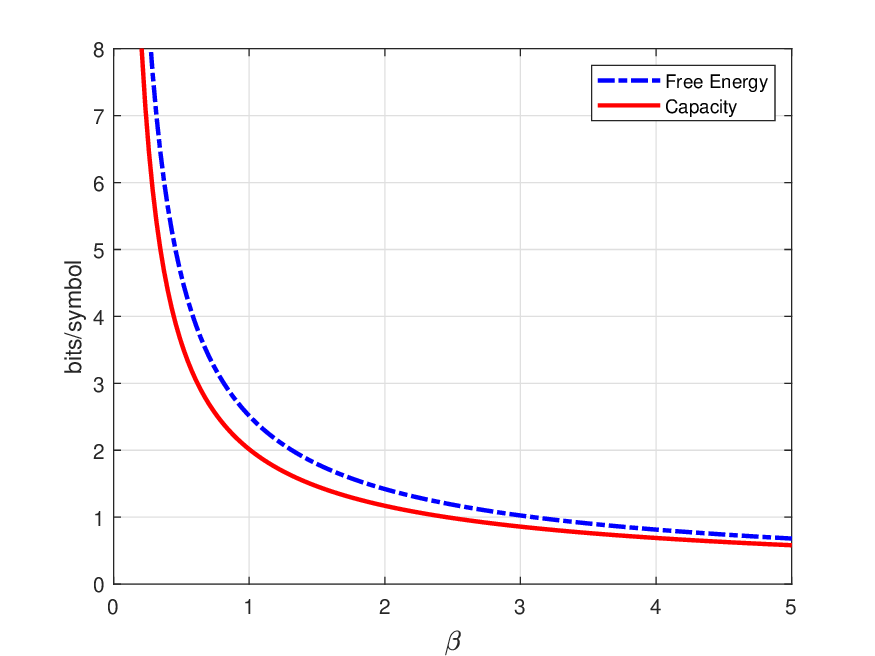}
	\caption{Free energy and average mutual information as functions of $\beta$ for the symmetric case $\alpha=\delta=0.3$.}
	\label{fig:DMK2}
\end{figure}
Solving the optimization problem in \eqref{FEoptm} is very challenging since $C_{\beta}(\alpha,\delta)$ may have more than one elements, which corresponds to multiple fixed points of the optimization problem in \eqref{FEoptm}. However, by observing that given $\alpha$ and $\delta$, then $\beta$, $\barG(-1,\eta,\alpha)$ and $\barG(1,\eta,\alpha)$ are functions of $\eta$. The multiple fixed-points happen if there exists at least two different values $\eta_1$ and $\eta_2$ such that $\beta(\eta_1)=\beta(\eta_2)$. In simulations, for a fixed $\beta$, we can estimate all the values of $\eta$ such that $|\beta(\eta)-\beta|<10^{-3}$ and then estimate the free energy as a functions of $\eta$ and find the minimum value among them as the free energy corresponding to $\beta(\eta)$. This procedure can avoid the multiple fixed-point problem.

For the symmetric case $\delta=\alpha=0.3$, $\beta$ is a monotone function in $\eta$ (cf. Fig.~\ref{fig:DMK1}), hence $\calC_{\beta}(\alpha,\delta)$ in \eqref{FEoptm} contains only one element $\eta$, at which we achieve the average mutual information and free energy. For example, in Fig. \ref{fig:DMK2}, we plot the free energy and the average mutual information for symmetric case $\delta=\alpha=0.3$. 

\begin{figure}[ht]
	\centering
	\includegraphics[width=0.7\linewidth]{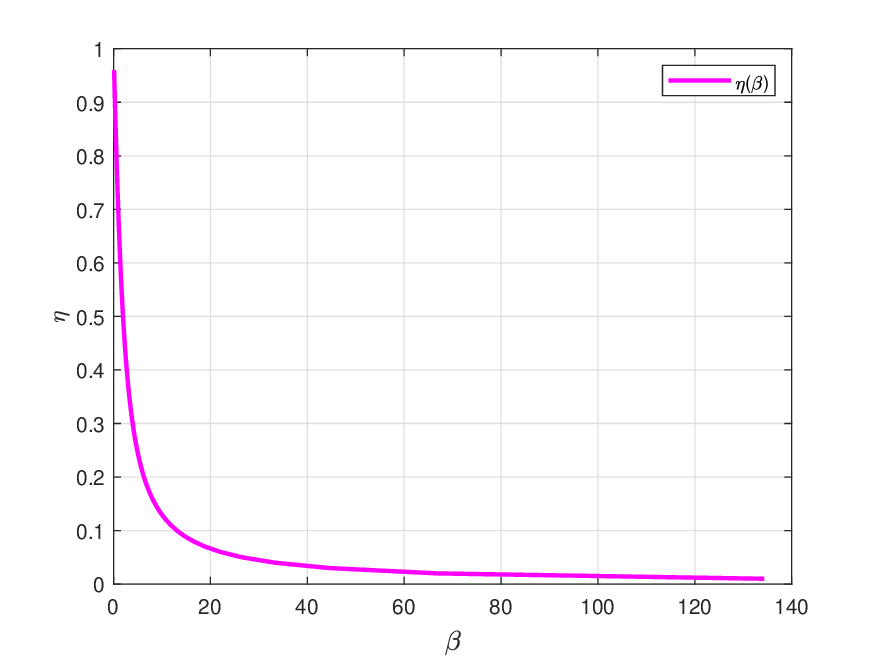}
	\caption{$\eta$ as a decreasing function in $\beta$ for the non-symmetric case $\alpha=0.2$ and $\delta=0.5$.}
	\label{fig:DMK1b}
\end{figure}

\begin{figure}[H]
	\centering
	\includegraphics[width=0.7\linewidth]{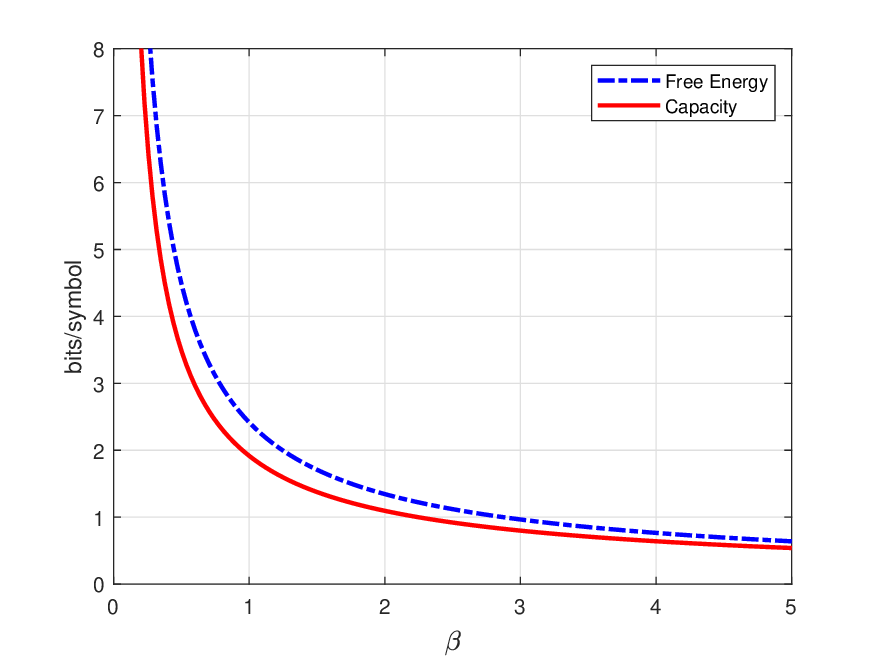}
	\caption{Free energy and average mutual information as functions of $\beta$ for the non-symmetric case $\alpha=0.2$ and $\delta=0.5$.}
	\label{fig:DMK3}
\end{figure}

For the non-symmetric case $\alpha=0.2$ and $\delta=0.5$, $\beta$ is also a monotone function in $\eta$ (cf. Fig. \ref{fig:DMK1b}), so $\calC_{\beta}(\alpha,\delta)$ contains only one element. Then, we obtain the free energy and the average mutual information for this case as in Fig. \ref{fig:DMK3}. 

\subsubsection{Markov Chain Monte Carlo (MCMC) vs. Replica Prediction} \label{sec:bemcmc}
\begin{figure}[H]
	\centering
	\includegraphics[width=0.7\linewidth]{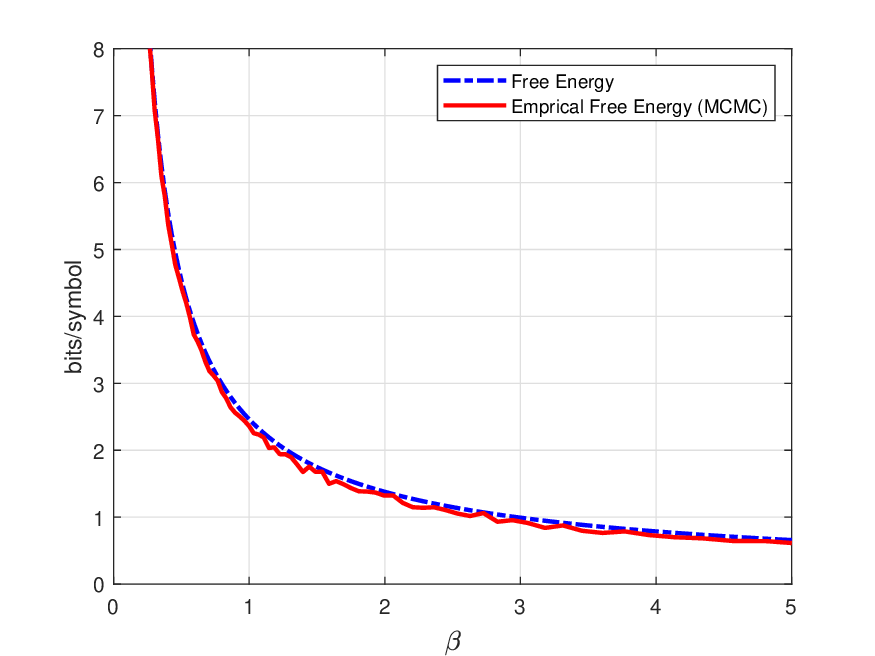}
	\caption{Free energy by Replica Method and MCMC as functions for the i.i.d. prior $\alpha=\delta=0.5$.}
	\label{fig:DMK1bc}
\end{figure}

\begin{figure}[H]
	\centering
	\includegraphics[width=0.7\linewidth]{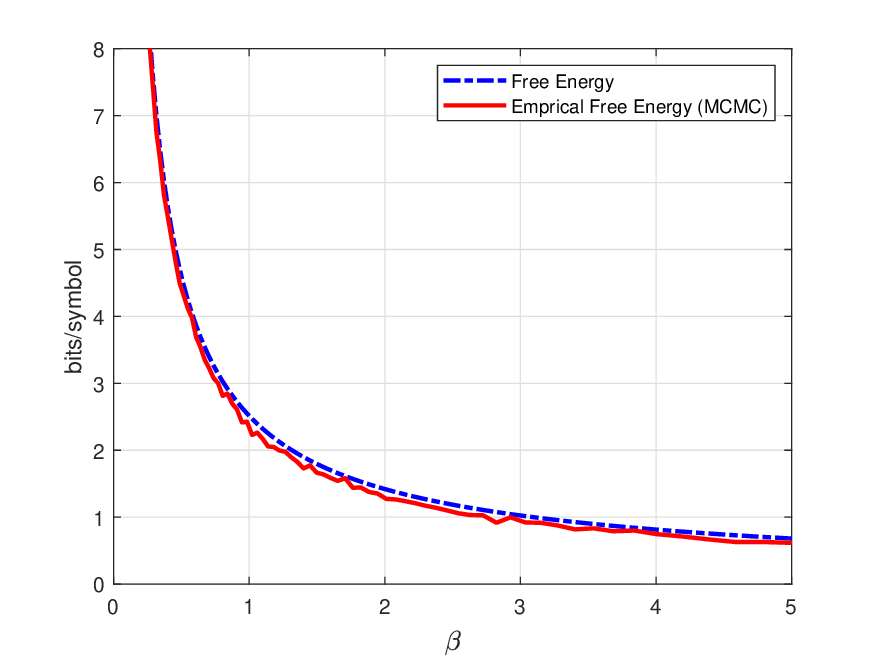}
	\caption{Free energy by Replica Method and MCMC as functions of $\beta$ for the symmetric case $\alpha=\delta=0.3$.}
	\label{fig:DMK2bc}
\end{figure}

\begin{figure}[H]
	\centering
	\includegraphics[width=0.7\linewidth]{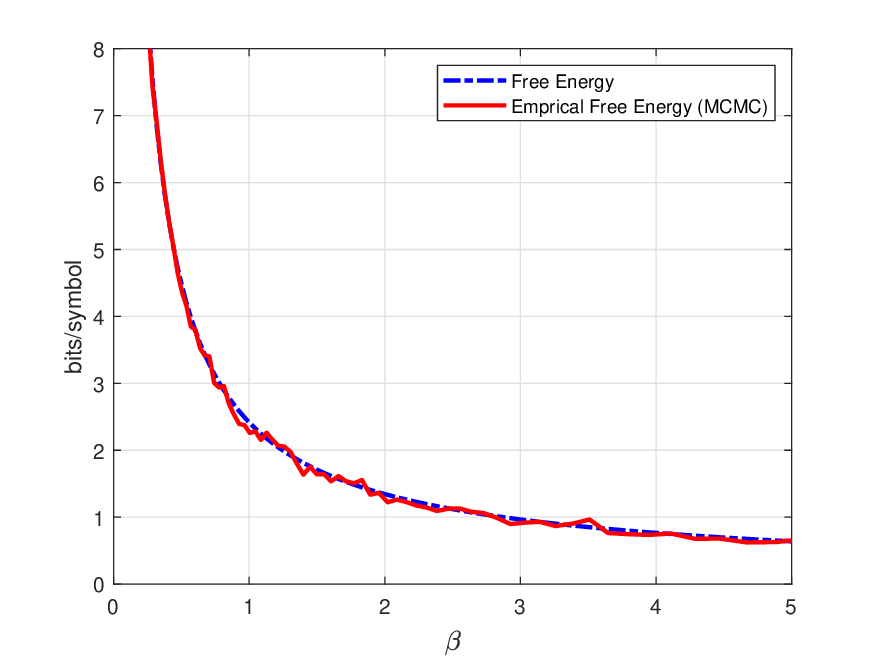}
	\caption{Free energy by Replica Method and MCMC as functions of $\beta$ for the non-symmetric case $\alpha=0.2$ and $\delta=0.5$.}
	\label{fig:DMK2b}
\end{figure}

\blue{In this subsection, we use the Markov Chain Monte-Carlo (MCMC) simulation method to estimate the density function $\bP_{\by|\bPhi}(\by|\bPhi)$ and verify our replica predictions in  Claims \ref{claim:maincontri1} and \ref{claim:jointmoments}. More specifically, we compare the free energies achieved by the replica prediction and MCMC for the linear model with binary-valued Markov prior defined in \eqref{exmarkov}. Our simulation shows that the free energy curves by the replica method and MCMC nearly coincide to each other for all three cases: (1) i.i.d. prior ($\alpha=\delta=0.5$), (2) symmetric Markov prior $\alpha=\delta=0.3$, (3) asymmetric Markov prior $(\alpha=0.2, \delta=0.5)$ (cf. Figs. \ref{fig:DMK1bc}, \ref{fig:DMK2bc}, and \ref{fig:DMK2b}). In those simulations, the Metropolis–Hastings algorithm is used where the state $\bx_t:=\rm{vec}(\Phi_{t+1},\by_{t+1})$ and the probability transition $g(\bx_{t+1}|\bx_t)\sim \calN(\bx_t, \bI_{mn+n})$. Our simulation results show that the replica prediction in Claim \ref{claim:maincontri1} for free energy is very closed to MCMC result.}
\blue{Since MMSE is fixed function of the free energy (or mutual information) \cite{GuoShamaiVerdu2005}, these simulation results also indicate that our replica prediction for MMSE in Claim \ref{claim:jointmoments} closely approximates the MMSE of the model.}
\subsection{Gauss-Markov Prior}
We consider a Gauss-Markov prior $\{X_n\}_{n=1}^{\infty}$ on $\calX=\bbR$, i.e.,  $X_n=\nu X_{n-1}+Z_n$, where $Z_n \sim \calN(0,\sigma_0^2)$ and $\nu \in (0,1)$. Then, the transition probability is
\begin{align}
\pi(x_0,x):=\frac{1}{\sigma_0 \sqrt{2\pi}} \exp\bigg[-\frac{1}{2\sigma_0^2}(x-\nu x_0)^2\bigg]\label{markv99}.
\end{align} 
This means that $X_n|X_{n-1}=x_0 \sim \calN(\nu x_0,\sigma_0^2)$ for all $n\in \bbZ_+$. \blue{This is not hard to show that the Markov chain in \eqref{markv99} is irreducible by using \cite[Definition 1.1]{Tuominen1979}. We even can show that this Markov chain is a Harris chain by using its definition in \cite{Durrett} or using \cite[Theorem 4.2]{Tuominen1979}. To guarantee the irreducible and recurrent properties of this continuous-space Markov chain, we show that $\bbP[\tau_A<\infty|X_0=x]=1$ for any $x \in \bbR$ and $A \in \calB(\bbR)$, where $\tau_A=\{\inf n\geq 1: X_n \in \calA\}$.}
\subsubsection{Free Energy and Average Mutual Information}
We assume that all postulated distributions are the same as their true ones for simplicity. Now, given $S=s_0 \in \bbR_+$\footnote{For example, in BPSK or QPSK modulation schemes in communications, all symbols in the constellation have a fixed energy $s_0$.}, for any $x_0 \in \calX=\bbR$ and $X_1 \sim \calN(\nu x_0,\sigma_0^2)$, from \eqref{singPME} and \eqref{estxq}, we have
\begin{align}
\rho_s&:=\frac{\bbE[\sfX_1 U|\sfX_0=x_0]-\bbE[\sfX_1|\sfX_0=x_0]\bbE[U|\sfX_0=x_0]}{\sigma_0\sqrt{s_0\sigma_0^2+\frac{1}{\eta}} }\\
&=\frac{\sqrt{s_0}(\nu^2 x_0^2+\sigma_0^2)-\nu^2 x_0^2\sqrt{s_0} }{\sigma_0\sqrt{s_0\sigma_0^2+\frac{1}{\eta}} }\\ 
&=\sqrt{\frac{s_0\sigma_0^2}{s_0\sigma_0^2+\frac{1}{\eta}}}.
\end{align}
From \eqref{condME}, \eqref{condMV}, and the standard result for MMSE of the bivariate Gaussian distribution (e.g.,\cite{Kay}), given any $s_0 \in \calS$ and $x_0 \in \calX=\bbR$, it holds that 
\begin{align}
\calV(s_0;\eta,\eta\mid x_0)&=\calE(s_0;\eta,\eta\mid x_0)\\
&=\bbE\bigg[\big(\sfX_1-\langle \sfX |\sfX_0=x_0 \rangle\big)^2|\sfX_0=x_0, s_0;\eta,\eta\bigg]\\
&=\var(\sfX_1|\sfX_0=x_0)(1-\rho_s^2)\\
&=\sigma_0^2 \bigg(1-\frac{s_0\sigma_0^2}{s\sigma_0^2+\frac{1}{\eta}}\bigg)\\
&=\frac{\sigma_0^2 \frac{1}{\eta}}{s_0\sigma_0^2+\frac{1}{\eta}}\label{eq192b}.
\end{align}
Hence, from \eqref{eqeta} and \eqref{eq192b}, $\eta$ is a solution of the following equation
\begin{align}
\eta^{-1}=1+ \beta \frac{s_0\sigma_0^2}{\eta s_0 \sigma_0^2+1} \label{drawbetax}.
\end{align}
In addition, since $\xi=\eta$ and $\sigma=1$, from \eqref{defFx0}, we have
\begin{align}
\calG(x_0)&=-\int p_{U|\sfX_0,s_0;\eta}(u\mid x_0,s_0;\eta)\log p_{U|\sfX_0,S;\eta}(u\mid x_0,S;\eta)du \nn\\
&\qquad + \frac{1}{2\beta}\bigg[(\eta-1)\log e -\log \eta \bigg]-\frac{1}{2}\log \frac{2\pi}{\eta}-\frac{1}{2}\log e\nn\\
&\qquad +\frac{1}{2\beta}\log(2\pi)+\frac{1}{2\beta}\log e\\
&=h(U|\sfX_0=x_0,S;\eta)\nn\\
&\qquad + \frac{1}{2\beta}\bigg[(\eta-1)\log e -\log \eta \bigg]-\frac{1}{2}\log \frac{2\pi}{\eta}-\frac{1}{2}\log e\nn\\
&\qquad +\frac{1}{2\beta}\log(2\pi)+\frac{1}{2\beta}\log e\\
&=\frac{1}{2}\log\bigg[2\pi e \bigg(s_0\sigma_0^2+\frac{1}{\eta}\bigg)\bigg] + \frac{1}{2\beta}\bigg[(\eta-1)\log e -\log \eta \bigg]-\frac{1}{2}\log \frac{2\pi}{\eta}-\frac{1}{2}\log e\nn\\
&\qquad +\frac{1}{2\beta}\log(2\pi)+\frac{1}{2\beta}\log e.
\end{align}
Since $\calG(x_0)$ does not depend on $x_0$, hence we also have a tight bound for this case, and the free energy is equal to
\begin{align}
\calF_q \bigg|_{\sigma=1,\tilde{\pi}=\pi}&=\frac{1}{2}\log\bigg(2\pi e \bigg(s_0 \sigma_0^2+\frac{1}{\eta}\bigg)\bigg) + \frac{1}{2\beta}\bigg[(\eta-1)\log e -\log \eta \bigg]\nn\\
&\qquad -\frac{1}{2}\log \frac{2\pi}{\eta}-\frac{1}{2}\log e +\frac{1}{2\beta}\log(2\pi)+\frac{1}{2\beta}\log e \label{eqF}, 
\end{align}  where $\eta$ is a solution of \eqref{drawbetax}, which is chosen to minimize $\calF_q$.

Now, for a fixed $\beta$, \eqref{drawbetax} is equivalent to
	\begin{align}
	s_0\sigma_0^2 \eta^2 +((\beta-1) s_0 \sigma_0^2 +1 )\eta-1=0,
	\end{align} which has solution
	\begin{align}
	\eta \in \bigg\{\frac{-((\beta-1)s_0 \sigma_0^2+1)\pm \sqrt{((\beta-1) s_0\sigma_0^2+1)^2+4s_0\sigma_0^2}}{2s_0\sigma_0^2} \bigg\}.
	\end{align}
	Note that $\eta \in (0,1)$ (cf. \eqref{drawbetax}), it follows that
	\begin{align}
	\eta=\frac{-((\beta-1)s_0 \sigma_0^2+1)+ \sqrt{((\beta-1)s_0 \sigma_0^2+1)^2+4s_0\sigma_0^2}}{2s_0\sigma_0^2} .
	\end{align} 
	
	Hence, by \eqref{eqF}, the free energy satisfies
	\begin{align}
	\calF_q\bigg|_{S=s_0,\sigma=1,\tilde{\pi}=\pi}&= \frac{1}{2}\log\bigg(2\pi e  \bigg(s_0\sigma_0^2+\frac{1}{\eta}\bigg)\bigg) + \frac{1}{2\beta}\bigg[(\eta-1)\log e -\log \eta \bigg]\nn\\
	&\qquad -\frac{1}{2}\log \frac{2\pi}{\eta}-\frac{1}{2}\log e +\frac{1}{2\beta}\log(2\pi)+\frac{1}{2\beta}\log e \label{gh}
	\end{align}
	where 
	\begin{align}
	\eta =\frac{-((\beta-1)s_0 \sigma_0^2+1)+ \sqrt{((\beta-1) s_0\sigma_0^2+1)^2+4s_0\sigma_0^2}}{2s_0\sigma_0^2}.
	\end{align}
		 
\subsubsection{Markov Chain Monte Carlo (MCMC) vs. Replica Prediction} \label{simu:sec}
\begin{figure}[H]%[ht]
	\centering
	\includegraphics[width=0.7\linewidth]{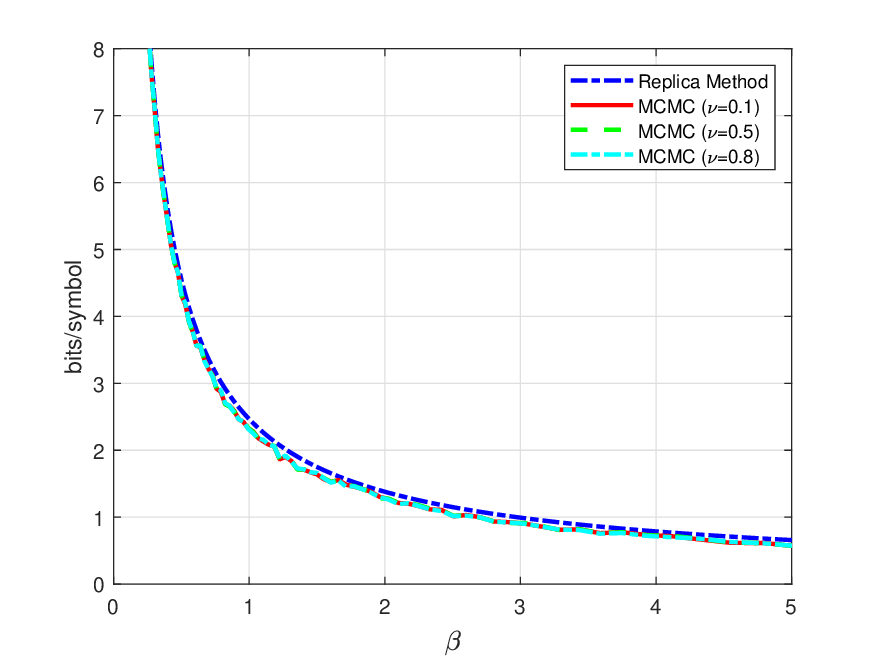}
	\caption{Free energy by replica method and empirical MCMC as functions of $\beta$ for $\sigma_0^2=1$ and $s_0=1$.}
	\label{fig:CMK1MCMC}
\end{figure}

\blue{In this subsection, we use the same MCMC algorithm as  Subsection \ref{sec:bemcmc}, which is the Metropolis–Hastings algorithm. In the Fig. \ref{fig:CMK1MCMC}, we plot the free energy curves for the linear model with Markov prior in \eqref{markv99} for three cases $\nu=0.1$, $\nu=0.5$, and $\nu=0.8$. The curves suggest that the free energy does not depend on $\nu$ as we can observe from \eqref{gh}. In these plots, we set $X_1 \sim \calN(0, \frac{\sigma_0^2}{1-\nu^2})$ to force the state distribution of the Markov (Harris) chain $X_n \sim \calN(0, \frac{\sigma_0^2}{1-\nu^2})$ for all $n\geq 1$. The plot also shows that the replica prediction for the free energy is very closed to the MCMC simulation result. Since the MMSE is a fixed function of the free energy (or mutual information) \cite{GuoShamaiVerdu2005}, this also means that the MMSE curve by replica method closely approaches the MMSE of the model.}
 \subsection{Hidden Markov Prior}
  
  %\RV{One example with a hidden Markov prior}
  In this section, we estimate free energy and mutual information for the linear model in Section \ref{sec:setting} with hidden Markov sources defined in \cite[Sect. 7]{Philip}. The sequence $\{X_n\}_{n=1}^{\infty}$ which takes values on $\bbR$ is generated via 
  \begin{align}
  p_{X_n|\Upsilon_n}(x_n\mid \upsilon_n)&=\upsilon_n \calN(x_n;0,1)+(1-\upsilon_n)\delta (x_n)\\
  &= \frac{\upsilon_n}{\sqrt{2\pi}} \exp\bigg(-\frac{x_n^2}{2}\bigg)+ (1-\upsilon_n) \delta(x_n)
  \end{align} using a time-homogeneous irreducible Markov chain-generated sparsity pattern $\{\Upsilon_n\}_{n=1}^{\infty}$ on $\calS_{\gamma}=\{0,1\}$. Such a Markov chain is fully described by the following transition stochastic matrix
  \begin{align}
  P_{\Upsilon}=\begin{bmatrix}
  1-\kappa \gamma& \gamma \kappa \\ (1-\kappa) \gamma & 1-(1-\kappa)\gamma
  \end{bmatrix}
  \end{align} for some $\gamma \in (0,1]$ called the Markov independence parameter. This irreducible Markov chain yields a stationary distribution with activity rate $P(\Upsilon_n=1)=\kappa$ for all $n\in \bbZ^+$.
   
  \subsubsection{Free Energy and Average Mutual Information}\label{sec:spec1HM}
  
  First, it is easy to see that the left Perron-Frobenius eigenvector of the stochastic matrix $P_{\Upsilon}$ with unit Manhattan norm is
  \begin{align}
  \lambda_0=(1-\kappa,\kappa)^T. 
  \end{align}
  Observe that
  \begin{align}
  P_{\sfX_1,\Upsilon_1|\sfX_0,\Upsilon_0}(x_1,\upsilon_1\mid x_0,\upsilon_0)&=P_{X|\Upsilon}(x_1\mid \upsilon_1) \pi_{\Upsilon}(\upsilon_0,\upsilon_1)
  \end{align} where $P_{X|\Upsilon}(\cdot|\cdot)$ is the emission probability of the hidden Markov process.
  
  Hence, the left Perron-Frobenius eigenvector of the stochastic matrix $P_{\pi_{\Upsilon},\calX}$ in Subsection \ref{freEnergyx0_hmm} has the following form: 
  \begin{align}
  \lambda^{(\pi_{\Upsilon})}=((1-\kappa)\{l_0(x)\}_{x \in \calX}, \kappa \{l_1(x)\}_{x \in \calX})^T,
  \end{align} where $l_0: \calX \to \bbR_+$ and $l_1: \calX \to \bbR_+$ and
  \begin{align}
  \int_{\calX} l_i(x) dx=1, \quad \forall i \in \{0,1\} \label{condPF2}.
  \end{align}
  By setting $\lambda_{x,0}^{(\pi_{\Upsilon})}:=(1-\kappa)l_0(x)$ and $\lambda_{x,1}^{(\pi_{\Upsilon})}=\kappa l_1(x)$, it follows that
  \begin{align}
  \int_{\calX} \lambda_{x,0}^{(\pi_{\Upsilon})}dx&=1-\kappa \label{fact1E},\\
  \int_{\calX} \lambda_{x,1}^{(\pi_{\Upsilon})}dx&=\kappa \label{fact2E}.
  \end{align}
  
  First, we estimate $\tilG(x_0,0)$ as a function of $\kappa,\gamma$, and $x_0$ for $x_0 \in \calX$. We assume that all postulated distributions are the same as their true ones for simplicity. We also assume that $S=1$ with probability $1$. Now from \eqref{estq0HM}, we have
  \begin{align}
  q_0(u,x_0,0,1;\eta)&=\bbE\bigg[q_{U|\sfX_0,\Upsilon_0, \sfX_1,S;\xi}\big(u\mid x_0,0, X,1;\xi\big)\bigg|S=1\bigg]\\
  &=\bbE\bigg[\sqrt{\frac{\eta}{2\pi}} \exp\bigg[-\frac{\eta}{2}(u-\sfX)^2\bigg]\bigg|\sfX_0=x_0, \Upsilon_0=0\bigg]\\
  &=\int_{\bbR} \sum_{\upsilon \in \{0,1\}} \sqrt{\frac{\eta}{2\pi}} \exp\bigg[-\frac{\eta}{2}(u-x)^2\bigg]P_{X_1|\Upsilon_1}(x|\upsilon)\pi_{\Upsilon}(0,\upsilon) dx \\
  &=\int_{\bbR} \sqrt{\frac{\eta}{2\pi}} \exp\bigg[-\frac{\eta}{2}(u-x)^2\bigg]\delta(x) \pi_{\Upsilon}(0,0) dx \nn \\
  &\qquad + \int_{\bbR} \sqrt{\frac{\eta}{2\pi}} \exp\bigg[-\frac{\eta}{2}(u-x)^2\bigg] \frac{1}{\sqrt{2\pi}}\exp\bigg(-\frac{x^2}{2} \bigg)  \pi_{\Upsilon}(0,1) dx\\
  &=(1-\gamma \kappa)\int_{\bbR} \sqrt{\frac{\eta}{2\pi}} \exp\bigg[-\frac{\eta}{2}(u-x)^2\bigg]\delta(x)  dx \nn \\
  &\qquad + \kappa \gamma \int_{\bbR} \sqrt{\frac{\eta}{2\pi}} \exp\bigg[-\frac{\eta}{2}(u-x)^2\bigg] \frac{1}{\sqrt{2\pi}}\exp\big(-\frac{x^2}{2} \bigg) dx\\
  &=(1-\gamma \kappa) \sqrt{\frac{\eta}{2\pi}} \exp\bigg[-\frac{\eta u^2}{2}\bigg] + \kappa \gamma  \sqrt{\frac{\eta}{2\pi(1+\eta)}}\exp\bigg(-\frac{\eta u^2}{2(1+\eta)}\bigg)  \label{estq0estHM},
  \end{align} which does not depend on $x_0$.
  
  Similarly, from \eqref{estq1HM}, for all $x_0 \in \calX$, we also have
  \begin{align}
  q_1(u,x_0,0, 1;\eta)&=\bbE\bigg[\sfX q_{U|\sfX_0,\Upsilon_0, \sfX_1,S;\xi}\big(u\mid x_0,0, \sfX,1;\xi\big)\bigg|S=1\bigg]\\
  &=\bbE\bigg[\sfX \sqrt{\frac{\eta}{2\pi}} \exp\bigg[-\frac{\eta}{2}(u-\sfX)^2\bigg]\bigg|\sfX_0=x_0, \Upsilon_0=0\bigg]\\
  &=\int_{\bbR} \sum_{\upsilon \in \{0,1\}} x\sqrt{\frac{\eta}{2\pi}} \exp\bigg[-\frac{\eta}{2}(u-x)^2\bigg]P_{X_1|\Upsilon_1}(x|\upsilon)\pi_{\Upsilon}(0,\upsilon) dx \\
  &=\int_{\bbR} \sqrt{\frac{\eta}{2\pi}}x \exp\bigg[-\frac{\eta}{2}(u-x)^2\bigg]\delta(x) \pi_{\Upsilon}(0,0) dx \nn \\
  &\qquad + \int_{\bbR} \sqrt{\frac{\eta}{2\pi}}x \exp\bigg[-\frac{\eta}{2}(u-x)^2\bigg] \frac{1}{\sqrt{2\pi}}\exp\bigg(-\frac{x^2}{2} \bigg)  \pi_{\Upsilon}(0,1) dx\\
  &=(1-\gamma \kappa)\int_{\bbR} \sqrt{\frac{\eta}{2\pi}} x\exp\bigg[-\frac{\eta}{2}(u-x)^2\bigg]\delta(x)  dx \nn \\
  &\qquad + \kappa \gamma \int_{\bbR} \sqrt{\frac{\eta}{2\pi}} x\exp\bigg[-\frac{\eta}{2}(u-x)^2\bigg] \frac{1}{\sqrt{2\pi}}\exp\bigg(-\frac{x^2}{2} \bigg) dx\\
  &=\kappa \gamma \int_{\bbR} \sqrt{\frac{\eta}{2\pi}} x\exp\bigg[-\frac{\eta}{2}(u-x)^2\bigg] \frac{1}{\sqrt{2\pi}}\exp\bigg(-\frac{x^2}{2} \bigg) dx\\
  &=   \kappa \gamma \sqrt{\frac{\eta}{2\pi(1+\eta)}}\exp\bigg(-\frac{\eta u^2}{2(1+\eta)}\bigg)\bigg(\frac{\eta u}{1+\eta}\bigg) \label{estq1estHM}.
  \end{align}
  Therefore, from \eqref{singPMEHM}, \eqref{estxqHM}, \eqref{estq0estHM}, and \eqref{estq1estHM}, we have
  \begin{align}
  \langle \sfX|x_0,0\rangle_q&=\bbE_{q}\bigg[\sfX\big|\sfX_0=x_0,\Upsilon_0=0, U,1;\eta\bigg]\\
  &=\frac{q_1(U,x_0,0,1;\eta)}{q_0(U,x_0,0,1;\eta)}\\
  &=\frac{ \kappa \gamma \sqrt{\frac{\eta}{2\pi(1+\eta)}}\exp\bigg(-\frac{\eta u^2}{2(1+\eta)}\bigg)\bigg(\frac{\eta u}{1+\eta}\bigg)}{(1-\gamma \kappa) \sqrt{\frac{\eta}{2\pi}} \exp\bigg(-\frac{\eta u^2}{2}\bigg) + \kappa \gamma  \sqrt{\frac{\eta}{2\pi(1+\eta)}}\exp\bigg(-\frac{\eta u^2}{2(1+\eta)}\bigg)}.
  \end{align} 
  It follows from \eqref{condMEHM} and \eqref{condMVHM} that
  \begin{align}
  \calV(1;\eta,\eta\mid x_0,0)&=\calE(1;\eta,\eta\mid x_0,0)\\
  &=\bbE\bigg[\big(\sfX_1-\langle \sfX |x_0,0\rangle_q\big)^2|\sfX_0=x_0,\Upsilon_0=0, 1;\eta,\eta\bigg]\\
  &=\bbE\bigg[\sfX_1^2\big|\sfX_0=x_0,\Upsilon_0=0, 1;\eta,\eta\bigg]-\langle \sfX |\sfX_0=x_0,\Upsilon_0=0 \rangle_q^2 \\
  &=\int_{\bbR} \sum_{\upsilon \in \{0,1\}} x^2 P_{X_1|\Upsilon_1}(x\mid \upsilon)\pi_{\Upsilon}(0,\upsilon)dx \nn\\
  & \qquad -\left(\frac{ \kappa \gamma  \sqrt{\frac{\eta}{2\pi(1+\eta)}}\exp\bigg(-\frac{\eta u^2}{2(1+\eta)}\bigg)\bigg(\frac{\eta u}{1+\eta}\bigg)}{(1-\gamma \kappa) \sqrt{\frac{\eta}{2\pi}} \exp\bigg[-\frac{\eta u^2}{2}\bigg] + \kappa \gamma  \sqrt{\frac{\eta}{2\pi(1+\eta)}}\exp\bigg(-\frac{\eta u^2}{2(1+\eta)}\bigg)}\right)^2\\
  &=(1-\kappa \gamma)\int_{\bbR} x^2 P_{X_1|\Upsilon_1}(x\mid 0)dx + \kappa \gamma \int_{\bbR} x^2 P_{X_1|\Upsilon_1}(x\mid 1)dx \nn\\
  & \qquad -\left(\frac{ \kappa \gamma  \sqrt{\frac{\eta}{2\pi(1+\eta)}}\exp\bigg(-\frac{\eta u^2}{2(1+\eta)}\bigg)\bigg(\frac{\eta u}{1+\eta}\bigg)}{(1-\gamma \kappa) \sqrt{\frac{\eta}{2\pi}} \exp\bigg[-\frac{\eta u^2}{2}\bigg] + \kappa \gamma \sqrt{\frac{\eta}{2\pi(1+\eta)}}\exp\bigg(-\frac{\eta u^2}{2(1+\eta)}\bigg)}\right)^2\\
  &=(1-\kappa \gamma)\int_{\bbR} x^2  \delta(x) dx + \kappa \gamma \int_{\bbR} x^2 \frac{1}{\sqrt{2\pi}}\exp\bigg(-\frac{x^2}{2}\bigg) dx \nn\\
  & \qquad -\left(\frac{ \kappa \gamma  \sqrt{\frac{\eta}{2\pi(1+\eta)}}\exp\bigg(-\frac{\eta u^2}{2(1+\eta)}\bigg)\bigg(\frac{\eta u}{1+\eta}\bigg)}{(1-\gamma \kappa) \sqrt{\frac{\eta}{2\pi}} \exp\bigg[-\frac{\eta u^2}{2}\bigg] + \kappa \gamma \sqrt{\frac{\eta}{2\pi(1+\eta)}}\exp\bigg(-\frac{\eta u^2}{2(1+\eta)}\bigg)}\right)^2\\
  &=\kappa \gamma -\left(\frac{ \kappa \gamma  \sqrt{\frac{\eta}{2\pi(1+\eta)}}\exp\bigg(-\frac{\eta u^2}{2(1+\eta)}\bigg)\bigg(\frac{\eta u}{1+\eta}\bigg)}{(1-\gamma \kappa) \sqrt{\frac{\eta}{2\pi}} \exp\bigg(-\frac{\eta u^2}{2}\bigg) + \kappa \gamma \sqrt{\frac{\eta}{2\pi(1+\eta)}}\exp\bigg(-\frac{\eta u^2}{2(1+\eta)}\bigg)}\right)^2 \label{eq242eqHM},
  \end{align} which does not depend on $x_0$.
  
  Similarly, by symmetry, we also have
  \begin{align}
  &\calV(1;\eta,\eta\mid x_0,1)=\calE(1;\eta,\eta\mid x_0,1)\\
  &=1-(1-\kappa)\gamma  - \bbE\left[\left(\frac{ (1-(1-\kappa) \gamma)  \sqrt{\frac{\eta}{2\pi(1+\eta)}}\exp\bigg(-\frac{\eta U^2}{2(1+\eta)}\bigg)\bigg(\frac{\eta U}{1+\eta}\bigg)}{(1-\gamma) \kappa \sqrt{\frac{\eta}{2\pi}} \exp\bigg(-\frac{\eta U^2}{2}\bigg) + (1-(1-\kappa) \gamma) \sqrt{\frac{\eta}{2\pi(1+\eta)}}\exp\bigg(-\frac{\eta U^2}{2(1+\eta)}\bigg)}\right)^2\right]\label{eq242eqHMcr},
  \end{align} which holds for any $x_0 \in \calX$.
  
  Now, by \eqref{singPMEHM}, the single-symbol PME for this special case is
  \begin{align}
  U=\sfX_1+\frac{1}{\sqrt{\eta}}W. 
  \end{align} 
  Observe that
  \begin{itemize}
  	\item Under the condition $\Upsilon_0=0$, we have
  	\begin{align}
  	P_{X_1}(x) &= \sum_{\upsilon \in \{0,1\}}P_{X_1|\Upsilon_1}(x\mid \upsilon) \pi_{\Upsilon}(0,\upsilon)\\
  	&= (1-\kappa \gamma)\delta(x)+  \frac{\kappa \gamma}{\sqrt{2\pi}}\exp\bigg(-\frac{x^2}{2}\bigg). 
  	\end{align}
  	It follows that:
  	\begin{align}
  	F_U(u)&=\bbP(U \leq u)\\
  	&=\int_{\bbR} \bbP(U \leq u|X_1=x)P_{X_1}(x)dx\\
  	&=\int_{\bbR} \bbP(W \leq (u-1)\sqrt{\eta})P_{X_1}(x)dx\\
  	&=(1-\kappa \gamma)\int_{\bbR} \bbP(W \leq (u-x)\sqrt{\eta})\delta(x)dx  + \kappa \gamma \int_{\bbR} \bbP(W \leq (u-x)\sqrt{\eta})\frac{1}{\sqrt{2\pi}}\exp\bigg(-\frac{x^2}{2}\bigg)dx \\
  	&=(1-\kappa \gamma )\int_{\bbR}\Phi((u-x) \sqrt{\eta})\delta(x)dx+ \frac{ \kappa \gamma}{2\pi} \int_{-\infty}^{\infty} \int_{-\infty}^{(u-x)\sqrt{\eta}} \exp\bigg(-\frac{x^2+t^2}{2}\bigg)dt dx\\
  	&=(1-\kappa \gamma )\Phi(u \sqrt{\eta})+ \frac{ \kappa \gamma}{2\pi} \int_{-\infty}^{\infty} \int_{-\infty}^{(u-x)\sqrt{\eta}} \exp\bigg(-\frac{x^2+t^2}{2}\bigg)dt dx
  	\label{cupet1HM}
  	\end{align}  where $\Phi(x):=\frac{1}{\sqrt{2\pi}}\int_{-\infty}^x \exp(-t^2/2)dt$. From \eqref{cupet1HM}, we obtain\footnote{We can derive it use the convolution since $W$ and $X_1$ are independent random variables.}
  	\begin{align}
  	f_U^{(\rm{HM},1)}(u)&=\frac{(1-\gamma \kappa)\sqrt{\eta}}{\sqrt{2\pi}} \exp\bigg(-\frac{u^2 \eta}{2}\bigg)+ \frac{\kappa \gamma \sqrt{\eta}}{2\pi}  \int_{\bbR}\exp\bigg(-\frac{x^2+(u-x)^2 \eta}{2}\bigg)dx\\
  	&=(1-\gamma \kappa)\sqrt{\frac{\eta}{2\pi}} \exp\bigg(-\frac{u^2 \eta}{2}\bigg)+ \kappa \gamma \sqrt{\frac{\eta}{2\pi(1+\eta)}} \exp\bigg(-\frac{u^2 \eta}{2(1+\eta)}\bigg)
  	\label{deffzHM}.
  	\end{align}
  	\item Similarly, under the condition $\Upsilon_0=1$, we have
  	\begin{align}
  	f_U^{(\rm{HM},2)}(u)=(1- \kappa)\gamma\sqrt{\frac{\eta}{2\pi}} \exp\bigg(-\frac{u^2 \eta}{2}\bigg)+ (1-(1-\kappa) \gamma) \sqrt{\frac{\eta}{2\pi(1+\eta)}} \exp\bigg(-\frac{u^2 \eta}{2(1+\eta)}\bigg)\label{deffzHMz}.
  	\end{align}
  \end{itemize}
  Hence, from \eqref{eqetaHM}, $\eta$ is a solution of the following equation
  \begin{align}
  \eta^{-1}&=1+\beta\bigg(\int_{\calX} \lambda_{x,0}^{(\pi_{\Upsilon})} \bbE[1 \calE(1;\eta,\eta|x,0)]dx + \int_{\calX} \lambda_{x,1}^{(\pi_{\Upsilon})} \bbE[1 \calE(1;\eta,\eta|x,1)]dx \bigg),\\
  &=1 + \beta \bigg(\int_{\calX} \lambda_{x,0}^{(\pi_{\Upsilon})}dx\bigg) \bigg(\gamma \kappa -  \bbE\left[\left(\frac{ \kappa \gamma  \sqrt{\frac{\eta}{2\pi(1+\eta)}}\exp\bigg(-\frac{\eta U^2}{2(1+\eta)}\bigg)\bigg(\frac{\eta U}{1+\eta}\bigg)}{(1-\gamma \kappa) \sqrt{\frac{\eta}{2\pi}} \exp\bigg(-\frac{\eta U^2}{2}\bigg) + \kappa \gamma \sqrt{\frac{\eta}{2\pi(1+\eta)}}\exp\bigg(-\frac{\eta U^2}{2(1+\eta)}\bigg)}\right)^2\right]\bigg)\nn\\
  &\quad +  \beta \bigg(\int_{\calX} \lambda_{x,1}^{(\pi_{\Upsilon})}dx\bigg) \nn\\
  &\quad \times \bigg(1-(1- \kappa)\gamma -  \bbE\left[\left(\frac{ (1-(1-\kappa) \gamma)  \sqrt{\frac{\eta}{2\pi(1+\eta)}}\exp\bigg(-\frac{\eta U^2}{2(1+\eta)}\bigg)\bigg(\frac{\eta U}{1+\eta}\bigg)}{(1- \kappa)\gamma \sqrt{\frac{\eta}{2\pi}} \exp\bigg(-\frac{\eta U^2}{2}\bigg) + (1-(1-\kappa) \gamma) \sqrt{\frac{\eta}{2\pi(1+\eta)}}\exp\bigg(-\frac{\eta U^2}{2(1+\eta)}\bigg)}\right)^2\right]\bigg)\\
  &=1 + \beta (1-\kappa) \bigg(\gamma \kappa -  \bbE\left[\left(\frac{ \kappa \gamma  \sqrt{\frac{\eta}{2\pi(1+\eta)}}\exp\bigg(-\frac{\eta U^2}{2(1+\eta)}\bigg)\bigg(\frac{\eta U}{1+\eta}\bigg)}{(1-\gamma \kappa) \sqrt{\frac{\eta}{2\pi}} \exp\bigg(-\frac{\eta U^2}{2}\bigg) + \kappa \gamma \sqrt{\frac{\eta}{2\pi(1+\eta)}}\exp\bigg(-\frac{\eta U^2}{2(1+\eta)}\bigg)}\right)^2\right]\bigg)\nn\\
  & +  \beta \kappa
  \bigg(1-(1- \kappa)\gamma -  \bbE\left[\left(\frac{ (1-(1-\kappa) \gamma)  \sqrt{\frac{\eta}{2\pi(1+\eta)}}\exp\bigg(-\frac{\eta U^2}{2(1+\eta)}\bigg)\bigg(\frac{\eta U}{1+\eta}\bigg)}{(1- \kappa)\gamma \sqrt{\frac{\eta}{2\pi}} \exp\bigg(-\frac{\eta U^2}{2}\bigg) + (1-(1-\kappa) \gamma) \sqrt{\frac{\eta}{2\pi(1+\eta)}}\exp\bigg(-\frac{\eta U^2}{2(1+\eta)}\bigg)}\right)^2\right]\bigg),
  \label{drawbetaHM}
  \end{align} where \eqref{drawbetaHM} follows from \eqref{fact1E} and \eqref{fact2E}.
  
  Now, since $S=1, \xi=\eta, \sigma=1, \tilde{\pi}_{\Upsilon}=\pi_{\Upsilon}$, from \eqref{defFx0HM}, we have
  \begin{align}
  \tilG(x_0,0)\bigg|_{\sigma=1,\tilde{\pi}_{\Upsilon}=\pi_{\Upsilon}}&=-\int p_{U|X_0,\Upsilon_0, S;\eta}(u\mid x_0,0,1;\eta)\log p_{U|X_0,\Upsilon_0, S;\eta}(u\mid x_0,0,1;\eta)du \nn\\
  &\qquad + \frac{1}{2\beta}\bigg[(\eta-1)\log e -\log \eta \bigg]-\frac{1}{2}\log \frac{2\pi}{\eta}-\frac{1}{2}\log e\nn\\
  &\qquad +\frac{1}{2\beta}\log(2\pi)+\frac{1}{2\beta}\log e.
  \end{align}
  On the other hand, 
  \begin{align}
  p_{U|\sfX_0,\Upsilon_0, 1;\eta}(u\mid x_0,0,1;\eta)&=\int_{\bbR} p_{U|\sfX_0,\Upsilon_0, \sfX_1, 1;\eta}(u\mid x_0,0,x,1;\eta) P_{X_1|X_0,\Upsilon_0}(x\mid x_0,0) dx \\
  &=\int_{\bbR} \sqrt{\frac{\eta}{2\pi}} \exp\bigg[-\frac{\eta}{2}(u-x)^2\bigg]\bigg(\sum_{\upsilon \in \{0,1\}} p_{X_1|X_0,\Upsilon_0}(x\mid x_0,0)\bigg) dx \\
  &=\int_{\bbR} \sqrt{\frac{\eta}{2\pi}} \exp\bigg[-\frac{\eta}{2}(u-x)^2\bigg]\bigg(\sum_{\upsilon \in \{0,1\}} p_{X_1|\Upsilon_1}(x\mid \upsilon) \pi_{\Upsilon}(0,\upsilon)\bigg) dx \\
  &=(1-\kappa \gamma )\int_{\bbR} \sqrt{\frac{\eta}{2\pi}} \exp\bigg[-\frac{\eta}{2}(u-x)^2\bigg]\delta(x) dx \\
  & \qquad+ \kappa \gamma \int_{\bbR} \sqrt{\frac{\eta}{2\pi}} \exp\bigg[-\frac{\eta}{2}(u-x)^2\bigg]\frac{1}{\sqrt{2\pi}}\exp\bigg(-\frac{x^2}{2}\bigg) dx \\
  &=(1-\kappa\gamma )\sqrt{\frac{\eta}{2\pi}}\exp\bigg[-\frac{\eta u^2}{2}\bigg]+ \kappa \gamma \sqrt{\frac{\eta}{2\pi(1+\eta)}}\exp\bigg[-\frac{\eta u^2}{2(1+\eta)}\bigg],
  \end{align} which does not depend on $x_0$.
  
  It follows that
  \begin{align}
  \tilG(x_0,0)\bigg|_{S=1,\sigma=1,\tilde{\pi}_{\Upsilon}=\pi_{\Upsilon}}= \hatG(0,\eta,\kappa, \gamma), \quad \forall x_0 \in \calX \label{optim1HM},
  \end{align} where
  \begin{align}
  \hatG(0,\eta,\kappa, \gamma)&:=-\int_{-\infty}^{\infty}\bigg((1-\kappa\gamma )\sqrt{\frac{\eta}{2\pi}}\exp\bigg[-\frac{\eta u^2}{2}\bigg]+ \kappa \gamma \sqrt{\frac{\eta}{2\pi(1+\eta)}}\exp\bigg[-\frac{\eta u^2}{2(1+\eta)}\bigg]\bigg)\nn\\
  &\qquad \times  \log\bigg((1-\kappa \gamma )\sqrt{\frac{\eta}{2\pi}}\exp\bigg[-\frac{\eta u^2}{2}\bigg]+ \kappa \gamma \sqrt{\frac{\eta}{2\pi(1+\eta)}}\exp\bigg[-\frac{\eta u^2}{2(1+\eta)}\bigg]\bigg)du \nn\\
  &\qquad + \frac{1}{2\beta}\bigg[(\eta-1)\log e -\log \eta \bigg]-\frac{1}{2}\log \frac{2\pi}{\eta}-\frac{1}{2}\log e\nn\\
  &\qquad +\frac{1}{2\beta}\log(2\pi)+\frac{1}{2\beta}\log e.
  \end{align}
  
  By the symmetry, it is not hard to see that
  \begin{align}
  \tilG(x_0,1)\bigg|_{S=1,\sigma=1,\tilde{\pi}_{\Upsilon}=\pi_{\Upsilon}}= \hatG(1,\eta,\kappa, \gamma), \quad \forall x_0 \in \calX \label{optim2HM},
  \end{align} where
  \begin{align}
  \hatG(1,\eta,\kappa, \gamma)&:=-\int_{-\infty}^{\infty}\bigg((1-\kappa)\gamma \sqrt{\frac{\eta}{2\pi}}\exp\bigg[-\frac{\eta u^2}{2}\bigg]+ (1-(1-\kappa) \gamma) \sqrt{\frac{\eta}{2\pi(1+\eta)}}\exp\bigg[-\frac{\eta u^2}{2(1+\eta)}\bigg]\bigg)\nn\\
  &\qquad \times  \log\bigg((1-\kappa)\gamma \sqrt{\frac{\eta}{2\pi}}\exp\bigg(-\frac{\eta u^2}{2}\bigg)+ (1-(1-\kappa) \gamma) \sqrt{\frac{\eta}{2\pi(1+\eta)}}\exp\bigg[-\frac{\eta u^2}{2(1+\eta)}\bigg]\bigg)du \nn\\
  &\qquad + \frac{1}{2\beta}\bigg[(\eta-1)\log e -\log \eta \bigg]-\frac{1}{2}\log \frac{2\pi}{\eta}-\frac{1}{2}\log e\nn\\
  &\qquad +\frac{1}{2\beta}\log(2\pi)+\frac{1}{2\beta}\log e.
  \end{align} 
  Now, let $\hatC_{\beta}(\kappa,\gamma)$ is the set of all solutions $\eta$ of equation \eqref{drawbetaHM} given $\beta$ and $\kappa$ and $\gamma$.
  
  From Claim \ref{claim:hiddenMarkov} and \eqref{fact1E} and \eqref{fact2E}, we have
  \begin{align}
  \calF\bigg|_{S=1,\sigma=1,\tilde{\pi}_{\Upsilon}=\pi_{\Upsilon}} &=\min_{\eta \in \hatC_{\beta}(\kappa,\gamma)}\bigg[\int_{\calX} \lambda_{x,0}^{(\pi_{\Upsilon})} \tilG(x,0)dx +\int_{\calX} \lambda_{x,1}^{(\pi_{\Upsilon})} \tilG(x,1)dx\bigg]\\
  &=\min_{\eta \in \hatC_{\beta}(\kappa,\gamma)} \bigg[(1-\kappa)\hatG(0,\eta,\kappa, \gamma)+ \kappa \hatG(1,\eta,\kappa, \gamma)\bigg].
  \end{align}
  \begin{figure}[H]
  	\centering
  	\includegraphics[width=0.7\linewidth]{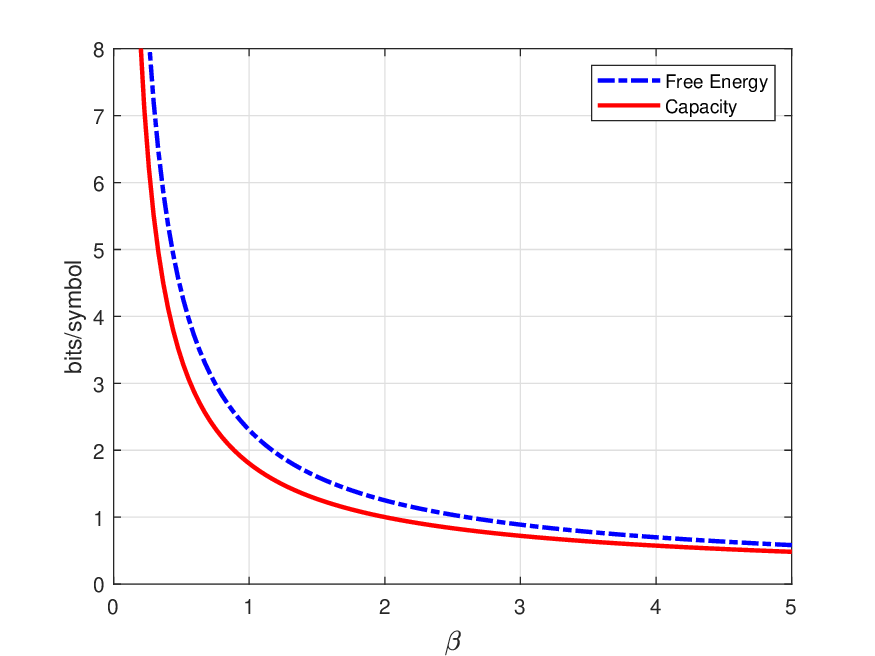}
  	\caption{Free energy and average mutual information as functions of $\beta$ for the symmetric i.i.d. case $\lambda=0.5$ and $\gamma=1$.}
  	\label{fig:DMK2HM}
  \end{figure}
  Solving the optimization problems in \eqref{drawbetaHM} is very challenging. However, by observing that given $\kappa$ and $\gamma$, $\beta$, $\hatG(0,\eta,\kappa,\gamma)$ and $\hatG(1,\eta,\kappa,\gamma)$ are functions of $\eta$. Hence, we can plot lower and upper bounds for the free energy $\calF$ and the average mutual information as functions of $(\kappa,\gamma)$. In Fig. \ref{fig:DMK2HM}, we plot the free energy and the average mutual information for $\kappa=0.5$ and $\gamma=1$, i.e., the sequence $\{X_n\}_{n=1}^{\infty}$ is i.i.d. generated. 
  
\begin{figure}[H]
  	\centering
  	\includegraphics[width=0.7\linewidth]{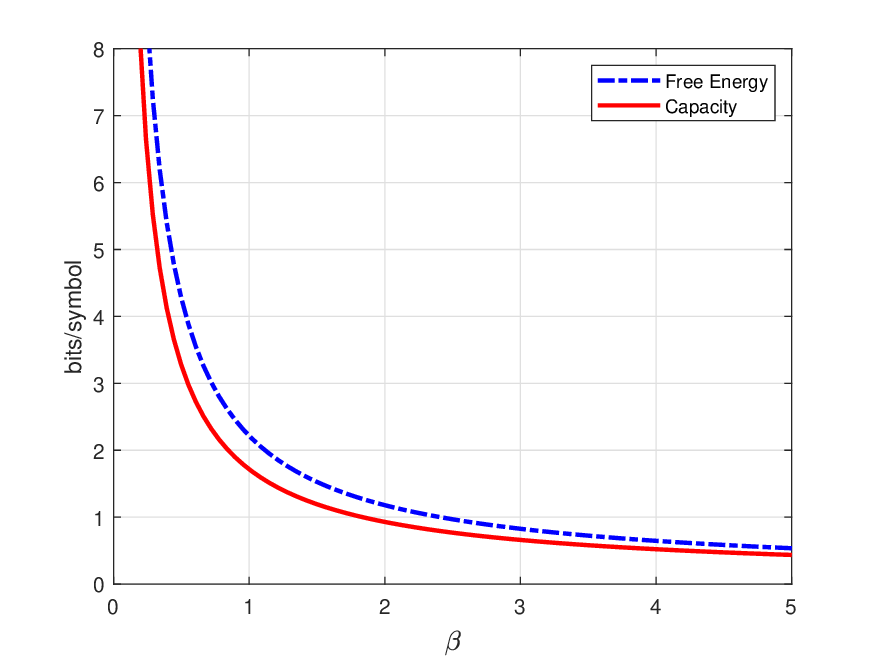}
  	\caption{Free energy and average mutual information as functions of $\beta$ for the non-symmetric case $\lambda=0.3$ and $\gamma=0.8$.}
  	\label{fig:DMK3HM}
  \end{figure}

  For the non-symmetric case where $\kappa=0.3$ and $\gamma=0.8$, we obtain the free energy and the average mutual information as in Fig. \ref{fig:DMK3HM}.  
  
  \subsubsection{Approximate Message Passing Algorithm vs. Replica Prediction} \label{sub:amp3}
 % As in Section \ref{sec:spec1HM}, we consider the linear model in Section \ref{sec:setting} with hidden Markov sources defined in \cite[Sect. 7]{Philip}. 
 
 Observe that
  \begin{align}
  &\bbE\bigg[\sfX_1^2\bigg|\sfX_0=x_0,\Upsilon_0=\upsilon_0\bigg]\nn \\
  &= \int_{\bbR} \sum_{\upsilon \in \{0,1\}} x^2 P_{X_1,\Upsilon_1|X_0,\Upsilon_0}(x,\upsilon\mid x_0,\upsilon_0) dx\\
  &=\sum_{\upsilon \in \{0,1\}} \int_{\bbR} x^2 P_{X_1|\Upsilon_1}(x\mid \upsilon)\pi_{\Upsilon}(\upsilon_0,\upsilon) dx\\
  &=\int_{\bbR} x^2 \delta(x)\pi_{\Upsilon}(\upsilon_0,0) dx + \int_{\bbR} x^2 \frac{1}{\sqrt{2\pi}} \exp\bigg(-\frac{x^2}{2}\bigg)\pi_{\Upsilon}(\upsilon_0,1) dx\\
  &=\pi_{\Upsilon}(\upsilon_0,1).
  \end{align}
  Hence, we have
    \begin{align}
  \bbE[\sfX_1^2]&=\bbE\bigg[\bbE\bigg[\sfX_1^2\bigg|\sfX_0,\Upsilon_0\bigg]\bigg]\\
  &=\bbE[\pi_{\Upsilon}(\Upsilon_0,1)]\\
  &=P_{\Upsilon}(0) \pi_{\Upsilon}(0,1)+ P_{\Upsilon}(1) \pi_{\Upsilon}(1,1)\\
  &=(1-\kappa) \kappa \gamma+ \kappa \big(1-(1-\kappa)\gamma\big)\\
  &=\kappa \label{horse1}.
  \end{align}
  On the other hand, given $\Upsilon_0=0$, for any $x_0 \in \calX$, we also have
  \begin{align}
  &\bbE\big[\langle \sfX_1\big|\sfX_0=x_0,\Upsilon_0=0\rangle^2\big]=\bbE\bigg[\sfX_1^2\bigg|\sfX_0=x_0,\Upsilon_0=0\bigg]-\bbE\bigg[(\sfX_1-\langle \sfX_1|\sfX_0=x_0,\Upsilon_0\rangle)^2\bigg|\sfX_0=x_0,\Upsilon_0=0\bigg]\\
  &\qquad =\pi_{\Upsilon}(0,1)- \bbE[\calE(1;\eta,\eta\mid x_0,0)]\\
  &\qquad =\kappa \gamma- \bbE[\calE(1;\eta,\eta\mid x_0,0)]\\
  &\qquad = \bbE\left[\left(\frac{ \kappa \gamma  \sqrt{\frac{\eta}{2\pi(1+\eta)}}\exp\bigg(-\frac{\eta U^2}{2(1+\eta)}\bigg)\bigg(\frac{\eta U}{1+\eta}\bigg)}{(1-\gamma \kappa) \sqrt{\frac{\eta}{2\pi}} \exp\bigg(-\frac{\eta U^2}{2}\bigg) + \kappa \gamma \sqrt{\frac{\eta}{2\pi(1+\eta)}}\exp\bigg(-\frac{\eta U^2}{2(1+\eta)}\bigg)}\right)^2\right]:=R_1  \label{eq242eqbHM}
  \end{align} which does not depend on $x_0$, where \eqref{eq242eqbHM} follows from \eqref{eq242eqHM}. Here, the expectation in \eqref{eq242eqbHM} is taken over $U$ with distribution (cf. \eqref{deffzHM})
  \begin{align}
  f_U^{(\rm{HM},1)}(u)&=\frac{(1-\gamma \kappa)\sqrt{\eta}}{\sqrt{2\pi}} \exp\bigg(-\frac{u^2 \eta}{2}\bigg)+ \frac{\kappa \gamma \sqrt{\eta}}{2\pi}  \int_{\bbR}\exp\bigg(-\frac{x^2+(u-x)^2 \eta}{2}\bigg)dx\\
  &=(1-\gamma \kappa)\sqrt{\frac{\eta}{2\pi}} \exp\bigg(-\frac{u^2 \eta}{2}\bigg)+ \kappa \gamma \sqrt{\frac{\eta}{2\pi(1+\eta)}} \exp\bigg(-\frac{u^2 \eta}{2(1+\eta)}\bigg).
  \end{align}

  Similarly, by symmetry, given $\Upsilon_0=1$, for any $x_0 \in \calX$, we also have:
  \begin{align}
  \bbE[\langle &\sfX_1|\sfX_0=1,\Upsilon_0=1\rangle^2]=\bbE\bigg[\sfX_1^2\bigg|\sfX_0=1,\Upsilon_0=1\bigg]-\bbE\bigg[(\sfX_1-\langle \sfX_1|\sfX_0=1\rangle)^2\bigg|\sfX_0=1,\Upsilon_0=1\bigg]\\
  &\qquad =1-(1-\kappa)\gamma-\bbE\bigg[\calE(1;\eta,\eta\mid 1,1)\bigg]\\
  &\qquad =\bbE\left[\left(\frac{ (1-(1-\kappa) \gamma)  \sqrt{\frac{\eta}{2\pi(1+\eta)}}\exp\bigg(-\frac{\eta U^2}{2(1+\eta)}\bigg)\bigg(\frac{\eta U}{1+\eta}\bigg)}{(1-\kappa)\gamma \sqrt{\frac{\eta}{2\pi}} \exp\bigg(-\frac{\eta U^2}{2}\bigg) + (1-(1-\kappa) \gamma) \sqrt{\frac{\eta}{2\pi(1+\eta)}}\exp\bigg(-\frac{\eta U^2}{2(1+\eta)}\bigg)}\right)^2\right]:=R_2  \label{eq243eqbHM}
  \end{align} which does not depend on $x_0$, where the expectation is taken over $U$ with distribution (cf. \eqref{deffzHMz})
  \begin{align}
  f_U^{(\rm{HM},2)}(u)=(1- \kappa)\gamma\sqrt{\frac{\eta}{2\pi}} \exp\bigg(-\frac{u^2 \eta}{2}\bigg)+ (1-(1-\kappa) \gamma) \sqrt{\frac{\eta}{2\pi(1+\eta)}} \exp\bigg(-\frac{u^2 \eta}{2(1+\eta)}\bigg) \label{deffzcu}.
  \end{align}
  
  Hence, from \eqref{horse1}, \eqref{eq242eqbHM}, and \eqref{eq243eqbHM}, we obtain from Claim \ref{claim:hiddenMarkov} that
  \begin{align}
  \rm{MMSE}_{\rm{HM}}&:= \lim_{n\to \infty} \frac{1}{n} \bbE[\|\bX-[\bX]\|_2^2]\\
  &=\kappa- \bigg[\bigg(\int_{\calX} \lambda_{x,0}^{(\pi_{\Upsilon})}\bigg)R_1+ \bigg(\int_{\calX} \lambda_{x,1}^{(\pi_{\Upsilon})}\bigg)R_2\bigg]\\
  &=\kappa- ((1-\kappa)R_1 + \kappa R_2) \label{eqbupi},
  \end{align} where $R_1$ and $R_2$ are defined in \eqref{eq242eqbHM} and \eqref{eq243eqbHM}, respectively. Here, \eqref{eqbupi} follows from \eqref{fact1E} and \eqref{fact2E}.

  \begin{figure}[H]
  	\centering
  	\includegraphics[width=0.7\linewidth]{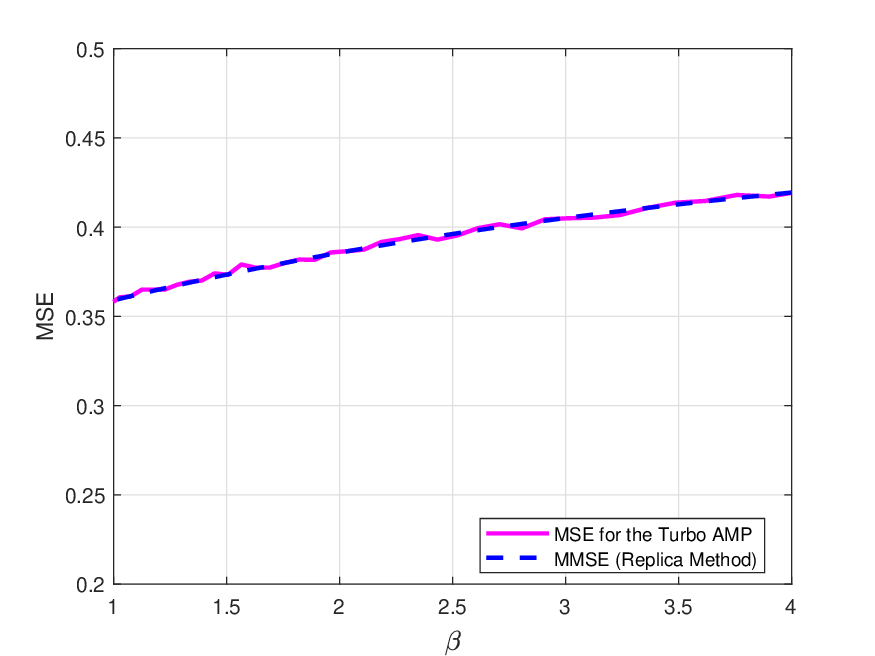}
  	\caption{MSE by Turbo AMP algorithm and MMSE by the Replica Method as functions of $\beta$ for the symmetric case $\kappa=0.5,\gamma=1$, i.e., $\{X_n\}_{n=1}^{\infty}$ is an i.i.d. sequence.}
  	\label{fig:HDMK4}
  \end{figure}
    \begin{figure}[H]
  	\centering
  	\includegraphics[width=0.7\linewidth]{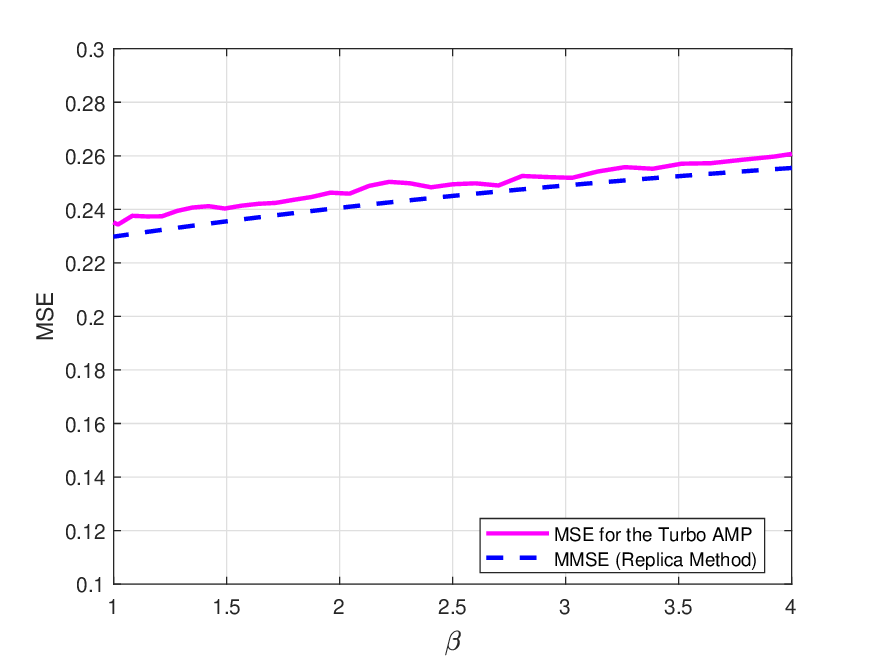}
  	\caption{MSE by Turbo AMP algorithm and MMSE by the Replica Method as functions of $\beta$ for the asymmetric case $\kappa=0.3$ and $\gamma=0.8$.}
  	\label{fig:HDMK6}
  \end{figure}
    
  In this section, we compare the MMSE in Claim \ref{claim:hiddenMarkov} with the MSE achieved by the AMP algorithm in \cite{Philip} for $n=1000$ (signal dimension) and $m=\lceil \frac{n}{\beta} \rceil$ (observations). We assume that $S=1$ and $\bA$ is a random matrix where each element is normal distributed $\calN(0,1/m)$ as Section \ref{sec:setting}. However, this algorithm assumes some level of sparsity in signal $\bX$. Before introducing the algorithm, we define some new functions:
  \begin{align}
  \alpha_l(c)&=\frac{1}{c+1},\\
  \beta_l(c)&=\bigg(\frac{1-\kappa}{\kappa}\bigg)\bigg(\frac{c+1}{c}\bigg),\\ \zeta_l(c)&=\frac{1}{c(c+1)},\\
  F_l(\theta;c)&=\frac{\alpha_l(c) \theta}{1+\beta_n(c)e^{-\zeta_l(c)|\theta|^2}},\\
  G_l(\theta;c)&=\beta_n(c)e^{-\zeta_n(c)|\theta|^2}|F_n(\theta;c)|^2+\frac{c}{\theta}F_l(\theta;c)\\
  F_l'(\theta;c)&=\frac{\alpha_l(c)}{1+\beta_l(c)e^{-\zeta_l(c)|\theta|^2}}\bigg[1+ \frac{\zeta_l(c)|\theta^2|}{1+(\beta_l(c)e^{-\zeta_l(c)|\theta|^2})^{-1}} \bigg], \quad \forall l \in \{1,2,\cdots,n\}.
  \end{align}
  We call this algorithm Turbo AMP since it is based on an approximation of a loopy BP which has demonstrated very accurate results in LDPC and Turbo decoding \cite{Philip}. The algorithm for our setting is as follows:
  \begin{enumerate}
  	\item Initialize 
  	\begin{align}
  	c^{0}=10;\quad  \mu_l^{0}=0 \quad \forall l \in \{1,2,\cdots,n\}; \quad z_k^{0}=y_k \quad \forall k \in \{1,2,\cdots,m\}.
  	\end{align}
  	\item Repeat the following for all $i=0,1,2,\cdots$ (we use $10$ iterations in our simulations):
  	\begin{align}
  	\theta_l^i&=\frac{1}{\sqrt{n}}\sum_{k=1}^m  A_{kl} z_k^i+\mu_l^i, \quad \forall l \in \{1,2,\cdots,n\},\\
  	\mu_l^{i+1}&=F_l(\theta_l^i;c^i), \quad \forall l \in \{1,2,\cdots,n\},\\
  	\upsilon_l^{i+1}&=G_l(\theta_l^i;c^i),  \quad \forall l \in \{1,2,\cdots,n\},\\
  	c^{i+1}&=1+ \frac{\beta}{n}\sum_{l=1}^n \upsilon_l^{i+1},\\
  	z_k^{i+1}&=y_k-\sum_{l=1}^n A_{kl}\mu_l^{i+1}+ \frac{z_k^i}{m}\sum_{l=1}^n F_l'(\theta_l^i;c^i), \quad \forall k \in \{1,2,\cdots,m\}.
  	\end{align}
  \end{enumerate}

  Our obtained results are as follows.
  \begin{itemize}
  	\item For the symmetric case $\kappa=0.5$ and $\gamma=1$, the Markov model in Section \ref{sec:setting} becomes the linear model with i.i.d. sequence $\{X_n\}_{n=1}^{\infty}$ in \cite[Sect. II]{GuoVerdu2005}. Fig. \ref{fig:HDMK4} shows that Turbo AMP works well for this case. The gap between the MSE of AMP and the MSE of the Replica Method in Claim \ref{claim:hiddenMarkov}  is very small. 
  	\item For the non-symmetric case $\kappa=0.3$ and $\gamma=0.8$, the Markov model in Section \ref{sec:setting} is very different from the linear model with i.i.d. sequence $\{X_n\}_{n=1}^{\infty}$  in \cite[Sect. II]{GuoVerdu2005}. However, Fig. \ref{fig:HDMK6} shows that Turbo AMP also works well for this case. The gap between the MSE of Turbo AMP and the upper bound of MSE by using the Replica Method in Claim \ref{claim:hiddenMarkov} is also still small. However, the gap is bigger than the symmetric case. The multiple fixed points (multiple solutions) of the equation \eqref{drawbetaHM} can be a reason for this gap.  Besides, Turbo AMP may not be optimal for this given model although it exploits the Markov structure of the sequence $\{X_n\}_{n=1}^{\infty}$ quite well. %More specially, this algorithm makes use of the knowledge of the stationary distribution (i.e., $1-\kappa,\kappa$) in its algorithm.  
  \end{itemize}
    
  \section{Proofs of main results} \label{sec:mainproof}
	
This section proves Claims \ref{claim:maincontri1}--\ref{claim:hiddenMarkov} using the replica method. We first state some preliminary results which are required to estimate the free energy of the linear model with Markov sources. Then, we obtain the joint moments for the linear model with Markov sources. Finally, we obtain the free energy and joint moments for the linear model with hidden Markov sources based on the results of the linear model with Markov sources.

	\begin{lemma}\cite[p. 1998]{GuoVerdu2005}\label{lem:gv1}
	 Let $X_n^{(a)}$ be replicated vectors with distribution $q_{\bX}$. Define a sequence of $(\nu+1) \times (\nu+1)$ random matrices $\{\bQ_n\}_{n=1}^{\infty}$ such that
	\begin{align}
	Q_n^{(a,b)}=S_n X_n^{(a)} X_n^{(b)}
	\end{align} for all $a,b \in \{0,1,\cdots,\nu\}$ and $n=1,2,\cdots$. Let
	\begin{align}
	\bT_n=\frac{1}{n}\sum_{k=1}^n \bQ_k, \quad n=1,2,\cdots.
	\end{align}
	Then, the following holds:
	\begin{align}
	\frac{1}{n}\log \bbE[Z^{\nu}(\bY,\bPhi)]=\frac{1}{n}\log \bbE\bigg\{ \exp\bigg[m \bigg(G^{(\nu)}(\bT_n)+O(n^{-1})\bigg)\bigg]\bigg\} \label{lieubieu},
	\end{align}
	where
	\begin{align}
	G^{(\nu)}(Q):=-\frac{1}{2}\log\det(I+\Sigma Q)-\frac{1}{2}\log\bigg(1+\frac{\nu}{\sigma^2}\bigg)-\frac{\nu}{2}\log(2\pi \sigma^2)\label{def:GQ},
	\end{align} and $\Sigma$ is a $(\nu+1)\times (\nu+1)$ matrix
	\begin{align}
	\Sigma=\frac{\beta}{\sigma^2+\nu}\begin{bmatrix}\nu & -e^T\\-e & (1+\frac{\nu}{\sigma^2})I-\frac{1}{\sigma^2}e e^T\end{bmatrix},
	\end{align} where $e$ is a $\nu \times 1$ column vector whose entries are all $1$.
\end{lemma} 

The following two lemmas state some new results on large deviations for Markov chains induced by the channel setting. The proofs of these results can be found in Appendix \ref{sec:largedev}. 
\begin{lemma}\label{simlemmod} 
	 Let $\{S_n\}_{n=1}^{\infty}$ be an i.i.d. sequence of random variable on a finite set $\calS \subset \bbR^+$. Let $\bX:=\{X_n\}_{n=1}^{\infty}$ be a Markov chain with states on a Polish space $\calX$ with the transition matrix $P=\{\pi(x,x')\}_{x, x' \in \calX}$. Assume this Markov chain is irreducible. Set $\bX^{(0)}=\bX$. Let $\bX^{(a)}:=\{X_{n}^{(a)}\}_{n=1}^{\infty}$ be a set of $\nu$ replica sequences with (postulated) distribution $q_{\bX}$ for each $a=1,2,\cdots,\nu$. This means that
	 \begin{align}
	 p_{\bX^{(0)}\bX^{(1)}\bX^{(2)} \cdots \bX^{(\nu)}}(x^{(0)},x^{(1)},x^{(2)},\cdots, x^{(\nu)}) \sim p_{\bX}(x^{(0)})\prod_{i=1}^{\nu} q_{\bX}(x^{(i)}),
	 \end{align}
	 where
	 \begin{align}
	 p_{\bX}(x^{(0)})&=\prod_{i=1}^{\infty}\pi(x_i^{(0)}, x_{i+1}^{(0)})\\
	 q_{\bX}(x^{(a)})&=\prod_{i=1}^{\infty}\tilde{\pi}(x_i^{(a)}, x_{i+1}^{(a)}), \qquad \forall a \in [\nu].
	 \end{align}
	 Define a new sequence of $(\nu+1) \times (\nu+1)$ random matrices $\{\bQ_n\}_{n=1}^{\infty}$ such that
	 \begin{align}
	 Q_n^{(a,b)}=S_n X_n^{(a)} X_{n}^{(b)}
	 \end{align} for all $a \in [\nu]$ and $b \in [\nu]$ and for all $n=1,2,\cdots$. Then, $\{\bQ_n\}_{n=1}^{\infty}$ is also an irreducible Markov chain  with states on $\calQ$, \blue{where $\calQ$ is defined in \eqref{defspaceQ}. In addition, the transition probability, namely $P(Q|Q')$, of this Markov chain satisfies:}
	 \blue{\begin{align}
	 	P(Q|Q')=\frac{\sum_{(s,x_0,x_1,\cdots,x_{\nu}, s',x'_0,x'_1,\cdots,x'_{\nu}) \in  \calA_Q \times \calA_{Q'}} P_S(s')P_S(s)  p_{X_{n-1}}(x'_0)\pi(x_0',x_0)\prod_{i=1}^{\nu} q_{X_{n-1}}(x'_i)  \tilde{\pi} (x'_i,x_i)}{\sum_{(s',x'_0,x'_1,\cdots,x'_{\nu}) \in  \calA_{Q'}} P_S(s') p_{X_{n-1}}(x'_0)\prod_{i=1}^{\nu} q_{X_{n-1}}(x'_i)},
	 	\end{align}
	 	where $p_{X_{n-1}}(\cdot)$ is the state distribution at time $n-1$ of the Markov chain $\{X_n\}_{n=1}^{\infty}$ with the transition probability $\pi$ defined in \eqref{defmc1}, and $q_{X_{n-1}}(\cdot)$ is the state distribution at time $n-1$ of the Markov chain $\{X_n\}_{n=1}^{\infty}$ with the (postulated) transition probability $\tilde{\pi}(\cdot,\cdot)$ defined in \eqref{priorq}, and
	 	\begin{align}
	 	\calA_Q:= \big\{(s,x) \in \calS \times \calX^{\nu+1}: sxx^T=Q\big\}, \quad \forall Q \in \calQ.
	 	\end{align}
	 }
\end{lemma}
\begin{lemma} \label{markovvaradmod}  Let $\calX$ be a Polish space with finite cardinality and a irreducible Markov chain $\bX:=\{X_n\}_{n=1}^{\infty}$ defined on $\calX$ and $\nu$ be a positive integer number. Let $X^{(a)}_n$ for $a \in [\nu]$ be replicas of the Markov process $\bX$. Recall the definition of the sequence $\bQ_n$ in Lemma~\ref{simlemmod} and $\bT_n=\frac{1}{n}\sum_{j=1}^n \bQ_j$. Let $P_n(U):=\bbP(\bT_n \in U)$ for any measurable set $U$ on the $\sigma$-algebra generated by $\{\bQ_n\}_{n=1}^{\infty}$. Then, for and bounded and continuous function $F: \calQ \to \bbR$
	\begin{align}
	\lim_{n\to \infty} \frac{1}{n}\log \bbE\big[e^{nF(\bT_n)}\big]&=\lim_{n\to \infty} \frac{1}{n}\log \int e^{nF(Q)} dP_n(Q) \label{eq34bmod} \\
	&=\sup_{Q}\bigg[F(Q)-I(Q)\bigg] \label{eq35bmod}
	\end{align} where $I(Q)=\sup_{\tilQ}(\mathrm{tr}(\tilQ Q) -\log \rho(P_{
		\tilQ}))$ and $\rho (P_{\tilQ})$ is the Perron-Frobenius eigenvalue of the matrix \blue{$P_{\tilQ}=\{e^{\mathrm{tr}(\tilQ \barQ_j)} P_{\barQ_j|\barQ_i}\}_{0 \leq i,j\leq M}$}  and $M=|\calQ|-1$,
	\blue{where $\calQ$ and $\{\barQ_i\}_{i=0}^M$ are defined in Subsection \ref{sub:nota}.}
\end{lemma}

\begin{lemma} \label{lem18mod} \blue{Recall the definitions of  $\{\barQ_i\}_{i=0}^M$ in Subsection \ref{sub:nota}. Recall the definition of the sequence $\bQ_n$ in Lemma~\ref{simlemmod}.} Then, the following holds:
	\begin{align}
	\frac{\partial \log \rho(P_{\tilQ})}{\partial \tilQ}(\tilQ)=\frac{1}{\rho(P_{\tilQ})}\sum_{i=0}^M \lambda_i(\tilQ)\sum_{j=0}^M \psi_j(\tilQ)\barQ_j P(\barQ_j|\barQ_i)e^{\mathrm{tr}(\tilQ \barQ_j)},\label{colremod}
	\end{align} 
	where $\lambda(\tilQ)$ and $\psi(\tilQ)$ are left and right eigenvectors associated with the Perron-Frobenius eigenvalue $\rho(P_{\tilQ})$ which are normalized such that $\lambda(\tilQ)^T \psi(\tilQ)=1$. 
\end{lemma}
\begin{IEEEproof}
Refer to Appendix \ref{sec:perronfrobenius} for a detailed proof.
\end{IEEEproof}

\begin{theorem} \label{mainthm1} Recall the definition of $G^{(\nu)}(Q)$ in Lemma \ref{lem:gv1}. In the large system limit, given any initial state $x_0$, the free energy satisfies:
	\begin{align}
	\calF_q\big|_{X_0=x_0}=-\lim_{\nu \to 0}\frac{\partial}{\partial \nu}\sup_Q \bigg[\beta^{-1}G^{(\nu)}(Q)-I^{(\nu)}(Q)\bigg]\label{eq32},
	\end{align} where
	\begin{align}
	I^{(\nu)}(Q):=\sup_{\tilQ}\bigg[\mathrm{tr}(\tilQ Q) -\log \rho(P_{
		\tilQ})] \label{eq33},
	\end{align} and $\rho (P_{\tilQ})$ is the Perron-Frobenius eigenvalue of the matrix \blue{$P_{\tilQ}=\{e^{\mathrm{tr}(\tilQ \barQ_j)} P_{\barQ_j|\barQ_i}\}_{0\leq i,j\leq M}$} and  $M=|\calQ|-1$ where $\calQ:=\{s x x^T \enspace \mbox{for some} \enspace s \in \calS, \enspace x \in \calX^{\nu+1}\}$.
\end{theorem}
\begin{IEEEproof}
	The proof follows the same idea as \cite[Part A, Sect. IV]{GuoVerdu2005} with some important changes to account for the Markov setting. 
	\begin{enumerate}
		\item By applying Lemma \ref{markovvaradmod}, from \eqref{lieubieu}, we obtain
		\begin{align}
		\lim_{n\to \infty} \frac{1}{n}\log \bbE[Z^{\nu}(\bY,\bPhi)]&=	\lim_{n\to \infty}\frac{1}{n}\log \bbE\bigg\{ \exp\bigg[\frac{n}{\beta} \bigg(G^{(\nu)}(\bT_n)+O(n^{-1})\bigg)\bigg]\bigg\}\\
		&=\sup_{Q} \bigg[\frac{1}{\beta}G^{(\nu)}(Q)-I^{(\nu)}(Q)\bigg] \label{bana1}.
		\end{align}
		\item Estimate the free energy.\\
		
		Now, observe that
		\begin{align}
		\calF_q\big|_{X_0=x_0}&=-\lim_{n\to \infty}\frac{1}{n} \lim_{\nu \to 0} \frac{\partial}{\partial\nu}\log \bbE[Z^{\nu}(\bY,\bPhi)]\label{culi1}\\
		&=- \lim_{\nu \to 0}\frac{\partial}{\partial\nu}\lim_{n\to \infty} \frac{1}{n}\log \bbE[Z^{\nu}(\bY,\bPhi)]\label{culi2}\\
		&=- \lim_{\nu \to 0}\frac{\partial}{\partial\nu}\sup_{Q} \bigg[\frac{1}{\beta}G^{(\nu)}(Q)-I^{(\nu)}(Q)\bigg]\bigg|_{X_0=x_0}\label{culi3}
		\end{align} where \eqref{culi1} follows from the assumption (A1), (A2), and the fact (F1), \eqref{culi2} follows from the assumption (A3), and \eqref{culi3} follows from  \eqref{bana1}.
	\end{enumerate}
\end{IEEEproof}
\begin{theorem} \label{thmQstar} Recall the definitions of $\Sigma$ in Lemma \ref{lem:gv1}, the matrix $P_{\tilQ}$ in Theorem \ref{mainthm1}, and \blue{$\{\barQ_i\}_{i=0}^M$ in Subsection \ref{sub:nota}}. The optimal matrix $Q^*$ of equation \eqref{eq32} in Theorem \ref{mainthm1} must satisfy the following constraints:
	\begin{align}
	Q^*&=\frac{\partial \log \rho(P_{\tilQ^*})}{\partial \tilQ^*},\label{core0}\\
	\tilQ^*&=-(2\beta)^{-1}(I+\Sigma Q^*)^{-1}\Sigma \label{core1},\\
	\frac{\partial \log \rho(P_{\tilQ^*})}{\partial \tilQ^*}&=\frac{1}{\rho(P_{\tilQ^*})}\sum_{i=0}^M \lambda_i(\tilQ^*)\sum_{j=0}^M \psi_j(\tilQ^*)\barQ_j P(\barQ_j|\barQ_i)e^{\mathrm{tr}(\tilQ^* \barQ_j)},\label{colre2}
	\end{align} 
	where $\lambda(\tilQ^*)$ and $\psi(\tilQ^*)$ are left and right eigenvectors associated with the Perron-Frobenius eigenvalue $\rho(P_{\tilQ^*})$ which are normalized such that $\lambda(\tilQ^*)^T \psi(\tilQ^*)=1$. 
\end{theorem}
\begin{IEEEproof}
	Recall the definition of $I^{(\nu)}$ in Theorem \ref{mainthm1}. It is easy to see that the optimization problem in \eqref{bana1} is equivalent to the following optimization problem:
	\begin{align}
	\sup_Q \inf_{\tilQ} T^{(\nu)}(Q,\tilQ) \label{bana2}
	\end{align} 
	where
	\begin{align}
	T^{(\nu)}(Q,\tilQ):=-\frac{1}{2\beta}\log\det(I+\Sigma Q)-\mathrm{tr}(\tilQ Q)+\log \rho(P_{\tilQ})-\frac{1}{2\beta}\log\big(1+\frac{\nu}{\sigma^2}\big)-\frac{\nu}{2\beta}\log(2\pi \sigma^2).
	\end{align}
	For an arbitrary $Q$, we first seek critical points with respect to $\tilQ$ and find that for any given $Q$, the extremum in $\tilQ$ satisfies
	\begin{align}
	Q=\frac{\partial \log \rho(P_{\tilQ})}{\partial \tilQ} \label{eq112verdu}
	\end{align}
	Let $\tilQ(Q)$ be a solution to \eqref{eq112verdu}. We then seek the critical point of $T^{(\nu)}(Q,\tilQ(Q))$ with respect to $Q$. 
	
	\begin{align}
	K_{Q,\tilQ}:=\begin{bmatrix}\frac{\partial \tilQ_{0,0}}{\partial Q_{0,0}} & \frac{\partial \tilQ_{0,1}}{\partial Q_{0,1}}& \cdots & \frac{\partial \tilQ_{0,\nu}}{\partial Q_{0,\nu}}\\
	\frac{\partial \tilQ_{1,0}}{\partial Q_{1,0}} & \frac{\partial \tilQ_{1,1}}{\partial Q_{1,1}}& \cdots & \frac{\partial \tilQ_{1,\nu}}{\partial Q_{1,\nu}}\\ \vdots & \vdots & \ddots& \vdots \\ \frac{\partial \tilQ_{\nu,0}}{\partial Q_{\nu,0}} & \frac{\partial \tilQ_{\nu,1}}{\partial Q_{\nu,1}}& \cdots & \frac{\partial \tilQ_{\nu,\nu}}{\partial Q_{\nu,\nu}}  \end{bmatrix} \in \bbR^{\nu+1 \times \nu+1}.
	\end{align}
	
	Observe that
	\begin{align}
	\frac{\partial \mathrm{tr}(\tilQ Q)}{\partial Q}&=\frac{\partial \mathrm{tr}( Q\tilQ)}{\partial Q}\\
	&=\tilQ + Q\odot K_{Q,\tilQ} \label{eq105},
	\end{align} where $\odot$ is the Hadamard product.
	
	It follows that
	\begin{align}
	\frac{\partial T^{(\nu)}(Q,\tilQ)}{\partial Q}&=-\frac{1}{2\beta}(I+\Sigma Q)^{-1}\Sigma- \bigg(\tilQ +  Q\odot K_{Q,\tilQ}\bigg)+ \frac{\partial \log \rho(P_{\tilQ})}{\partial Q} \label{eq107}\\
	&=-\frac{1}{2\beta}(I+\Sigma Q)^{-1}\Sigma- \bigg(\tilQ +  Q\odot K_{Q,\tilQ}\bigg)+ \frac{\partial \log \rho(P_{\tilQ})}{\partial \tilQ} \odot K_{Q,\tilQ} \\
	&=-\frac{1}{2\beta}(I+\Sigma Q)^{-1}\Sigma-\tilQ -\bigg[ Q-\frac{\partial \log \rho(P_{\tilQ})}{\partial \tilQ} \bigg]\odot K_{Q,\tilQ} \\
	&=-\frac{1}{2\beta}(I+\Sigma Q)^{-1}\Sigma-\tilQ,
	\label{eq109}
	\end{align} where~\eqref{eq107} follows from \eqref{eq105}, and \eqref{eq109} follows from \eqref{eq112verdu}. Hence, the optimal value of the Theorem \ref{mainthm1} is the solution of the following equation systems:
	\begin{align}
	Q&=\frac{\partial \log \rho(P_{\tilQ})}{\partial \tilQ},\\
	\tilQ&=-(2\beta)^{-1}(I+\Sigma Q)^{-1}\Sigma.
	\end{align}
	Finally, from Lemma \ref{lem18mod}, we also obtain an additional constraint in \eqref{colre2}.
\end{IEEEproof}

Observe that the matrix $\Sigma$ defined in Lemma \ref{lem:gv1} is invariant if two non-zero indices are interchanged, i.e., $\Sigma$ is symmetric in replicas. Now, we use the RS assumption (A4) to simplify the result in Theorem \ref{mainthm1}. More specifically, we use the following RS assumption:
\begin{definition} \cite[p. 1999]{GuoVerdu2005} \label{defrs} An solution $(\tilQ^*,Q^*)$ of the optimization problem in Theorem \ref{mainthm1}, i.e.,
	\begin{align}
	&\sup_Q \bigg[\beta^{-1}G^{(\nu)}(Q)-I^{(\nu)}(Q)\bigg]\nn\\
	&=\sup_Q \inf_{\tilQ} \bigg[-\frac{1}{2\beta}\log\det(I+\Sigma Q)-\mathrm{tr}(\tilQ Q)+\log \rho(P_{\tilQ})-\frac{1}{2\beta}\log\bigg(1+\frac{\nu}{\sigma^2}\bigg)-\frac{\nu}{2\beta}\log(2\pi \sigma^2)\bigg] \label{keyQ},
	\end{align} is called to satisfy the Replica Symmetry (RS) if both $Q^*$ and $\tilQ^*$ are invariant under the exchange of any two (nonzero) replica indices. In other words, the extrema can be written as
	\begin{align}
	Q^*&=\begin{bmatrix}r & m&m &\cdots &m \\
	m &p &q &\cdots& q\\
	m& q & p& \ddots& \vdots\\
	\vdots& \vdots& \ddots& \ddots& q\\
	m&q&\cdots&q&p \end{bmatrix},\\
	\tilQ^*&=\begin{bmatrix}c & d&d &\cdots &d \\
	d &g &f &\cdots& f\\
	d& f & g& \ddots& \vdots\\
	\vdots& \vdots& \ddots& \ddots& f\\
	d&f&\cdots&f&g \end{bmatrix} \label{tilQmax},
	\end{align} where $r,m,p,q,c,d,f,g$ are some real numbers which are not dependent on $\nu$.
\end{definition}

Next, we show the following results:
\begin{lemma}\label{repre1} Let $\{\barQ_i\}_{i=0}^M$ be states of the Markov chain $\{\bQ_n\}_{n=1}^{\infty}$ in Lemma \ref{simlemmod}. Assume that
	\begin{align}
	\rho(P_{\tilQ^*}) \to 1 \quad \mbox{and} \quad \sum_{j=0}^M  P(\barQ_j|\barQ_i)e^{\mathrm{tr}(\tilQ^* \barQ_j)} \to 1  \label{eqkool2}
	\end{align}  for all $i \in [M]$ as $\nu \to 0$. 
	Then, under the RS assumption in Definition \ref{defrs}, the following holds:
	\begin{align}
	Q^* =\lim_{\nu \to 0} \sum_{i=0}^M \lambda_i(\tilQ^*)\bbE[\bQ_1 e^{\mathrm{tr}(\tilQ^* \bQ_1)}\big|\bQ_0=\barQ_i]
	\end{align} where $Q^*$ is defined in Theorem \ref{thmQstar} and  $\lambda(\tilQ^*)$ is a left (positive) eigenvector associated with the Perron-Frobenius eigenvalue $\rho(P_{\tilQ^*})$ such that $\|\lambda(\tilQ^*)\|_1 = 1$. 
	\red{In addition, we have
		\begin{align}
		\rho(P_{\tilQ^*})=\sum_{i=1}^M \lambda_i(\tilQ^*)\bbE[e^{\rm{tr}(\tilQ^* \bQ_1)}|\bQ_0=\barQ_i]  \label{taggy}.
		\end{align}}
\end{lemma}
\begin{IEEEproof} Since $\psi(\tilQ^*)$ is the right eigenvector associated with the Perron-Frobenius eigenvalue of the matrix $P_{\tilQ^*}$, it holds that
	\begin{align}
	\sum_{j=0}^M P(\barQ_j|\barQ_i)e^{\mathrm{tr}(\tilQ^* \barQ_j)}\psi_j(\tilQ^*)=\rho(P_{\tilQ^*})\psi_i(\tilQ^*) \label{eqkool1}
	\end{align} for all $i \in [M]$. From \eqref{eqkool1} and \eqref{eqkool2}, we can set $\psi(\tilQ^*)=(1,1,\cdots, 1)^T$ is a right eigenvector associated with the eigenvalue $ \rho(P_{\tilQ^*})$ as $\nu \to 0$.
	
	Hence, from Theorem \ref{thmQstar}, we have
	\begin{align}
	Q^*=\lim_{\nu \to 0}\sum_{i=0}^M \lambda_i(\tilQ^*)\bbE[\bQ_1 e^{\mathrm{tr}(\tilQ^* \bQ_1)}\big|\bQ_0=\barQ_i].
	\end{align}
	Now, since by Theorem \ref{thmQstar}, it holds that
	\begin{align}
	\sum_{j=0}^M \psi_j(\tilQ^*)\lambda_j(\tilQ^*)=1,
	\end{align} so we have
	\begin{align}
	\|\lambda(\tilQ^*)\|_1 = 1 \label{ca1}.
	\end{align}
	Now, since $\lambda(\tilQ^*):=(\lambda_0(\tilQ^*),\lambda_1(\tilQ^*),\cdots, \lambda_M(\tilQ^*))$ is the left (positive) eigenvector associated with $\rho(P_{\tilQ^*})$, it holds that
	\begin{align}
	\lambda_j(\tilQ^*)\rho(P_{\tilQ^*})=\sum_{i=0}^M \lambda_i(\tilQ^*) e^{\rm{tr}(\tilQ^* \barQ_j)}P(\barQ_j|\barQ_i) \label{ca3}.
	\end{align} 
	Then, it follows that
	\begin{align}
	\rho(P_{\tilQ^*})&=\sum_{j=0}^M \lambda_j(\tilQ^*)\rho(P_{\tilQ^*}) \label{ca2}\\
	&=\sum_{j=0}^M\sum_{i=1}^M \lambda_i(\tilQ^*) e^{\rm{tr}(\tilQ^* \barQ_j)}P(\barQ_j|\barQ_i)  \label{ca4}\\
	&=\sum_{i=0}^M \lambda_i(\tilQ^*) \sum_{j=0}^M e^{\rm{tr}(\tilQ^* \barQ_j)}P(\barQ_j|\barQ_i)\\
	&=\sum_{i=0}^M \lambda_i(\tilQ^*)\bbE[e^{\rm{tr}(\tilQ^* \bQ_1)}|\bQ_0=\barQ_i] \label{ca5},
	\end{align} where \eqref{ca2} follows from \eqref{ca1}, \eqref{ca4} follows from \eqref{ca3}.
\end{IEEEproof}
\begin{lemma}\label{rhoto1} Under the RS assumption in Definition \ref{defrs}, as $\nu \to 0$, the following hold:
	\begin{align}
	\rho(P_{\tilQ^*}) \to 1 \quad \mbox{and} \quad \sum_{j=0}^M  P(\barQ_j|\barQ_i)e^{\mathrm{tr}(\tilQ^* \barQ_j)} \to 1  \label{eqkool2b}.
	\end{align}  
Furthermore, it holds that
	\begin{align}
	 \frac{\partial \log \rho(P_{\tilQ^*})}{\partial \nu}\bigg|_{\nu=0}&=-\frac{\xi}{2}\bigg(\bbE[S] \bbE_{\sfX_0 \sim \lambda^{(\pi)}}\bigg[\bbE[\sfX_1^2|\sfX_0]\bigg]+\frac{1}{\eta}\bigg)\log e+ \frac{1}{2}\log \frac{2\pi}{\xi}\nn\\
	&\qquad \qquad + \bbE_{\sfX_0 \sim \lambda^{(\pi)}}\bigg[\bbE_{S}\bigg\{\int_{\bbR} p_{U|\sfX_0,S;\eta}(u|x_0,S;\eta)  \log q_{U|\sfX_0,S;\eta}(u|x_0,S;\eta) du \bigg\}\bigg]\label{latax}.
	\end{align}
\end{lemma}
%\begin{remark} For the i.i.d. case, Guo and Verd\'{u} \cite{GuoVerdu2005} already pointed out that 
%	{\begin{align}
%	\lim_{\nu \to 0}\frac{\partial \log \calM(\tilQ^*)}{\partial \nu} = \label{latax},
%	\end{align}} where $\calM(\tilQ^*)=\bbE[e^{\mathrm{tr}(\tilQ^* \bQ)}]$. Lemma \ref{rhoto1} shows that these facts still hold in our more general setting. The following proof follows the same steps as in \cite{GuoVerdu2005}.
%\end{remark}
\begin{IEEEproof}
By Lemma \ref{simlemmod}, $\bQ_0- \bQ_1- \cdots - \bQ_n$ forms a Markov chain \blue{on the state-space  $\{\barQ_i\}_{i=0}^M$ defined in Subsection \ref{sub:nota}} with the following transition matrix:
	\begin{align}
	P_{\tilQ}=\left[\begin{array}{cccc} P(\barQ_0|\barQ_0)e^{\mathrm{tr}(\tilQ \barQ_0)}& P(\barQ_1|\barQ_0)e^{\mathrm{tr}(\tilQ \barQ_1)}& \cdots & P(\barQ_M|\barQ_0)e^{\mathrm{tr}(\tilQ \barQ_M)}\\ P(\barQ_0|\barQ_1)e^{\mathrm{tr}(\tilQ \barQ_0)}& P(\barQ_1|\barQ_1)e^{\mathrm{tr}(\tilQ \barQ_1)}& \cdots & P(\barQ_M|\barQ_1)e^{\mathrm{tr}(\tilQ \barQ_M)} \\ \vdots & \vdots & \vdots & \vdots \\ P(\barQ_0|\barQ_M)e^{\mathrm{tr}(\tilQ \barQ_0)} & P(\barQ_1|\barQ_M)e^{\mathrm{tr}(\tilQ \barQ_1)}& \cdots & P(\barQ_M|\barQ_M)e^{\mathrm{tr}(\tilQ \barQ_M)}\end{array}\right].
	\end{align}
	where
	\begin{align}
	P(\barQ_j|\barQ_i):=\bbP(\bQ_1=\barQ_j|\bQ_0=\barQ_i)
	\end{align} and $\bQ_0$ and $\bQ_1$ are random (state) matrices at time $0$ and $1$, respectively.
	
	By \cite{Lancaster1985b}, we have
	\begin{align}
	\min_{i\in [M]}\sum_{j=0}^M P(\bQ_1=\barQ_j|\bQ_0=\barQ_i)e^{\mathrm{tr}(\tilQ^* \barQ_j)} \leq \rho(P_{\tilQ^*}) \leq \max_{i \in [M]} \sum_{j=0}^M P(\bQ_1=\barQ_j|\bQ_0=\barQ_i)e^{\mathrm{tr}(\tilQ^* \barQ_j)}.
	\end{align} It follows that
	\begin{align}
	\min_{i\in [M]}\bbE\bigg[e^{\mathrm{tr}(\tilQ^* \bQ_1)}|\bQ_0=\barQ_i\bigg]\leq \rho(P_{\tilQ^*}) \leq \max_{i\in [M]}\bbE\bigg[e^{\mathrm{tr}(\tilQ^* \bQ_1)}|\bQ_0=\barQ_i\bigg] \label{tofact}.
	\end{align}	
	\blue{First, we show that 
	\begin{align}
	\lim_{\nu \to 0} \rho(P_{\tilQ^*})=1 \label{cufact1}.
	\end{align}}
	To show \eqref{cufact1}, it is enough to show that
	\begin{align}
	\bbE\bigg[e^{\mathrm{tr}(\tilQ^* \bQ_1)}\big|\bQ_0=\barQ_i\bigg]\to 1
	\end{align} for all $i \in [M]$. Indeed, by the definition of $\{\bQ_n\}_{n=1}^{\infty}$ in Lemma \ref{simlemmod}, we have $\bQ_1=S_1 \bX_1 \bX_1^T$. Hence, we have
	\begin{align}
	&\bbE\bigg[e^{\mathrm{tr}(\tilQ^* \bQ_1)}\big|\bQ_0=\barQ_i\bigg]=\bbE\bigg[e^{S_1 \bX_1\tilQ^* \bX_1^T}\bigg|\bQ_0=\barQ_i\bigg] \label{eq259}\\
	&=\bbE\bigg[e^{S_1 \bX_1\tilQ^* \bX_1^T}\bigg|\bQ_0=\barQ_i\bigg]\\
	&=\bbE\bigg[\exp\bigg(S_1\bigg[2d \sum_{a=1}^{\nu}X_1^{(0)}X_1^{(a)}+2f \sum_{1\leq a<b\leq \nu}X_1^{(a)}X_1^{(b)}+c\big(X_1^{(0)}\big)^2 +g \sum_{a=1}^{\nu}\big(X_1^{(a)}\big)^2\bigg]\bigg)\bigg|\bQ_0=\barQ_i\bigg]\label{bunhi1},
	\end{align} where \eqref{bunhi1} follows from RS assumption in Definition \ref{defrs}.
	
	Now, the eight parameters $(r,m,p,q,f,g)$ that define $Q^*$ and $\tilQ^*$ are the solution to the joint equations \eqref{core0} and \eqref{core1} in Theorem \ref{thmQstar}. Using \eqref{core1}, it can be shown that \cite[Eq. (123)]{GuoVerdu2005}
	\begin{align}
	c&=0 \label{factc},\\
	d&=\frac{1}{2[\sigma^2+\beta(p-q)]},\\
	f&=\frac{1+\beta(r-2m+q)}{2[\sigma^2+\beta(p-q)]^2},\\
	g&=f-d.
	\end{align} 
	Now, define
	\begin{align}
	\eta=\frac{2d^2}{f}, \quad \xi=2d \label{factxi}.
	\end{align}
In addition, for the simplicity of presentation, let $S:=S_1$. Then, by using some algebraic calculation and using the following interesting identity
	\begin{align}
	e^{x^2}=\sqrt{\frac{\eta}{2\pi}}\int \exp\bigg[-\frac{\eta}{2}u^2+\sqrt{2\eta}xu\bigg]du, \quad \forall x,\eta,
	\end{align}
	from \eqref{bunhi1}, we have (cf. a similar formula in \cite[Eq. (125)]{GuoVerdu2005}):
	\begin{align}
	&\bbE\bigg[e^{\mathrm{tr}(\tilQ^* \bQ_1)}\big|\bQ_0=\barQ_i\bigg]\\
	&\qquad=\bbE\bigg[\sqrt{\frac{\eta}{2\pi}}\int \exp\bigg[-\frac{\eta}{2}(u-\sqrt{S} X_1)^2\bigg]\nn\\
	&\qquad \qquad \times  \bigg[\bbE_{q}\bigg\{\exp\bigg[-\frac{\xi}{2}u^2-\frac{\xi}{2}\big(u-\sqrt{S} X\big)^2\bigg]\bigg|\bQ_0=\barQ_i\bigg\}\bigg]^{\nu} du\bigg|\bQ_0=\barQ_i\bigg] \label{eq308} \\
	& \qquad \to \bbE\bigg[\sqrt{\frac{\eta}{2\pi}}\int \exp\bigg[-\frac{\eta}{2}(u-\sqrt{S} X_1)^2 \bigg]du\bigg|\bQ_0=\barQ_i\bigg] \label{cubw}\\
	&\qquad =\bbE\big[1\big|\bQ_0=\barQ_i\big]\\
	&\qquad =1, 
	\end{align} where \eqref{cubw} follows from the dominated convergence theorem \cite{Royden}. Here, as above, we note that the conditional event $\{\bQ_0=\barQ_i\}$ only affects the distributions of $S$, $\sfX$, and $\sfX_1$.

Next, we prove that 
	\begin{align}
	\lim_{\nu \to 0}\frac{\partial \rho(\tilQ^*)}{\partial \nu}&=\bigg(\frac{1}{\log e}\bigg)\bigg(-\frac{\xi}{2}\bigg(\bbE[S] \bbE_{\sfX_0 \sim \lambda^{(\pi)}}\bigg[\bbE[\sfX_1^2|\sfX_0]\bigg]+\frac{1}{\eta}\bigg)\log e+ \frac{1}{2}\log \frac{2\pi}{\xi}\nn\\
	&\qquad \qquad + \bbE_{\sfX_0 \sim \lambda^{(\pi)}}\bigg[\bbE_{S}\bigg\{\int_{\bbR} p_{U|\sfX_0,S;\eta}(u|x_0,S;\eta)  \log q_{U|\sfX_0,S;\eta}(u|x_0,S;\eta) du \bigg\}\bigg]\bigg) \label{cufact2a}.
	\end{align}
Indeed, at $\nu=0$, it holds from \eqref{eq308} that
	\begin{align}
	\min_{i\in [M]}\bbE\bigg[e^{\mathrm{tr}(\tilQ^* \bQ_1)}|\bQ_0=\barQ_i\bigg]\bigg|_{\nu=0}=1 = \max_{i\in [M]}\bbE\bigg[e^{\mathrm{tr}(\tilQ^* \bQ_1)}|\bQ_0=\barQ_i\bigg]\bigg|_{\nu=0} \label{eq312}.
	\end{align}
Therefore, from \eqref{tofact} and \eqref{eq312}, it holds that
\begin{align}
\rho(P_{\tilQ^*})\big|_{\nu=0} = 1 \label{tofact2}.
\end{align}

On the other hand, observe that 
\begin{align}
\bbE\bigg[e^{\mathrm{tr}(\tilQ^* \bQ_1)}\bigg|\bQ_0=\barQ_i\bigg]&=\bbE\bigg[e^{S \bX_1\tilQ^* \bX_1^T}\bigg|\bQ_0=\barQ_i\bigg] \label{eq59b}\\
&=\bbE\bigg\{\sqrt{\frac{\eta}{2\pi}} \int_{\bbR} \exp\bigg[-\frac{\eta}{2}\big(u-\sqrt{S} \sfX_1\big)^2\bigg]\nn\\
&\qquad \qquad \times \bigg[\bbE_{q}\bigg\{\exp\bigg[-\frac{\xi}{2}u^2-\frac{\xi}{2}(u-\sqrt{S}\sfX)^2\bigg] \bigg|S\bigg\}\bigg]^{\nu} du \bigg|\bQ_0=\barQ_i\bigg\} \label{eq60b}
\end{align} where \eqref{eq59b} follows from \eqref{eq259}, \eqref{eq60b} follows by using \eqref{factc}--\eqref{factxi} (see \cite[Eq.~(125)]{GuoVerdu2005}). Hence, we have
\begin{align}
&\frac{\partial}{\partial \nu} \bbE\bigg[e^{\mathrm{tr}(\tilQ^* \bQ_1)}\bigg|\bQ_0=\barQ_i\bigg]\nn\\
&\qquad =\bigg(\frac{1}{\log e}\bigg)\bbE\bigg\{\sqrt{\frac{\eta}{2\pi}} \int_{\bbR} \exp\bigg[-\frac{\eta}{2}\big(u-\sqrt{S} \sfX_1\big)^2\bigg]\nn \\
&\qquad \qquad \times \bigg[\bbE_{q}\bigg\{\exp\bigg[-\frac{\xi}{2}u^2-\frac{\xi}{2}(u-\sqrt{S}\sfX)^2\bigg] \bigg|S\bigg\}\bigg]^{\nu} \nn\\
&\qquad \qquad \qquad  \times \log \bigg(\bbE_{q}\bigg\{\exp\bigg[-\frac{\xi}{2}u^2-\frac{\xi}{2}(u-\sqrt{S}\sfX)^2\bigg] \bigg|S\bigg\}\bigg)du \bigg|\bQ_0=\barQ_i\bigg\}
\end{align}

Hence, given $\sfX_0=x_0$ and $S_0=s_0$, it holds from \eqref{eq60b} that
\begin{align}
&\lim_{\nu \to 0} \frac{\partial}{\partial \nu}\bbE\bigg[e^{\mathrm{tr}(\tilQ^* \bQ_1)}\bigg|\bQ_0=\barQ_i\bigg] \nn\\
&\qquad =\lim_{\nu \to 0}\bigg(\frac{1}{\log e}\bigg)\bbE\bigg\{\sqrt{\frac{\eta}{2\pi}} \int_{\bbR} \exp\bigg[-\frac{\eta}{2}\big(u-\sqrt{S} \sfX_1\big)^2\bigg]\nn \\
&\qquad \qquad \times \bigg[\bbE_{q}\bigg\{\exp\bigg[-\frac{\xi}{2}u^2-\frac{\xi}{2}(u-\sqrt{S}\sfX)^2\bigg] \bigg|S\bigg\}\bigg]^{\nu} \nn\\
&\qquad \qquad \qquad  \times \log \bigg(\bbE_{q}\bigg\{\exp\bigg[-\frac{\xi}{2}u^2-\frac{\xi}{2}(u-\sqrt{S}\sfX)^2\bigg] \bigg|S\bigg\}\bigg)du \bigg|\bQ_0=\barQ_i\bigg\} \\
&\qquad =\bigg(\frac{1}{\log e}\bigg)\bbE\bigg\{\sqrt{\frac{\eta}{2\pi}} \int_{\bbR} \exp\bigg[-\frac{\eta}{2}\big(u-\sqrt{S} \sfX_1\big)^2\bigg]\nn\\
&\qquad \qquad \times \log \bigg(\bbE_{q}\bigg\{\exp\bigg[-\frac{\xi}{2}u^2-\frac{\xi}{2}(u-\sqrt{S}\sfX)^2\bigg] \bigg|S\bigg\}\bigg)du \bigg|\sfX_0=x_0\bigg\} \label{culi} \\
&\qquad= \bigg(\frac{1}{\log e}\bigg)\bigg(-\frac{\xi}{2} \bbE\bigg\{\sqrt{\frac{\eta}{2\pi}} \int_{\bbR} \exp\bigg[-\frac{\eta}{2}\big(u-\sqrt{S} \sfX_1\big)^2\bigg] u^2 du\bigg|\sfX_0=x_0\bigg\}\nn\\
&\qquad \qquad+ \bbE\bigg\{\int_{\bbR} p_{U|\sfX_0,\sfX_1,S;\eta}(u|\sfX_0,\sfX_1,S;\eta) \nn\\
&\qquad \qquad \times \log \bigg(\bbE_{q}\bigg\{\sqrt{\frac{2\pi}{\xi}} q_{U|\sfX_0,\sfX_1,S;\eta}(u|\sfX_0,\sfX_1,S;\eta) \bigg|S\bigg\}\bigg)du \bigg|\sfX_0=x_0\bigg\}\bigg)\\
&\qquad=  \bigg(\frac{1}{\log e}\bigg)\bigg(-\frac{\xi}{2}\bigg(\bbE[S] \bbE[\sfX_1^2|\sfX_0=x_0]+\frac{1}{\eta}\bigg)\log e+ \frac{1}{2}\log \frac{2\pi}{\xi}\\
&\qquad \qquad+ \bbE_{S}\bigg\{\int_{\bbR} \bbE_{\pi(x_0,\cdot)}\bigg[p_{U|\sfX_0,\sfX_1,S;\eta}(u|x_0,\sfX_1,S;\eta)\bigg|S\bigg] \nn\\
&\qquad \qquad \times \log \bigg(\bbE_{\tilde{\pi}(x_0,\cdot)}\bigg[q_{U|\sfX_0,\sfX_1,S;\eta}(u|x_0,\sfX_1,S;\eta) \bigg|S\bigg]\bigg)du \bigg|\sfX_0=x_0\bigg\}\bigg)\\
&\qquad =\bigg(\frac{1}{\log e}\bigg)\bigg( -\frac{\xi}{2}\bigg(\bbE[S \sfX_1^2|\sfX_0=x_0]+\frac{1}{\eta}\bigg)\log e+ \frac{1}{2}\log \frac{2\pi}{\xi}\\
&\qquad \qquad +\bbE_{S}\bigg\{\int_{\bbR} \bbE_{\pi(x_0,\cdot)}\big[p_{U|\sfX_0,\sfX_1,S;\eta}(u|\sfX_0,\sfX_1,S;\eta)\big] \nn\\
&\qquad \qquad \times \log \bigg(\bbE_{\tilde{\pi}(x_0,\cdot)}\bigg\{q_{U|\sfX_0,\sfX_1,S;\eta}(u|\sfX_0,\sfX_1,S;\eta) \bigg|S\bigg\}\bigg)du \bigg|\sfX_0=x_0\bigg\}\bigg)\\
& \qquad =\bigg(\frac{1}{\log e}\bigg)\bigg(-\frac{\xi}{2}\bigg(\bbE[S] \bbE[\sfX_1^2|\sfX_0=x_0]+\frac{1}{\eta}\bigg)\log e+ \frac{1}{2}\log \frac{2\pi}{\xi}\nn\\
&\qquad \qquad + \bbE_{S}\bigg\{\int_{\bbR} p_{U|\sfX_0,S;\eta}(u|x_0,S;\eta)  \log q_{U|\sfX_0,S;\eta}(u|x_0,S;\eta) du \bigg\}\bigg)
 \label{eq287},
\end{align} which is a constant which does not depend on $\barQ_i$, where \eqref{culi} follows from the dominated convergence theorem \cite{Royden}. Here, we note that the conditional event $\{\bQ_0=\barQ_i\}$ only affects the distribution of $\sfX$ and $\sfX_1$.

Now, from Lemma \ref{repre1}, it holds that
	\begin{align}
	\frac{\partial \rho(P_{\tilQ^*})}{\partial \nu}\bigg|_{\nu=0}&= \lim_{\nu \to 0} \frac{\rho(P_{\tilQ^*})\big|_{\nu}-1}{\nu} \label{cub}\\
	%&\qquad=\lim_{\nu \to 0} \sum_{i=1}^M\lambda_i(\tilQ^*)\frac{ \bbE[e^{\rm{tr}(\tilQ^* \bQ_1)}|\bQ_0=\barQ_i]-1\big)}{ \nu} \label{taggy10}\\
	&=\lim_{\nu \to 0} \sum_{i=1}^M\lambda_i(\tilQ^*)\bigg(\frac{ \bbE[e^{\rm{tr}(\tilQ^* \bQ_1)}|\bQ_0=\barQ_i]-1\big)}{ \nu}\bigg) \label{tag12}
	\end{align} where \eqref{cub} follows from \eqref{tofact2}.

Finally, as $\nu \to 0$, it holds that $\tilQ^* \to c=0$ by \eqref{factc} and \eqref{tilQmax} of Definition \ref{defrs}. Therefore, we have $P_{\tilQ^*} \to P_S \otimes P_{\pi}$ and $M \to \big|\{sx^2: (x,s) \in \calX \times \calS\}\big|:=M_0$, where $\otimes$ is denoted as the Kronecker product. It follows that for each fixed $S=s$, $\lambda(\tilQ) \to \tilde{\lambda}^{(\pi)}$ where $\tilde{\lambda}^{(\pi)}$ is the left Perron-Frobenius eigenvector of the stochastic matrix $P_S \otimes P_{\pi}$ such that $\|\lambda^{(\pi)}\|_1=1$. By Lemma \ref{best1}, the left Perron-Frobenius eigenvector exists, and it is unique up to a positive scaling factor, so $\tilde{\lambda}^{(\pi)}$ exists uniquely. 

Let $\lambda^{(\pi)}$ be the marginal distribution of $\tilde{\lambda}^{(\pi)}$. Then, from \eqref{eq287} and \eqref{tag12}, we obtain
\begin{align}
\frac{\partial \rho(P_{\tilQ^*})}{\partial \nu}\bigg|_{\nu=0}&=\sum_{s \in \calS} \sum_{x_0 \in \calX_0} \tilde{\lambda}^{(\pi)}_{s,x_0} \bigg(\frac{1}{\log e}\bigg)\bigg(-\frac{\xi}{2}\bigg(\bbE[S] \bbE[\sfX_1^2|\sfX_0=x_0]+\frac{1}{\eta}\bigg)\log e+ \frac{1}{2}\log \frac{2\pi}{\xi}\nn\\
&\qquad \qquad + \bbE_{S}\bigg\{\int_{\bbR} p_{U|\sfX_0,S;\eta}(u|x_0,S;\eta)  \log q_{U|\sfX_0,S;\eta}(u|x_0,S;\eta) du \bigg\}\bigg)\\
&=\sum_{x_0 \in \calX_0} \lambda^{(\pi)}_{x_0} \bigg(\frac{1}{\log e}\bigg)\bigg(-\frac{\xi}{2}\bigg(\bbE[S] \bbE[\sfX_1^2|\sfX_0=x_0]+\frac{1}{\eta}\bigg)\log e+ \frac{1}{2}\log \frac{2\pi}{\xi}\nn\\
&\qquad \qquad + \bbE_{S}\bigg\{\int_{\bbR} p_{U|\sfX_0,S;\eta}(u|x_0,S;\eta)  \log q_{U|\sfX_0,S;\eta}(u|x_0,S;\eta) du \bigg\}\bigg)\\
&=\bigg(\frac{1}{\log e}\bigg)\bigg(-\frac{\xi}{2}\bigg(\bbE[S] \bbE_{\sfX_0 \sim \lambda^{(\pi)}}\bigg[\bbE[\sfX_1^2|\sfX_0]\bigg]+\frac{1}{\eta}\bigg)\log e+ \frac{1}{2}\log \frac{2\pi}{\xi}\nn\\
&\qquad \qquad + \bbE_{\sfX_0 \sim \lambda^{(\pi)}}\bigg[\bbE_{S}\bigg\{\int_{\bbR} p_{U|\sfX_0,S;\eta}(u|x_0,S;\eta)  \log q_{U|\sfX_0,S;\eta}(u|x_0,S;\eta) du \bigg\}\bigg]\bigg) \label{xto}.
\end{align}
This concludes our proof of Lemma \ref{rhoto1}. 
\end{IEEEproof}

Then, we obtain our first main result as follows.
\begin{theorem} \label{thm:maincontri1} The free energy of the linear model with Markov sources in Section \ref{sec:setting} satisfies
	\begin{align}
	\calF_q=\calG, 
	\label{freeeq}
	\end{align} where $\calG$ is defined in \eqref{defFx0}. In addition, the average mutual information of this model satisfies:
	\begin{align}
	C=\lim_{n \to \infty}\frac{1}{n}I(\bX^n;\bY^m)=\calF_q\bigg|_{\sigma=1}-\frac{1}{2\beta} \label{tanakaap}.
	\end{align}
\end{theorem}
\begin{IEEEproof} \blue{Recall the definitions of $\{\barQ_i\}_{i=1}^M$ in Subsection \ref{sub:nota}.} 
	From Lemma \ref{repre1}, it holds that 
	\begin{align}
	Q^* =\lim_{\nu \to 0} \sum_{i=0}^M \lambda_i(\tilQ^*)\bbE[\bQ_1 e^{\mathrm{tr}(\tilQ^* \bQ_1)}\big|\bQ_0=\barQ_i]\label{babana},
	\end{align} where $\|\lambda(\tilQ^*)\|_1=1$ and all its components are positive.
	
	By Lemma \ref{lem:gv1}, we have $\bQ_1=S_1\bX_1 \bX_1^T$ and $\bQ_0=S_0\bX_0 \bX_0^T$ where  $\bX_1:=(X_1^{(0)},X_1^{(1)},\cdots, X_1^{(\nu)} )^T$ and $\bX_0:=(X_0^{(0)},X_0^{(1)},\cdots, X_0^{(\nu)} )^T$. It follows that for any $\tilQ \in \calQ$ and $\barQ_i \in \calQ$, we have
	\begin{align}
	\hatQ_i(\tilQ):&=\bbE[\bQ_1 e^{\mathrm{tr}(\tilQ \bQ_1)}\big|\bQ_0=\barQ_i] \label{defbat}\\
	&=\bbE\bigg[S_1\bX_1 \bX_1^T\exp\big[\bX_1^T\tilQ \bX_1 \big]\bigg|S_0=s_i,\bX_0=x_i \bigg] \label{tulupre} \\
	&=\bbE\bigg[S_1\bX_1 \bX_1^T\exp\big[\bX_1^T\tilQ \bX_1 \big]\bigg|\bX_0=x_i \bigg] 
	\label{tulu2} 
	\end{align} for some $s_i \in \calS$ and $x_i \in \calX^{\nu+1}$ such that $s_i x_i x_i^T=\barQ_i$, where \eqref{tulupre} follows from the uniqueness of the $x_i$ and $s_i$ by the definition of $\calQ$ in \eqref{defspaceQ}, and \eqref{tulu2} follows from the fact that $S_0$ is independent of $\bX_1,\bX_0$. 
	
	%It follows that
	%\begin{align}
	%\hatQ_i(\tilQ):=\bbE[\bQ_1 e^{\mathrm{tr}(\tilQ \bQ_1)}\big|\bQ_0=\barQ_i]=\bbE\bigg[S_1\bX_1 \bX_1^T\exp\big[\bX_1^T\tilQ \bX_1 \big]\bigg|\bX_0=x_i  \bigg] \label{tulu2}. 
	%\end{align} T
This means that for each fixed $i \in [M]$, $\hatQ_i^{(a,b)}(\tilQ) $ is in the same form as \cite[Eq. (127)]{GuoVerdu2005} for each $(a,b) \in [\nu+1] \times [\nu+1]$. Hence, by setting $S:=S_1 \sim P_S$ as above, we have
	\begin{align}
	\hatQ_i^{(0,1)}(\tilQ)&=\bbE\bigg[S X_1^{(0)} X_1^{(1)}\exp\big[\bX_1^T\tilQ \bX_1 \big]\bigg|\bX_0=x_i\bigg]\\
	&=\bbE\bigg[S \sfX_1 \langle \sfX\big|\bX_0=x_i \rangle_q\bigg|\bX_0=x_i\bigg] \label{totalsm},
	\end{align} where \eqref{totalsm} follows from \cite[Eq.~(131)]{GuoVerdu2005}.
	
	Similarly, we also have
	\begin{align}
	r_i&:=\hatQ_i^{(0,0)}=\bbE\big[S\big|\bX_0=x_i\big], \label{Q1}\\
	m_i&:=\hatQ_i^{(0,1)}=\bbE\big[S \sfX_1 \langle \sfX\big|\bX_0=x_i \rangle_q\big|\bX_0=x_i\big], \label{Q2}\\
	p_i&:=\hatQ_i^{(1,1)}=\bbE\big[S \sfX^2\big|\bX_0=x_i\big] \label{Q3},\\
	q_i&:=\hatQ_i^{(1,2)}=\bbE\big[S \langle \sfX\big|\bX_0=x_i\rangle_q^2\big|\bX_0=x_i\big] \label{Q4},
	\end{align}	 for all $i \in [M]$. Since $\bX^{(a)} \sim q_{\bX}$ for all $a=1,2,\cdots,\nu$ and mutually independent to each other, it follows from \eqref{Q1}--\eqref{Q4} that $\hatQ_i(\tilQ)$ has the RS form as defined in  Lemma \ref{defrs}, i.e.,
	\begin{align}
	\hatQ_i(\tilQ)&=\begin{bmatrix}r_i & m_i&m_i &\cdots &m_i \\
	m_i &p_i &q_i &\cdots& q_i\\
	m_i& q_i & p_i& \ddots& \vdots\\
	\vdots& \vdots& \ddots& \ddots& q_i\\
	m_i&q_i&\cdots&q_i&p_i \end{bmatrix}\label{banav2}.
	\end{align} for all $i \in [M]$.

	It follows from Theorem \eqref{babana} and \eqref{defbat} that
	\begin{align}
	Q^*(\tilQ) = &\lim_{\nu \to 0}\sum_{i=0}^M \lambda_i(\tilQ) \hatQ_i(\tilQ)\\
	&= \lim_{\nu \to 0}\sum_{i=0}^M \lambda_i(\tilQ) \begin{bmatrix}r_i & m_i&m_i &\cdots &m_i \\
	m_i &p_i &q_i &\cdots& q_i\\
	m_i& q_i & p_i& \ddots& \vdots\\
	\vdots& \vdots& \ddots& \ddots& q_i\\
	m_i&q_i&\cdots&q_i&p_i \end{bmatrix} \label{factQstar}.
	\end{align}
	
	Hence, from the RS assumption in Definition \ref{defrs} and \eqref{factQstar}, we obtain
	\begin{align}
	r&=\lim_{\nu \to 0}\sum_{i=0}^M \lambda_i(\tilQ) r_i\\
	&=\lim_{\nu \to 0}\sum_{i=0}^M\lambda_i(\tilQ) \bbE\big[S\big|\bX_0=x_i\big] \\ 
	&=\lim_{\nu \to 0}\sum_{i=0}^M\lambda_i(\tilQ) \bbE\big[S\big|\sfX_0=x_i^{(0)}\big] \label{factr},
	\end{align}
	where $x_i^{(0)}$ is the first element of the vector $x_i$. In addition, we also have
	\begin{align}
	m&=\lim_{\nu \to 0} \sum_{i=0}^M \lambda_i(\tilQ) m_i\\
	&=\lim_{\nu \to 0}\sum_{i=0}^M\lambda_i(\tilQ) \bbE\big[S \sfX_1 \langle \sfX\big|\bX_0=x_i \rangle_q\big|\bX_0=x_i\big] \\
	&=\lim_{\nu \to 0}\sum_{i=0}^M\lambda_i(\tilQ) \bbE\big[S \sfX_1 \langle \sfX\big|\sfX_0=x_i^{(0)} \rangle_q\big|\sfX_0=x_i^{(0)}\big] \label{factm},\\ 
	p&=\lim_{\nu \to 0}\sum_{i=0}^M \lambda_i(\tilQ) p_i\\
	&=\lim_{\nu \to 0}\sum_{i=0}^M\lambda_i(\tilQ) \bbE\big[S \sfX^2\big|\bX_0=x_i\big] \\
	&=\lim_{\nu \to 0}\sum_{i=0}^M\lambda_i(\tilQ) \bbE\big[S \sfX^2\big|\sfX_0=x_i^{(0)}\big]  \label{factp},\\
	q&=\lim_{\nu \to 0} \sum_{i=0}^M \lambda_i(\tilQ) q_i\\
	&=\lim_{\nu \to 0}\sum_{i=0}^M\lambda_i(\tilQ)\bbE\big[S \langle \sfX\big|\bX_0=x_i\rangle_q^2\big|\bX_0=x_i\big]\\ 
	&=\lim_{\nu \to 0}\sum_{i=0}^M\lambda_i(\tilQ)\bbE\big[S \langle \sfX\big|\sfX_0=x_i^{(0)}\rangle_q^2\big|\sfX_0=x_i^{(0)}\big]  \label{factq}. 
	\end{align} 
	From these facts, we obtain
	\begin{align}
	r-2m+q&=\lim_{\nu \to 0} \sum_{i=0}^M \lambda_i(\tilQ)  \bbE\bigg[S\bigg(\sfX_1^2-2 \sfX_1  \langle \sfX\big|\sfX_0=x_i^{(0)} +  \langle \sfX\big|\sfX_0=x_i^{(0)}\rangle_q^2\bigg) \bigg|\sfX_0=x_i^{(0)}\bigg]\\
	&=\lim_{\nu \to 0}\sum_{i=0}^M \lambda_i(\tilQ)\bbE\bigg[S\bigg(\sfX_1-\langle \sfX\big|\sfX_0=x_i^{(0)}\rangle_q\bigg)^2\bigg|\sfX_0=x_i^{(0)}\bigg] \label{Qfact1cure},\\
	\end{align}
	and similarly,
	\begin{align}
	p-q&=\lim_{\nu \to 0} \sum_{i=0}^M \lambda_i(\tilQ)\bbE\bigg[S\bigg(\sfX-\langle \sfX\big|\sfX_0=x_i^{(0)}\rangle_q\bigg)^2\bigg|\sfX_0=x_i^{(0)}\bigg] \label{Qfact2cure}.
	\end{align}
	On the other hand, from \eqref{factc}--\eqref{factxi}, we also have
	\begin{align}
	r-2m+q&=\frac{1}{\beta}\bigg(\frac{1}{\eta}-1\bigg), \label{Qfactcure3}\\
	p-q&=\frac{1}{\beta}\bigg(\frac{1}{\xi}-\sigma^2\bigg) \label{Qfactcure4}.
	\end{align}
	From \eqref{Qfact1cure}--\eqref{Qfactcure4}, $(\eta, \xi)$ is a solution of the following equation system:
	\begin{align}
	\eta^{-1}&=1+ \beta \lim_{\nu \to 0}\sum_{i=0}^M \lambda_i(\tilQ)\bbE\bigg[S\bigg(\sfX_1-\langle \sfX\big|\sfX_0=x_i^{(0)}\rangle_q\bigg)^2\bigg|\sfX_0=x_i^{(0)}\bigg], \\
	&=1+ \beta \lim_{\nu \to 0}\sum_{i=0}^M \lambda_i(\tilQ) \bbE\bigg[S\calE(S;\eta,\xi|\sfX_0=x_i^{(0)})\bigg] \label{qb1},\\
	\xi^{-1}&=\sigma^2 +  \beta\lim_{\nu \to 0} \sum_{i=0}^M \lambda_i(\tilQ)\bbE\bigg[S\bigg(\sfX-\langle \sfX\big|\sfX_0=x_i^{(0)}\rangle_q\bigg)^2\bigg|\sfX_0=x_i^{(0)}\bigg]\\
	&=\sigma^2+ \beta \lim_{\nu \to 0}\sum_{i=0}^M \lambda_i(\tilQ) \bbE[S\calE(S;\eta,\xi|\sfX_0=x_i^{(0)})\bigg] \label{qb2}.
	\end{align}
	
	Now, from \eqref{def:GQ} in Lemma \ref{lem:gv1} and RS assumption on Definition \ref{defrs}, we obtain
	\begin{align}
	G^{(\nu)}(Q^*)&=-\frac{\nu}{2}\log (2\pi \sigma^2)-\frac{\nu-1}{2}\log \bigg[1+\frac{\beta}{\sigma^2}(p-q)\bigg]\nn\\
	&\qquad -\frac{1}{2}\log\bigg[1+\frac{\beta}{\sigma^2}(p-q)+\frac{\nu}{\sigma^2}\bigg(1+\beta(r-2m+q)\bigg)\bigg] \label{GfactQ}.
	\end{align}
	In addition, we also have
	\begin{align}
	I^{(\nu)}(Q^*)&=\mathrm{tr}(\tilQ^*Q^* )-\log \rho(P_{\tilQ^*})\\
	&=\mathrm{tr}(\tilQ^*Q^* )-\blue{\log \rho(P_{\tilQ^*})} \label{fbun1}, \\
	&=rc + \nu pg +2 \nu md + \nu(\nu-1)qf -\blue{\log \rho(P_{\tilQ^*})}\label{fbun2},
	\end{align} where \eqref{fbun1} follows from Lemma \ref{rhoto1}, and \eqref{fbun2} follows from assumptions $Q^*$ and $\tilQ^*$ in Definition \ref{defrs}.
	
	Now, by the RS assumption, the eight parameters $(r,m,p,q,c,d,f,g)$ have zero derivatives with respect to $\nu$ as $\nu \to 0$ \cite[p.1999]{GuoVerdu2005}. Let $\lambda^{(\pi)}$ is the left Perron-Frobenius eigenvector of the stochastic matrix $P_{\pi}$ such that $\|\lambda^{(\pi)}\|_1=1$, which is the stationary distribution of the stochastic matrix. By choosing the initial state at the state that the limit distribution of the Markov process $\{X_n\}_{n=1}^{\infty}$ converges to the stationary distribution. Then, from Theorem \ref{mainthm1}, we have
	\begin{align}
	\calF_q&=-\lim_{\nu \to 0}\frac{\partial}{\partial \nu}(\beta^{-1} G^{(\nu)}(Q^*)- I^{(\nu)}(Q^*))\\
	&=\lim_{\nu \to 0} \frac{\partial}{\partial \nu} \bigg(rc + \nu pg +2 \nu md + \nu(\nu-1)qf  -\beta^{-1}\bigg(-\frac{\nu}{2}\log (2\pi \sigma^2)-\frac{\nu-1}{2}\log \bigg[1+\frac{\beta}{\sigma^2}(p-q)\bigg]\nn\\
	& \qquad -\frac{1}{2}\log\bigg[1+\frac{\beta}{\sigma^2}(p-q)+\frac{\nu}{\sigma^2}\bigg(1+\beta(r-2m+q)\bigg)\bigg]  \bigg)\bigg)- \lim_{\nu \to 0} \frac{\partial}{\partial \nu} \log \rho(P_{\tilQ^*})\\
	&=pg+2md-qf+\beta^{-1} \bigg[\frac{1}{2}\log(2\pi \sigma^2)+\frac{1}{2}\log\bigg(1+\frac{\beta}{\sigma^2}(p-q)\bigg) +\frac{1+\beta(r-2m+q)}{2\sigma^2(1+\frac{\beta}{\sigma^2}(p-q))}\log e \bigg]\nn\\
	&\qquad\qquad  -\lim_{\nu \to 0} \frac{\partial}{\partial \nu} \log \rho(P_{\tilQ^*}) \label{eqfacF}\\
	&=p(f-d)+ 2md-qf +\beta^{-1} \bigg[\frac{1}{2}\log(2\pi \sigma^2)+\frac{1}{2}\log\bigg(1+\frac{\beta}{\sigma^2}(p-q)\bigg)+\frac{1+\beta(r-2m+q)}{2\sigma^2(1+\frac{\beta}{\sigma^2}(p-q))}\log e \bigg]\nn\\
	&\qquad \qquad -\lim_{\nu \to 0} \frac{\partial}{\partial \nu} \log \rho(P_{\tilQ^*}) \label{bat1} \\
	&=(p-q)f-p d+ 2md +\beta^{-1} \bigg[\frac{1}{2}\log(2\pi \sigma^2)+\frac{1}{2}\log\bigg(1+\frac{\beta}{\sigma^2}(p-q)\bigg)+\frac{1+\beta(r-2m+q)}{2\sigma^2(1+\frac{\beta}{\sigma^2}(p-q))}\log e \bigg]\nn\\
	&\qquad \qquad -\lim_{\nu \to 0} \frac{\partial}{\partial \nu} \log \rho(P_{\tilQ^*})\\
	&=\frac{1}{\beta}\bigg(\frac{1}{\xi}-\sigma^2\bigg)\frac{\xi^2}{2\eta}-\frac{p\xi}{2}+\xi m + \frac{1}{\beta}\bigg[\frac{1}{2}\log( 2\pi \sigma^2) -\frac{1}{2}\log\big(\xi \sigma^2\big) +\frac{\xi^2}{2\eta}\log e \bigg]- \lim_{\nu \to 0} \frac{\partial}{\partial \nu} \log \rho(P_{\tilQ^*}) \label{bat2}\\
	&=\xi m -\frac{p\xi}{2}+ \frac{1}{\beta}\bigg(\frac{1}{\xi}-\sigma^2\bigg)\frac{\xi^2}{2\eta}-\frac{1}{2\beta}\log \xi+\frac{1}{2\beta}\log(2\pi)+\frac{\xi^2}{2\beta \eta}\log e-\lim_{\nu \to 0} \frac{\partial}{\partial \nu} \log \rho(P_{\tilQ^*})\\
	&=\xi m -\frac{p\xi}{2}+ \frac{1}{\beta}\bigg(\frac{1}{\xi}-\sigma^2\bigg)\frac{\xi^2}{2\eta}-\frac{1}{2\beta}\log \xi+\frac{1}{2\beta}\log(2\pi)+\frac{\xi^2}{2\beta \eta}\log e\nn\\
	&\qquad +\frac{\xi}{2}\bigg(\bbE[S] \bbE[\sfX_1^2]+\frac{1}{\eta}\bigg)\log e -\frac{1}{2}\log \frac{2\pi}{\xi} \nn\\
	&\qquad \qquad-\bbE_{\sfX_0 \sim \lambda^{(\pi)} }\bigg\{\bbE_{S}\bigg\{\int_{\bbR} p_{U|\sfX_0,S;\eta}(u|X_0,S;\eta)\big\}  \log \bigg(q_{U|\sfX_0,S;\eta}(u|X_0,S;\eta) \bigg)du \bigg\}\bigg\}\bigg) \label{ufacto} \\
	&=\xi \lim_{\nu \to 0}\sum_{i=0}^M\lambda_i(\tilQ) \bbE\big[S \sfX_1 \langle \sfX\big|\sfX_0=x_i^{(0)} \rangle_q\big|\sfX_0=x_i^{(0)}\big] -\frac{\xi}{2}\lim_{\nu \to 0}\sum_{i=0}^M\lambda_i(\tilQ) \bbE\big[S \sfX^2\big|\sfX_0=x_i^{(0)}\big] \nn\\
	&\qquad + \frac{1}{\beta}\bigg(\frac{1}{\xi}-\sigma^2\bigg)\frac{\xi^2}{2\eta}-\frac{1}{2\beta}\log \xi+\frac{1}{2\beta}\log(2\pi)+\frac{\xi^2}{2\beta \eta}\log e+\frac{\xi}{2}\bigg(\bbE[S] \bbE[\sfX_1^2]+\frac{1}{\eta}\bigg)\log e \nn\\
	&\qquad-\frac{1}{2}\log \frac{2\pi}{\xi}-\bbE_{\sfX_0 \sim \lambda^{(\pi)} }\bigg\{\bbE_{S}\bigg\{\int_{\bbR} p_{U|\sfX_0,S;\eta}(u|X_0,S;\eta)\big]  \log \bigg(q_{U|\sfX_0,S;\eta}(u|X_0,S;\eta) \bigg)du \bigg\}\bigg\} \label{ufacto2} \\
	&=\xi \bbE_{X_0 \sim \lambda^{(\pi)}}\bigg[\bbE\big[S \sfX_1 \langle \sfX\big|\sfX_0 \rangle_q\big|\sfX_0\big]\bigg] -\frac{\xi}{2}\bbE_{X_0 \sim \lambda^{(\pi)}}\bigg[\bbE\big[S \sfX^2\big|\sfX_0\big]\bigg] \nn\\
	&\qquad + \frac{1}{\beta}\bigg(\frac{1}{\xi}-\sigma^2\bigg)\frac{\xi^2}{2\eta}-\frac{1}{2\beta}\log \xi+\frac{1}{2\beta}\log(2\pi)+\frac{\xi^2}{2\beta \eta}\log e+\frac{\xi}{2}\bigg(\bbE[S] \bbE[\sfX_1^2]+\frac{1}{\eta}\bigg)\log e  \nn\\
	&\qquad-\frac{1}{2}\log \frac{2\pi}{\xi}-\bbE_{\sfX_0 \sim \lambda^{(\pi)} }\bigg\{\bbE_{S}\bigg\{\int_{\bbR} p_{U|\sfX_0,S;\eta}(u|X_0,S;\eta)\big]  \log \bigg(q_{U|\sfX_0,S;\eta}(u|X_0,S;\eta) \bigg)du \bigg\}\bigg\} \label{ufactox} \\
	&=\sum_{x_0} \lambda_{x_0}^{(\pi)} \calG(x_0) \label{batF},
	\end{align} where \blue{\eqref{eqfacF} follows from Lemma \ref{rhoto1}}, \eqref{bat1} follows from \eqref{factc}--\eqref{factxi}, \eqref{bat2} follows from \eqref{Qfactcure3} and \eqref{Qfactcure4}, \eqref{ufacto} follows from Lemma \ref{rhoto1}, \eqref{ufacto2} follows from \eqref{factm} and \eqref{factp},  \eqref{ufactox} follows from $\lambda(\tilQ) \to \lambda^{(\pi)}$ since $P_{\tilQ} \to P_{\pi}$ as $\nu \to 0$ where $\lambda^{(\pi)}$ is the left Perron-Frobenius eigenvector of the stochastic matrix $P_{\pi}$ such that $\|\lambda^{(\pi)}\|_1=1$, which is the stationary distribution of the Markov chain $\{X_n\}_{n=1}^{\infty}$\footnote{By Lemma \ref{best1}, the left Perron-Frobenius eigenvector exists, and it is unique up to a positive scaling factor, so $\lambda^{(\pi)}$ exists uniquely.}, and \eqref{batF} follows from \cite[Sect. IV]{GuoVerdu2005}.
	
	 Hence, we obtain \eqref{freeeq} from \eqref{qb1}, \eqref{qb2}, and \eqref{batF}.
	
	Finally, \eqref{tanakaap} is an direct application of \cite[Prop. 5]{Tanaka2002a}.
\end{IEEEproof}

The following corollary recovers \cite[Sect. II-D]{GuoVerdu2005}:

\begin{corollary}\label{lemsymcorovera} For any i.i.d. sequence $\{X_n\}_{n=1}^{\infty}$ on the Polish space $\calX$ defined in Section \ref{sec:setting}, the free energy satisfies
	\begin{align}
	\calF_q=\calG(\emptyset) \label{verdufact1},
	\end{align} where $\calG(\emptyset)$ is the free-energy function estimated in Section \ref{freEnergyx0} when no state information appears in the corresponding single-symbol PME channel.
	
	In addition, the average mutual information of this model satisfies
	\begin{align}
	C=\lim_{n \to \infty}\frac{1}{n}I(\bX^n;\bY^m)=\calF_q\bigg|_{\sigma=1}-\frac{1}{2\beta} \label{verdufact2}.
	\end{align}
\end{corollary}
\begin{IEEEproof} Observe that an i.i.d. sequence $\{X_n\}_{n=1}^{\infty}$ can be considered as a Markov sequence with transition probability (function) $\pi(x,y)=p(y)$ for all $x,y \in \calX$. Hence, $\calG(x_0)$ is a constant, say $\calG(\emptyset)$, for all $x_0 \in \calX$. Here, $\calG(\emptyset)$ is the free energy function estimated in Section \ref{freEnergyx0} when there is no state information appeared in the correponding single-symbol PME channel, i.e. $X_0=\emptyset$. In addition, the left Perron-Frobenius eigenvector with unit Manhattan norm for this special case is $\{P_{X_1}(x)\}_{x \in \calX}$. 
	
	Hence, by Theorem \ref{thm:maincontri1}, we have
	\begin{align}
	\calF_q&=\sum_{x_0 \in \calX} \lambda_{x_0 }^{(\pi)} \calG(x_0)\\
	&=\bigg(\sum_{x_0 \in \calX}P_{X_1}(x_0) \bigg)\calG(\emptyset)\\
	&=\calG(\emptyset),
	\end{align} where the last equation follows from $\|\lambda^{(\pi)}\|_1=1$. Hence, we obtain \eqref{verdufact1}. Finally,  \eqref{verdufact2} is an direct application of \eqref{tanakaap} in Theorem \ref{thm:maincontri1}.
\end{IEEEproof}

To state our next main result, we recall Carleman theorem.
\begin{lemma}\cite[Theorem 3.1]{Chalendar2007a} \label{carleman} Denote $\calM(\bbR^n)$ be the set of all positive Borel measures $\mu$ on $\bbR^n$ such that
	\begin{align}
	\int_{\bbR^n} \|x\|_2^d d\mu(x)<\infty \quad \forall d\geq 0.
	\end{align} Suppose that $\mu_1,\mu_2 \in \calM(\bbR^n)$ satisfy
	\begin{align}
	s(\alpha):=\int_{\bbR^n} x^{\alpha}d\mu_1(x)=\int_{\bbR^n} x^{\alpha}d\mu_2(x) \quad \mbox{for all}\quad \alpha \in \bbN^n 
	\end{align}
	and that the conditions
	\begin{align}
	\sum_{m=1}^{\infty}s(2me_j)^{-1/(2m)}=\infty, \quad j=1,2,\cdots,n, \label{carcond}
	\end{align} hold, where $e_j$ is the $j$th canonical basis vector of $\bbR^n$. Then $\mu_1=\mu_2$.
\end{lemma}

\begin{claim} \label{thm:maincontri2}
	Recall the definition of $\{\lambda_{x_0}^{(\pi)}\}_{x_0 \in \calX}$ in Section \ref{freEnergyx0}. Assume that the generalized PME defined in \eqref{def1} is used for estimation. Then, for all $k \in \{1,2,\cdots,n\}$, the joint moments satisfy:
	\begin{align}
	\lim_{n\to \infty}\bbE\big[X_k^{i_0} \tilX_k^{j_0} \langle X_k \rangle_q^{l_0} \big]=\sum_{x_0 \in \calX} \lambda_{x_0}^{(\pi)} \bbE\big[\sfX_1^{i_0} \sfX^{j_0} \langle \sfX\big|\sfX_0 \rangle_q^{l_0}\big|\sfX_0=x_0\big] \quad \forall i_0,j_0,l_0 \in \bbZ_+ \label{jointmoment},
	\end{align} where $(\sfX_1,\sfX,\langle \sfX|\sfX_0=x_0\rangle_q)$ is the input and outputs defined in the (composite) single-symbol PME channel in Fig. \ref{fig:PME}, and $(X_k, \tilX_k, \langle X_k \rangle)$ is the $k$-th symbol in the vector $\bX \in \calX^n$, the $k$-th output of the vector retrochanel defined in \eqref{ch:gePME}, and its corresponding estimated symbol by using the PME estimate in \eqref{def1}.
	
	In addition, the average MMSE satisfies:
	\begin{align}
	\frac{1}{n}\bbE\big[\|\bX-[\bX]\|_2^2\big] = \bbE\big[\sfX_1^2\big]-\sum_{x_0 \in \calX}\lambda_{x_0}^{(\pi)} \bbE\big[\langle \sfX_1|\sf X_0 \rangle^2\big|\sfX_0=x_0\big] \label{uplowMMSE},
	\end{align} where $\sfX_1, \langle \sfX|\sfX_0=x_0 \rangle, \sfX_0$ are the input, output given channel state, and channel state in the single-symbol PME channel with available states at both encoder and decoder defined Section \ref{freEnergyx0}, and $\sfX_1 \sim \sum_{x_0 \in \calX} \pi(x_0,\cdot) \lambda_{x_0}^{(\pi)}$.
\end{claim}
\begin{remark} Some remarks are in order.
	\begin{itemize}
		\item For the i.i.d. case of the sequence $\{X_n\}_{n=1}^{\infty}$, we have a tight bound on \eqref{jointmoment}. It is not hard to check that the Carleman condition \eqref{carcond} holds for the joint Gaussian distribution on the composite single-symbol Gaussian channel in Fig. \ref{fig:PME}. Hence, from Carleman Theorem in Lemma \ref{carleman}, in the large system limit, the channel between the input $X_k$ and $\langle X_k \rangle_q$ for each symbol $k$ is equivalent to the Gaussian channel $p_{U|\sfX,\sfX_0,S;\eta}$ with available state $\sfX_0=\emptyset$ at both encoder and decoder concatenated with the one-to-one decision function with $S=S_k$. This result recovers \cite[Corrolary 1]{GuoVerdu2005} as a special case for the i.i.d. sequence $\{X_n\}_{n=1}^{\infty}$.
		\item From Theorem \ref{thm:maincontri2}, it can be inferred that for the generalized PME estimation problem, the channel (model) has been decoupled into AWGN channels with state information at both transmitters and receivers, where state vector distribution follows the left Perron-Frobenius eigenvector $\lambda^{(\pi)}$ of the stochastic matrix $P_{\pi}$. 
	\end{itemize} 
\end{remark}
\begin{IEEEproof}
	The result in \eqref{jointmoment} can be obtained by using the same ideas as in the proof of Theorem \ref{mainthm1}, \cite[Sec. IV-B]{GuoVerdu2005}, and the facts in \eqref{babana}, \eqref{tulu2}, and \eqref{batF}. The detailed proof can be found in Appendix \ref{proof:jointmoment}.
	
	Now, observe that by using the MMSE decoder defined in Section \ref{sec:PME}, we have
	\begin{align}
	\bbE\big[\|\bX-[\bX]\|_2^2\big]&=\sum_{k=1}^n \bbE\big[\big|\sfX_k-[\sfX_k]\big|^2\big] \label{IQ1}\\
	&=\sum_{k=1}^n \bbE\bigg[\bbE\big[\big|\sfX_k-\langle \sfX_k \rangle\big|^2\big|\bY,\bPhi\big]\bigg] \label{IQ2} \\
	&=\sum_{k=1}^n \bbE\bigg[\bbE\big[\sfX_k^2\big]-[ \sfX_k]^2\big|\bY,\bPhi\big]\bigg] \label{IQ3}\\
	&=\sum_{k=1}^n \bbE\big[\sfX_k^2 -[\sfX_k]^2\big]\\
	&=\sum_{k=1}^n \bbE\big[\sfX_k^2\big]-\sum_{k=1}^n \bbE\big[[\sfX_k]^2\big]
	\label{corofact1},
	\end{align} where \eqref{IQ1} follows from \eqref{MMSEEstimator}, \eqref{IQ2} follows from the tower property \cite{Billingsley}, and \eqref{IQ3} follows from the fact that
	\begin{align}
	[\sfX_k]=\bbE_p\big[\sfX_k\big|\bY,\bPhi\big]
	\end{align} which is drawn from \eqref{MMSEEstimator}.
	
	Now, by \eqref{jointmoment}, we have as $n \to \infty$,
	\begin{align}
	\bbE\big[[\sfX_k]^2\big] =\sum_{x_0 \in \calX} \lambda_{x_0}^{(\pi)} \bbE\big[\langle \sfX_1|\sfX_0=x_0\rangle^2\big], \quad \forall k \in \{1,2,\cdots,n\} \label{corofact2}.
	\end{align}
	In addition, for all $k \in \{1,2,\cdots,n\}$, we also have
	\begin{align}
	\bbE\big[\sfX_k^2\big]&=\bbE\big[\bbE\big[\sfX_k^2\big|\sfX_{k-1}\big]\big]\label{bunhiakop1}\\
	&=\bbE\big[\bbE\big[\sfX_1^2\big|\sfX_0\big]\big]\label{bunhiakop2}\\
	&=\bbE[\sfX_1^2] \label{corofact3},
	\end{align} where \eqref{bunhiakop1} follows from the tower property \cite{Billingsley}, and \eqref{bunhiakop2} follows from the time-homogeneity of Markov process $\{X_n\}_{n=1}^{\infty}$.
	
	From \eqref{corofact1}, \eqref{corofact2}, and \eqref{corofact3}, as $n\to \infty$, we have
	\begin{align}
	\bbE\big[\|\bX-[\bX]\|_2^2\big] = n \bigg(\bbE[\sfX_1^2]-\sum_{x_0 \in \calX}\lambda_{x_0}^{(\pi)}\bbE\big[\langle \sfX_1|\sfX_0=x_0\rangle^2\big]\bigg),
	\end{align} which leads to \eqref{uplowMMSE}.
\end{IEEEproof}

The following corollary also recovers \cite[Sect. II-D]{GuoVerdu2005}:

\begin{corollary}\label{lemsymcoromm} Let  $\{X_n\}_{n=1}^{\infty}$ be an i.i.d. sequence on the Polish space $\calX$ defined in Section \ref{sec:setting}. Assume that the generalized PME defined in \eqref{def1} is used for estimation. Then, for all $k \in \{1,2,\cdots,n\}$, the joint moments satisfy:
	\begin{align}
	\lim_{n\to \infty}\bbE\bigg[X_k^{i_0} \tilX_k^{j_0} [X_k]_q^{l_0} \bigg]=\bbE\bigg[\sfX_1^{i_0} \sfX^{j_0} \langle \sfX\big|\sfX_0 \rangle_q^{l_0}\big|\sfX_0=\emptyset \bigg] \quad \forall i_0,j_0,l_0 \in \bbZ_+ \label{jointmomentcor},
	\end{align} where $(\sfX_1,\sfX,\langle \sfX|\sfX_0=x_0\rangle_q)$ is the input and outputs defined in the (composite) single-symbol PME channel in Fig. \ref{fig:PME}, and $(X_k, \tilX_k, [X_k])$ is the $k$-th symbol in the vector $\bX \in \calX^n$, the $k$-th output of the vector retrochanel defined in \eqref{ch:gePME}, and its corresponding estimated symbol by using the PME estimate in \eqref{def1}. Here, $\sfX_0=\emptyset$ in the RHS of \eqref{jointmomentcor} means that the conditional joint moments is estimated when no state information $\sfX_0$ is assumed in the corresponding single-symbol PME channel in Section \ref{freEnergyx0}.
	
	In addition, the average MMSE satisfies:
	\begin{align}
	\frac{1}{n}\bbE\big[\|\bX-[\bX]\|_2^2\big] &= \bbE\big[\sfX_1^2\big]- \bbE\big[\langle \sfX_1|\sfX_0 \rangle^2\big|\sfX_0=\emptyset \big]
	\label{uplowMMSEcor},
	\end{align} where $\sfX_1, \langle \sfX|\sfX_0=x_0 \rangle, \sfX_0$ are the input, output given channel state, and state in the single-symbol PME channel with available states at both encoder and decoder defined Section \ref{freEnergyx0}, respectively.
\end{corollary}
\begin{IEEEproof}
	These results can be obtained by using the same arguments as Corollary \ref{lemsymcorovera}. They are  direct applications of Theorem \ref{thm:maincontri2}. 	
\end{IEEEproof}

\begin{claim}\label{HMMFreeEnergy} Assume that $\{X_n\}_{n=1}^{\infty}$ is the hidden states (outputs) of a hidden Markov model generated by a Markov chain $\{\Upsilon_n\}_{n=1}^{\infty}$ with transition probability (function) $\pi_{\Upsilon}(\cdot,\cdot)$ on some Polish space $\calS_{\Upsilon}$, i.e.,
	\begin{itemize}
		\item $\Upsilon_n$ is a Markov process and is not directly observable.
		\item $\bbP(X_n \in \calA|\Upsilon_1=\upsilon_1, \Upsilon_2=\upsilon_2,\cdots, \Upsilon_n=\upsilon_n)=\bbP(X_n\in \calA|\Upsilon_n=\upsilon_n)=P_{X|\Upsilon}(\calA|\upsilon_n)$,
	\end{itemize} for every $n\geq 1$, $\upsilon_1,\upsilon_2,\cdots, \upsilon_n$, and an arbitrary measurable set $\calA$, where $P_{X|\Upsilon}(\cdot|\cdot)$ is some probability measure called emission probability. Then, the following holds:
	\begin{itemize}
		\item $\{X_n,\Upsilon_n\}_{n=1}^{\infty}$ forms a Markov chain on $\calX \times \calS_{\Upsilon}$ with transition probability $P_{X_1,\Upsilon_1|X_0,\Upsilon_0}(x_1,\upsilon_1|x_0,\upsilon_0)=P_{X|\Upsilon}(x_1|\upsilon_1) \pi_{\Upsilon}(\upsilon_0,\upsilon_1)$.
		\item Recall the definitions of  $\{\lambda_{x_0,\upsilon_0}^{(\pi_{\Upsilon})}\}_{(x_0,\upsilon_0) \in \calX \times \calS_{\Upsilon}}$ and  $\tilG$ in Section \ref{freEnergyx0_hmm}. Then, the free energy, mutual information, joint moments, the average MMSE of the linear model with hidden Markov sources in \ref{sec:setting} satisfy:
		\begin{align}
		\calF&= \tilG, \label{eq246HM} \\
		&C=\calF\bigg|_{\sigma=1}- \frac{1}{2\beta}, \label{eq247HM} \\ 
		&\lim_{n\to \infty}\bbE\bigg[X_k^{i_0} \tilX_k^{j_0} [X_k]_q^{l_0} \bigg] = \sum_{x_0, \upsilon_0 \in \calX \times \calS_{\Upsilon}}\lambda_{x_0,\upsilon_0}^{(\pi_{\Upsilon})} \bbE\bigg[\sfX_1^{i_0} \sfX^{j_0} \langle \sfX\big|\sfX_0,\Upsilon_0 \rangle_q^{l_0}\big|\sfX_0=x_0,\Upsilon_0=\upsilon_0\bigg],  \forall i_0,j_0,l_0 \in \bbZ_+ \label{jointmomentHM},\\
		&\lim_{n \to \infty} \frac{1}{n}\bbE[\|\bX-[\bX]\|_2^2]= \bbE[\sfX_1^2]-\sum_{x_0,\upsilon_0 \in \calX_1 \times \calS_{\gamma}} \lambda_{x_0,\upsilon_0}^{(\pi_{\Upsilon})}\bbE[\langle \sfX|\sfX_0=x_0,\Upsilon_0=\upsilon_0\rangle^2 ] \label{uplowMMSEHM}, 
		\end{align}  where $(\sfX_1,\sfX, \langle \sfX|\sfX_0,\Upsilon_0\rangle_q)$ is the input and outputs defined in the (composite) single-symbol PME channel in Fig. \ref{fig:PMEHM}, and $(X_k, \tilX_k, [X_k])$ is the $k$-th symbol in the vector $\bX \in \calX^n$, the $k$-th output of the vector retrochanel defined in \eqref{ch:gePME}, and its corresponding estimated symbol by using the generalized PME estimate in \eqref{def1}. In addition, in \eqref{uplowMMSEHM}, $\sfX_1 \sim \sum_{\upsilon \in \calS_{\Upsilon}} P_{X|\Upsilon}(\cdot|\upsilon)\pi_{\Upsilon}(\upsilon_0,\upsilon)$, where $P_{X|\Upsilon}$ is the stationary emission probability of the hidden Markov process.
	\end{itemize}
\end{claim}
\begin{IEEEproof} First, we show that $\{(X_n,\Upsilon_n)\}_{n=1}^{\infty}$ forms a Markov chain with states on $\calX \times \calS_{\Upsilon}$. Indeed, for any $n\geq 2$, by using Markov chains such as $\Upsilon_n -\Upsilon_{n-1}-(X_{n-1},\{X_k,\Upsilon_k\}_{k=1}^{n-2})$ and $X_n-\Upsilon_n -(\{X_k,\Upsilon_k\}_{k=1}^{n-1})$, we have
	\begin{align}
	&\bbP\bigg(X_n=x,\Upsilon_n=\upsilon_n\bigg|\bigg\{X_k=x_k, \Upsilon_k=\upsilon_k\bigg\}_{k=1}^{n-1}\bigg)=\bbP\bigg(\Upsilon_n=\upsilon_n\bigg|\bigg\{X_k=x_k, \Upsilon_k=\upsilon_k\bigg\}_{k=1}^{n-1}\bigg)\nn\\
	&\qquad \times \bbP\bigg(X_n=x_n\bigg|\Upsilon_n=\upsilon_n,\bigg\{X_k=x_k, \Upsilon_k=\upsilon_k\bigg\}_{k=1}^{n-1}\bigg)\\
	&=\bbP\bigg(\Upsilon_n=\upsilon_n\bigg| \Upsilon_{n-1}=\upsilon_{n-1}\bigg)\bbP\bigg(X_n=x_n\bigg|\Upsilon_n=\upsilon_n\bigg)\\
	&=\bbP\bigg(\Upsilon_n=\upsilon_n\bigg| \Upsilon_{n-1}=\upsilon_{n-1}, X_{n-1}=x_{n-1}\bigg)\bbP\bigg(X_n=x_n\bigg|\Upsilon_n=\upsilon_n,\Upsilon_{n-1}=\upsilon_{n-1}, X_{n-1}=x_{n-1}\bigg)\\
	&=\bbP\bigg(X_n=x_n,\Upsilon_n=\upsilon_n\bigg|X_{n-1}=x_{n-1}, \Upsilon_{n-1}=\upsilon_{n-1}\bigg).
	\end{align}
	Hence, \eqref{eq246HM},\eqref{eq247HM}, and \eqref{jointmomentHM} are direct results of Theorem \ref{thm:maincontri1} and Theorem \ref{thm:maincontri2}. Now, by \eqref{corofact1}, we also have
	\begin{align}
	\bbE\big[\|\bX-[\bX]\|_2^2\big]&=\sum_{k=1}^n \bbE\big[\big|\sfX_k-[\sfX_k]\big|^2\big]\\
	&=\sum_{k=1}^n \bbE\big[\sfX_k^2\big]-\sum_{k=1}^n \bbE\big[[\sfX_k]^2\big]
	\label{corofact1HM}.
	\end{align} 
	Now, by \eqref{jointmomentHM}, we have as $n \to \infty$,
	\begin{align}
	\bbE\big[[\sfX_k]^2\big] = \sum_{x_0,\upsilon_0 \in \calX \times \calS_{\Upsilon}}\lambda_{x_0,\upsilon_0}^{(\pi_{\Upsilon})} \bbE\big[\langle X_1|X_0=x_0,\Upsilon_0=\upsilon_0 \rangle^2\big], \quad \forall k \in \{1,2,\cdots,n\} \label{corofact2HM}.
	\end{align}
	In addition, for all $k \in \{1,2,\cdots,n\}$, we also have
	\begin{align}
	\bbE\big[\sfX_k^2\big]&=\bbE\big[\bbE\big[\sfX_k^2\big|\sfX_{k-1},\Upsilon_{k-1}\big]\big]\label{bunhiakop1HM}\\
	&=\bbE\big[\bbE\big[\sfX_1^2\big|\sfX_0,\Upsilon_0\big]\big]\label{bunhiakop2HM}\\
	&=\bbE[\sfX_1^2]\label{corofact3HM},
	\end{align} where \eqref{bunhiakop1HM} follows from the tower property \cite{Billingsley}, and \eqref{bunhiakop2HM} follows from the time-homogeneity of Markov process $\{X_n\}_{n=1}^{\infty}$.
	
	From \eqref{corofact1HM}, \eqref{corofact2HM}, and \eqref{corofact3HM}, as $n\to \infty$, we have
	\begin{align}
	&\bbE\big[\|\bX-[\bX]\|_2^2\big] = n \big(\bbE[\sfX_1^2]-\sum_{x_0,\upsilon_0 \in \calX \times \calS_{\Upsilon}}\lambda_{x_0,\upsilon_0}^{(\pi_{\Upsilon})} \bbE\big[\langle \sfX_1|\sfX_0=x_0,\Upsilon_0=\upsilon_0 \rangle^2\big]\big),
	\end{align} which leads to \eqref{uplowMMSEHM}. Note that for the hidden Markov process with initial states $\nu_0$ and the emission probability $P_{X_1|\Upsilon_1}(\cdot|\cdot)$, we have
	\begin{align}
	P_{\sfX_1}(x_1)&=\sum_{\upsilon \in \calS_{\Upsilon}} P_{X_1,\Upsilon_1}(x_1,\upsilon)\\
	&=\sum_{\upsilon \in \calS_{\Upsilon}}P_{X|\Upsilon}(x_1\mid \upsilon)\pi_{\Upsilon}(\upsilon_0,\upsilon), \quad \forall x_1 \in \calX.
	\end{align}
\end{IEEEproof}

\appendices
\section{Some New Results on Large Deviations for Markov Chains induced by the Channel Setting} \label{sec:largedev}
We begin this section with some well-known results on large deviations theory. Based on these results, we develop some large deviations results for the purpose of asymptotic analysis in this paper. For brevity, we only state some existing results in their versions for finite state-space Markov chains such as Theorem \ref{PFthm}. However, the Perron-Frobenius eigenvalue concept still exists for Markov chains with infinitely countable-state or uncountable state space (e.g. \cite{Glynn2018}).   

Consider a general sequence of random vectors $\bY_n \in \bbR^d$. Let $\phi_n(\theta)=\frac{1}{n}\log \bbE[\exp(n\langle \theta, \bY_n\rangle)]$. 
Define the Legendre-Fenchel transform:
\begin{align}
I(x):=\sup_{\theta \in \bbR^d}(\langle \theta, x \rangle-\phi(\theta)).	\label{defI}
\end{align} 
\begin{theorem}[G\"{a}rtner-Ellis Theorem~\cite{Ellis84}] \label{Garliss}
	Given a sequence of random vectors $\bY_n$, suppose that
	\begin{align}
	\lim_{n\to \infty} \phi_n(\theta)=\phi(\theta), \label{limtheta} 
	\end{align} which exists for all $\theta \in \bbR^d$. Furthermore, suppose $\phi(\theta)$ is finite and differentiable everywhere on $\bbR^d$. Then the following large deviations bounds hold for $I$ defined by~\eqref{defI}  
	\begin{align}
	\limsup_{n}\frac{1}{n}\log \bbP(\bY_n \in \rvF) &\leq -\inf_{x \in \rvF} I(x), \quad \mbox{for any closed set}\quad \rvF\in \bbR^d,\\
	\liminf_{n}\frac{1}{n}\log \bbP(\bY_n \in \rvU)&\geq -\inf_{x \in \rvU} I(x), \quad \mbox{for all open set} \quad \rvU \in \bbR^d.
	\end{align}
\end{theorem}

\begin{theorem}[Varadhan Theorem~\cite{Varadhan2008}]\label{varadhanthm} Recall the definition of Legendre-Fenchel transform $I$ in~\eqref{defI}. Assume that a large deviation principle holds for a sequence of probability measures  $\{P_n\}_{n=1}^{\infty}$  defined on the Borel subsets of a Polish (complete separable metric) space $\calX$, with rate function $I(x)$. Then,
	\begin{align}
	\lim_{n\to \infty} \frac{1}{n}\log \int e^{nF(x)} dP_n(x)= \sup_{x \in \calX} \big[F(x)-I(x)\big]
	\end{align} for and bounded and continuous function $F: \calX \to \bR$.
\end{theorem}
Our goal is to derive the large deviations bounds for empirical means of states in a Markov chain. For this purpose, we need to recall the Perron-Frobeninus Theorem for non-negative irreducible matrices~\cite{Lancaster1985b}.
\begin{definition} A non-negative matrix $C \in \bbR^{N \times N}$ is a matrix in which all elements are equal to or greater than zero, that is,
	$C_{ij}\geq 0, \forall i,j$. If $C_{ij}>0, \forall i,j$, the $C$ is referred to as a positive matrix. 
\end{definition}
\begin{definition}\label{irredef} A $C \in \bbC^{N\times N}$ is said to be reducible if there exists an $N\times N$ permutation matrix $P$ such that
	\begin{align}
	P^TAP =\begin{bmatrix}\tilC_{11}& \tilC_{12}\\ 0& \tilC_{22} \end{bmatrix},
	\end{align}  where $\tilC_{11}$ and $\tilC_{22}$ are square matrices of order less than $N$. If no such $P$ exists then $C$ is irreducible. 
\end{definition}
\begin{definition} Let $P_1,P_2,\cdots,P_N$ be distinct points of the complex plane and let $C \in \bbC^{N \times N}$. For each non-zero element $C_{ij}$ of $C$, connect $P_i$ and $P_j$ with a directed line $\overline{P_iP_j}$. The resulting figure in the complex plane is a \emph{directed graph} for $C$. We say that a directed graph is \emph{strongly connected} if, for each pair of nodes $P_i,P_j$ with $i\neq j$, there is a \emph{direct} path
	\begin{align}
	\overline{P_iP_{k_1}}, \overline{P_{k_1}P_{k_2}},\cdots, \overline{P_{k_{r-1},j}}
	\end{align} connecting $P_i$ to $P_j$. Hence, the path consists of $r$ directed lines. Observe that nodes $i$ and $j$ may be connected by a directed path while $j$ and $i$ are not. 
\end{definition}
\begin{theorem}\cite[Sec. 15.1]{Lancaster1985b}\label{irrmatr}
	A square matrix $C$ is irreducible if the directed graph for matrix $C$ is strongly connected. 
\end{theorem}
\begin{remark}
	It is clear that the stochastic matrix of an irreducible Markov chain belongs to the class of all irreducible matrices. However, the class of all irreducible matrices are not limited to the class of all stochastic matrices of irreducible Markov chains. The following well-known theorem works for this general class of matrices.
\end{remark}
\begin{theorem}[Perron-Frobenius Theorem~\cite{Lancaster1985b}]\label{PFthm} If the matrix $C\in \bbR^{N \times N}$ is non-negative and irreducible, then 
	\begin{enumerate}
		\item The matrix $C$ has a positive eigenvalue, $\rho(C)>0$, equal to the spectral radius of $C$;
		\item The eigenvalue $\rho(C)$ has algebraic multiplicity $1$.
		\item There is a positive right eigenvector associated with the eigenvalue $\rho(C)$ which is unique up to a positive scaling factor;
		\item There is a positive left eigenvector associated with the eigenvalue $\rho(C)$ which is unique up to a positive scaling factor;
	\end{enumerate}
\end{theorem}
The positive eigenvalue $\rho(C)$ in Theorem~\ref{PFthm} is called Perron-Frobenius eigenvalue of the matrix $C$. The following corollary of the Perron-Frobenious Theorem shows that the essential rate of growth of the sequence of matrices $C^n$ is $(\rho(C))^n$.

\begin{corollary}\label{corimp}
	Assume that $C\in \bbR^{N \times N}$ is non-negative and irreducible. Then, for every positive vector $h=(h_1, h_2,\cdots, h_N)$, the following holds
	\begin{align}
	\lim_{n\to \infty} \frac{1}{n}\log \bigg[\sum_{j=1}^N C_{i,j}^n h_j \bigg]=\log \rho(C), \quad \forall i \in \{1,2,\cdots,N\}.
	\end{align}
\end{corollary}
\begin{IEEEproof}
	Sinc $C$ is non-negative and irreducible, it holds that $C_{i,j}\geq 0$ for all $i,j \in [N]\times [N]$ and there exists a positive integer $r$ such that all the elements of $C^r$ are strictly positive by Theorem \ref{irrmatr}. Let $\nu$ be an eigenvector associated with the Ferron-Frobenius eigenvalue of $C$.
	Let $\alpha=\max_j \nu_j$, $\beta=\min_j \nu_j$, $\gamma=\max_j h_j$, and $\delta=\min_j h_j$. We have
	\begin{align}
	\frac{\gamma}{\beta}C_{i,j}^n \nu_j \geq C_{i,j}^n h_j\geq \frac{\delta}{\alpha}C_{i,j}^n \nu_j.
	\end{align}
	Therefore, we have
	\begin{align}
	\lim_{n\to \infty} \frac{1}{n}\log \bigg[\sum_{j=1}^N C_{i,j}^n h_j \bigg]&=\lim_{n\to \infty} \frac{1}{n}\log \bigg[\sum_{j=1}^N C_{i,j}^n \nu_j \bigg]\\
	&=\lim_{n\to \infty} \frac{1}{n}\log\bigg((\rho(C))^n \sum_{i=1}^N \nu_i\bigg)\\
	&=\log \rho(C).
	\end{align}
\end{IEEEproof}
\begin{lemma}\label{simlem} 
	Let $\{S_n\}_{n=1}^{\infty}$ be an i.i.d. sequence of random variable on a finite set $\calS \subset \bbR^+$. Let $\bX:=\{X_n\}_{n=1}^{\infty}$ be a Markov chain with states on a Polish space $\calX$ with the transition matrix $P=\{\pi(x,x')\}_{x, x' \in \calX}$. Assume this Markov chain is irreducible. Set $\bX^{(0)}=\bX$. Let $\bX^{(a)}:=\{X_{n}^{(a)}\}_{n=1}^{\infty}$ be a set of $\nu$ replica sequences with (postulated) distribution $q_{\bX}$ for each $a=1,2,\cdots,\nu$. This means that
	\begin{align}
	p_{\bX^{(0)}\bX^{(1)}\bX^{(2)} \cdots \bX^{(\nu)}}(x^{(0)},x^{(1)},x^{(2)},\cdots, x^{(\nu)}) \sim p_{\bX}(x^{(0)})\prod_{i=1}^{\nu} q_{\bX}(x^{(i)}),
	\end{align}
	where
	\begin{align}
	p_{\bX}(x^{(0)})&=\prod_{i=1}^{\infty}\pi(x_i^{(0)}, x_{i+1}^{(0)})\\
	q_{\bX}(x^{(a)})&=\prod_{i=1}^{\infty}\tilde{\pi}(x_i^{(a)}, x_{i+1}^{(a)}), \qquad \forall a \in [\nu].
	\end{align}
	Define a new sequence of $(\nu+1) \times (\nu+1)$ random matrices $\{\bQ_n\}_{n=1}^{\infty}$ such that
	\begin{align}
	Q_n^{(a,b)}=S_n X_n^{(a)} X_{n}^{(b)}
	\end{align} for all $a \in [\nu]$ and $b \in [\nu]$ and for all $n=1,2,\cdots$. Then, $\{\bQ_n\}_{n=1}^{\infty}$ is also an irreducible Markov chain  with states on $\calQ$, \blue{where $\calQ$ is defined in \eqref{defspaceQ}. In addition, the transition probability, namely $P(Q|Q')$, of this Markov chain satisfies:}
	\blue{\begin{align}
	P(Q|Q')=\frac{\sum_{(s,x_0,x_1,\cdots,x_{\nu}, s',x'_0,x'_1,\cdots,x'_{\nu}) \in  \calA_Q \times \calA_{Q'}} P_S(s')P_S(s)  p_{X_{n-1}}(x'_0)\pi(x_0',x_0)\prod_{i=1}^{\nu} q_{X_{n-1}}(x'_i)  \tilde{\pi} (x'_i,x_i)}{\sum_{(s',x'_0,x'_1,\cdots,x'_{\nu}) \in  \calA_{Q'}} P_S(s') p_{X_{n-1}}(x'_0)\prod_{i=1}^{\nu} q_{X_{n-1}}(x'_i)} \label{def:transpro},
		\end{align}
where $p_{X_{n-1}}(\cdot)$ is the state distribution at time $n-1$ of the Markov chain $\{X_n\}_{n=1}^{\infty}$ with the transition probability $\pi$ defined in \eqref{defmc1}, and $q_{X_{n-1}}(\cdot)$ is the state distribution at time $n-1$ of the Markov chain $\{X_n\}_{n=1}^{\infty}$ with the (postulated) transition probability $\tilde{\pi}(\cdot,\cdot)$ defined in \eqref{priorq}, and
\begin{align}
\calA_Q:= \big\{(s,x) \in \calS \times \calX^{\nu+1}: sxx^T=Q\big\}, \quad \forall Q \in \calQ \label{defAQ}.
\end{align}
}
\end{lemma}
\begin{IEEEproof} Let $\sigma(\bQ_1,\bQ_2,\cdots, \bQ_{n-1})$ be the $\sigma$-algebra generated by random matrices $\bQ_1,\bQ_2,\cdots, \bQ_{n-1}$ and $\sigma(\bQ_k)$ be the sigma-algebra generated by $\bQ_k$ for all $k \in \bbZ^+$. Observe that
	\begin{align}
	\bQ_k=S_k \begin{bmatrix} X_k^{(0)}& X_k^{(1)}& \cdots & X_k^{(\nu)}\end{bmatrix}\begin{bmatrix} X_k^{(0)}\\ X_k^{(1)}\\ \cdots \\ X_k^{(\nu)}\end{bmatrix}, \quad \forall k.
	\end{align} Hence, it holds that
	\begin{align}
	\sigma(\bQ_k)=\sigma(S_k, X_k^{(0)}, X_k^{(1)}, \cdots, X_k^{(\nu)})  \label{eq50bat0}
	\end{align} since a countable union of Borel sets is a Borel set and there are only a countable number of tuples $(s,x) \in \calS \times \calX^{\nu+1}$ such that $s x x^T=Q$ for each $Q \in \calQ$. The existence of only a countable number of tuples $(s,x)$ above follows from the assumption that for each $q \in \bbR$, there are only a countable number of pair $(s,x) \in \calS \times \calX$ such that $sx^2=q$ (cf. Section \ref{sec:setting}) and the fact that for each symmetric matrix $Q \in \calQ$, there are only two different decompositions
	\begin{align}
	Q=yy^T=(-y) (-y)^T
	\end{align} for some $y \in \calX^{\nu+1}$ by the unique up to the sign of the Singular Value Decomposition (SVD) \cite{Lancaster1985b}.

	In addition, we also have
	\begin{align}
	\sigma(\bQ_1,\bQ_2,\cdots, \bQ_{n-1})=\sigma\bigg(\big\{S_k, X_k^{(0)}, X_k^{(1)}, \cdots, X_k^{(\nu)}\big\}_{k=0}^{n-1}\bigg).
	\end{align}
	
	Let $Q \in \calQ$, where $\calQ$ is defined in \eqref{defspaceQ}. Observe that
	\begin{align}
	&\bbP(\bQ_n=Q|\sigma(\bQ_1,\bQ_2,\cdots, \bQ_{n-1}))\nn\\
	&=\bbP((S_n, X_n^{(0)},X_n^{(1)},\cdots, X_n^{(\nu)}) \in \calA_Q |\sigma(\bQ_1,\bQ_2,\cdots, \bQ_{n-1}))\\
	&=\bbP((S_n, X_n^{(0)},X_n^{(1)},\cdots, X_n^{(\nu)}) \in \calA_Q |\sigma\big(\big\{S_k, X_k^{(0)}, X_k^{(1)}, \cdots, X_k^{(\nu)}\big\}_{k=0}^{n-1}\big))\\
	&=\bbP(S_n \in \{s: (x,s) \in \calS \times \calX^{\nu+1}, sxx^T=Q \})\nn\\
	&\qquad \times \bbP((X_n^{(0)},X_n^{(1)},\cdots, X_n^{(\nu)}) \in \calA_Q |\sigma\big(\big\{X_k^{(0)}, X_k^{(1)}, \cdots, X_k^{(\nu)}\big\}_{k=0}^{n-1}\big)) \label{eq51bat1}\\
	&=\bbP(S_n \in \{s: (x,s) \in \calS \times \calX^{\nu+1}, sxx^T=Q \})\nn\\
	&\qquad \times \bbP((X_n^{(0)},X_n^{(1)},\cdots, X_n^{(\nu)}) \in \calA_Q |\sigma\big( X_{n-1}^{(0)}, X_{n-1}^{(1)}, \cdots, X_{n-1}^{(\nu)}\big))\label{eq51bat2}\\
	&=\bbP((S_n, X_n^{(0)},X_n^{(1)},\cdots, X_n^{(\nu)}) \in \calA_Q |\sigma\big(X_{n-1}^{(0)}, X_{n-1}^{(1)}, \cdots, X_{n-1}^{(\nu)}\big))\label{eq51bat3}\\
	&=\bbP((S_n, X_n^{(0)},X_n^{(1)},\cdots, X_n^{(\nu)}) \in \calA_Q|\sigma\big(S_{n-1}, X_{n-1}^{(0)}, X_{n-1}^{(1)}, \cdots, X_{n-1}^{(\nu)}\big))\label{eq51bat4}\\
	&=\bbP(\bQ_n=Q|\sigma(\bQ_{n-1})) \label{eq51bat5}\\
	&=\bbP(\bQ_n=Q|\bQ_{n-1}) \label{eq51bat6},
	\end{align}  where \eqref{eq51bat1} follows from $\{S_n\}_{n=1}^{\infty}$ is independent of $\{X_n\}_{n=1}^{\infty}$ and $S_n$ is independent of $\sigma(S_1,\cdots,S_{n-1})$, \eqref{eq51bat2} follows from the Markov property of the sequence $\{X_n\}_{n=1}^{\infty}$,  \eqref{eq51bat3} and \eqref{eq51bat4} follows from the same reasons as \eqref{eq51bat1}, \eqref{eq51bat5} follows from \eqref{eq50bat0}, and \eqref{eq51bat6} follows from conventional definition in probability.
	
	From \eqref{eq51bat6}, $\{\bQ_n\}_{n=1}^{\infty}$ is a Markov process. \blue{Furthermore, the irreducible property follows from that of the Markov chain $\{X_n\}$. Now, for all pair of state $(Q, Q') \in \calQ\times \calQ$, the transition probability satisfies:}
	\blue{
	\begin{align}
	P(Q|Q')&=\bbP(\bQ_n=Q|\bQ_{n-1}=Q')\\
	&=\frac{\bbP((\bQ_n,\bQ_{n-1})=(Q,Q'))}{\bbP(\bQ_{n-1}=Q')}.
	\end{align} 
	Here,
	\begin{align}
	\bbP(\bQ_{n-1}=Q')&=\bbP((S_{n-1}, X_{n-1}^{(0)},X_{n-1}^{(1)},\cdots, X_{n-1}^{(\nu)}) \in \calA_{Q'})\\
	&= \sum_{(s',x'_0,x'_1,\cdots,x'_{\nu}) \in  \calA_{Q'}} P_S(s') p_{X_{n-1}}(x'_0)\prod_{i=1}^{\nu} q_{X_{n-1}}(x'_i),
	\end{align}
	}
	\blue{and
	\begin{align}
	&\bbP((\bQ_n,\bQ_{n-1})=(Q,Q'))\nn\\
	&\qquad =\bbP((S_n, X_n^{(0)},X_n^{(1)},\cdots, X_n^{(\nu)}, S_{n-1}, X_{n-1}^{(0)},X_{n-1}^{(1)},\cdots, X_{n-1}^{(\nu)}) \in \calA_Q \times \calA_{Q'})\\
	&\qquad =\sum_{(s,x_0,x_1,\cdots,x_{\nu}, s',x'_0,x'_1,\cdots,x'_{\nu}) \in  \calA_Q \times \calA_{Q'}} P_S(s') p_{X_{n-1}}(x'_0)\prod_{i=1}^{\nu} q_{X_{n-1}}(x'_i) P_S(s) \pi(x_0',x_0) \prod_{i=1}^{\nu} \tilde{\pi} (x'_i,x_i). 
	\end{align}
	}
	This concludes our proof of Lemma~\ref{simlem}.
\end{IEEEproof}
Now, consider the homogeneous Markov chain with states in the set $\calQ$ as mentioned in Lemma~\ref{simlem}. This Markov chains have $M$ states  $\barQ_0,\barQ_1,\cdots, \barQ_M$ where $M=|\calQ|-1$. We define
\begin{align}
P_{\tilQ}=\begin{bmatrix}e^{\mathrm{tr}(\tilQ \barQ_0)} P(\barQ_0|\barQ_0)& e^{\mathrm{tr}(\tilQ \barQ_1)} P(\barQ_1|\barQ_0)& \cdots& e^{\mathrm{tr}(\tilQ \barQ_M)} P(\barQ_M|\barQ_0) \\
e^{\mathrm{tr}(\tilQ \barQ_0)} P(\barQ_0|\barQ_1)& e^{\mathrm{tr}(\tilQ \barQ_1)} P(\barQ_1|\barQ_1)& \cdots& e^{\mathrm{tr}(\tilQ \barQ_M)} P(\barQ_M|\barQ_1)\\ \vdots & \vdots& \ddots & \vdots \\
e^{\mathrm{tr}(\tilQ \barQ_0)} P(\barQ_0|\barQ_M)& e^{\mathrm{tr}(\tilQ \barQ_1)} P(\barQ_1|\barQ_M)& \cdots& e^{\mathrm{tr}(\tilQ \barQ_M)} P(\barQ_M|\barQ_M) \end{bmatrix}\label{defPtilQ},
\end{align} \blue{where
$P(\barQ_j|\barQ_i)$ is the transition probability of the Markov chain $\{\bQ_n\}_{n=1}^{\infty}$, which is defined in \eqref{def:transpro} of Lemma \ref{simlem}.}

Then, $P_{\tilQ}$ is an irreducible non-negative matrix, since $P=\{P(\barQ_j|\barQ_i)\}_{0 \leq i,j\leq M}$ is such a matrix by the fact that $P_{\tilQ}=P \times \diag(e^{\mathrm{tr}(\tilQ \barQ_1)},e^{\mathrm{tr}(\tilQ \barQ_2)},\cdots, e^{\mathrm{tr}(\tilQ \barQ_M)})$ and Definition \ref{irredef}. Let $\rho(P_{\tilQ})$ denote the Perron-Frobenious eigenvalue of the non-negative irreducible matrix $P_{\tilQ}$.
\begin{theorem} \label{auxthm} Let $\{X_n\}_{n=1}^{\infty}$ be a Markov chain defined in Lemma \ref{simlem} and recall the definition of $\{\bQ_n\}_{n=1}^{\infty}$ in this lemma. Then,  $\{\bT_n:=\frac{1}{n}\sum_{k=1}^n \bQ_k\}_{n=1}^{\infty}$ satisfies the large deviation bounds with rate function $I(Q)=\sup_{\tilQ}(\mathrm{tr}(\tilQ Q) -\log \rho(P_{
		\tilQ}))$, where $\rho(P_{\tilQ})$ is the Perron-Frobenius eigenvalue of the matrix $P_{\tilQ}$ defined in \eqref{defPtilQ}. Specifically, for every initial state $\barQ_0 \in \calQ$, every closed set $\rvF \subset \calQ$ and every open set $\rvU \in \calQ$, the following hold:
	\begin{align}
	\limsup_{n}\frac{1}{n}\log \bbP\big(\bT_n\in \rvF|\barQ_0\big) \leq -\inf_{Q\in \rvF} I(Q),\\
	\liminf_{n}\frac{1}{n}\log \bbP \big(\bT_n\in \rvU|\barQ_0\big) \leq -\inf_{Q \in \rvU} I(Q).
	\end{align}
\end{theorem}
\begin{IEEEproof}
	We will show that the sequence of functions $\phi_n(\tilQ)=\frac{1}{n}\log \bbE[e^{\mathrm{tr}(n \tilQ \bT_n)}]$ has a limit $\phi(\tilQ)$ which is finite and differentiable everywhere. Recall the definition of the matrix $P_{\tilQ}$ in \eqref{defPtilQ}. Given the starting state $\barQ_0$, we have
	\begin{align}
	\log \bbE[e^{\mathrm{tr} (n\tilQ \bT_n)}]&=\log \sum_{\barQ_1,\barQ_2,\cdots, \barQ_n \in \calQ} P(\barQ_1|\barQ_0) P(\barQ_2|\barQ_1)\cdots P(\barQ_n|\barQ_{n-1})\prod_{k=1}^n e^{\mathrm{tr}(\tilQ\barQ_j)}\\
	&=\log \bigg[\sum_{\barQ_n\in \calQ } P_{\tilQ}^n(\barQ_n|\barQ_0)\bigg],
	\end{align} where $P_{\tilQ}^n(\barQ_j|\barQ_i)$ denotes the $(i,j)$-th entry of the matrix $P_{\tilQ}^n$.  Let $h_j=\underline{1}$ and apply Corollary~\ref{corimp}, we obtain
	\begin{align}
	\lim_{n} \phi_n(\tilQ)=\log \rho(P_{\tilQ}).
	\end{align} 
	Since $\log \rho(P_{\tilQ})$ is the spectral radius of $P_{\tilQ}$, hence it is differentiable with respect to $\tilQ$. Thus, the G\"{a}rtner-Ellis can be applied.
\end{IEEEproof}
\begin{corollary}\cite[Eq. (107)]{GuoVerdu2005}\label{cor7} Assume that $\{X_n\}_{n=1}^{\infty}$ is a memoryless source which, together with another i.i.d. sequence $\{S_n\}_{n=1}^{\infty}$, induces an i.i.d. sequence of random matrices $\{\bQ_n\}_{n=1}^{\infty} \sim P_Q$ as defined in Lemma~\ref{simlem}. 
	Then, the sequence of random matrices $\{\bT_n=\frac{1}{n}\sum_{k=1}^n Q_k\}_{n=1}^{\infty}$ satisfies the large deviations bounds with rate function 
	\begin{align}
	I(Q)=\sup_{\tilQ}\big(\mathrm{tr}(\tilQ Q)-\log \calM(\tilQ)\big)
	\end{align}
	where $\calM(\tilQ):=\bbE_{\bQ}[e^{\mathrm{tr}(\tilQ \bQ)}]$ is the moment generating function of the random matrix $\bQ$ on $\calQ$  under the distribution
	\begin{align}
	P(\bQ=Q)&=\bbP\left(S_1 \begin{bmatrix} X_1^{(0)}& X_1^{(1)}&\cdots& X_1^{(\nu)}\end{bmatrix}\begin{bmatrix} X_1^{(0)}\\ X_1^{(1)}\\ \cdots\\ X_1^{(\nu)}\end{bmatrix}=Q \right) \\
	&=\sum_{s,x \in \calS \times \calX^{\nu+1}: sxx^T=Q} \bbP\bigg( (S_1,X_1^{(0)}, X_1^{(1)},\cdots, X_1^{(\nu)})=(s,x)\bigg)
	\end{align} for all $Q \in \calQ$, where $X_1^{(a)}$ be the first sample of replicas $\bX^{(a)}$ defined in Lemma \ref{simlem}.
\end{corollary}
\begin{remark} For the i.i.d. case, the rate function $I(Q)$ can be estimated since $\log \calM(\tilQ)$ is in the form of an expectation. However, in the more general Markov setting as in Theorem \ref{auxthm}, the estimation of Perron-Frobenius is very challenging. In the next sections, we provide bounds on this eigenvalue by making use of the structure of the Markov chain.
\end{remark}
\begin{IEEEproof}
	\blue{Recall the definitions of $\{\barQ_i\}_{i=1}^M$ in Subsection \ref{sub:nota}.} For this special case, the matrix $P_{\tilQ}$, which is defined in \eqref{defPtilQ}, becomes
	\begin{align}
	P_{\tilQ}=\left[\begin{array}{cccc} P(\barQ_0)e^{\mathrm{tr}(\tilQ \barQ_0)}& P(\barQ_1)e^{\mathrm{tr}(\tilQ \barQ_1)}& \cdots & P(\barQ_M)e^{\mathrm{tr}(\tilQ \barQ_M)}\\ P(\barQ_0)e^{\mathrm{tr}(\tilQ \barQ_0)}& P(\barQ_1)e^{\mathrm{tr}(\tilQ \barQ_1)}& \cdots & P(\barQ_M)e^{\mathrm{tr}(\tilQ \barQ_M)} \\ \vdots & \vdots & \vdots & \vdots \\ P(\barQ_0)e^{\mathrm{tr}(\tilQ \barQ_0)} & P(\barQ_1)e^{\mathrm{tr}(\tilQ \barQ_1)}& \cdots & P(\barQ_M)e^{\mathrm{tr}(\tilQ \barQ_M)}\end{array}\right].
	\end{align}
	This matrix has the Perron-Frobenious eigenvalue 
	\begin{align}
	\rho(P_{\tilQ}) &= \mathrm{tr}(P_{\tilQ})\\
	&=\sum_{i =0}^M P(\barQ_i) e^{\mathrm{tr}(\tilQ \barQ_i)}\\
	&=\bbE_{\bQ}[e^{\mathrm{tr}(\tilQ \bQ)}]\\
	&=\calM(\tilQ).
	\end{align}
\end{IEEEproof}
\begin{theorem} \label{markovvarad}  Let $\calX$ be a Polish space with finite cardinality and a irreducible Markov chain $\bX:=\{X_n\}_{n=1}^{\infty}$ defined on $\calX$ and $\nu$ be a positive integer number. Let $X^{(a)}_n$ for $a \in [\nu]$ be replicas of the Markov process $\bX$. Recall the definition of the sequence $\bQ_n$ in Lemma~\ref{simlem} and $\bT_n=\frac{1}{n}\sum_{j=1}^n \bQ_j$. Let $P_n(U):=\bbP(\bT_n \in U)$ for any measurable set $U$ on the $\sigma$-algebra generated by $\{\bQ_n\}_{n=1}^{\infty}$. Then, for and bounded and continuous function $F: \calQ \to \bR$
	\begin{align}
	\lim_{n\to \infty} \frac{1}{n}\log \bbE\big[e^{nF(\bT_n)}\big]&=\lim_{n\to \infty} \frac{1}{n}\log \int e^{nF(Q)} dP_n(Q) \label{eq34b} \\
	&=\sup_{Q}\bigg[F(Q)-I(Q)\bigg] \label{eq35b}
	\end{align} where $I(Q)=\sup_{\tilQ}(\mathrm{tr}(\tilQ Q) -\log \rho(P_{
		\tilQ}))$ and $\rho (P_{\tilQ})$ is the Perron-Frobenius eigenvalue of the matrix $P_{\tilQ}=\{e^{\mathrm{tr}(\tilQ \barQ_j)} P_{\barQ_j|\barQ_i}\}_{0 \leq i,j\leq M}$ and $M=|\calQ|-1$,
	where \blue{$\calQ$  and $\{\barQ_i\}_{i=1}^M$ are defined in  in Subsection \ref{sub:nota}}.
\end{theorem}
\begin{IEEEproof}
	Equation \eqref{eq34b} is an application of the change of measures \cite{Billingsley}. Equation \eqref{eq35b} is a direct application of Theorem~\ref{varadhanthm} and Theorem~\ref{auxthm}.
\end{IEEEproof}
\section{Perron-Frobenius Eigenvalue Estimation} \label{sec:perronfrobenius}
To estimate $I(Q)=\sup_{\tilQ}(\mbox{tr}(\tilQ Q) -\log \rho(P_{\tilQ}))$, we need to find the maximizer $\tilQ$. By taking derivatives of the objective function, it is easy to see that $\tilQ$ must satisfy the following critial equation:
\begin{align}
Q=\frac{\partial \log \rho(P_{\tilQ})}{\partial \tilQ}.
\end{align}
Next, we find the value of $\frac{\partial\log \rho(P_{\tilQ})}{\partial \tilQ}$. To derive this quantity, we use the following theorems:
\begin{definition}\cite[p. 6]{Deutsch1984a} Let $A \in \bbR^{n\times n}$ and consider the matrix equations
	\begin{align}
	(1) \quad ABA=A,  \qquad  (2)\quad  BAB=B,  \qquad (3) \quad BA=AB.
	\end{align}
	Let $\mu =\{1,2,3\}$ and $\nu \subseteq \mu$. A matrix $B \in \bbR^n$ satisfying equations (i) for all $i\in \nu$ is called a (generalized) $\nu$-inverse of $A$. Any matrix $A \in \bbR^{n\times n}$ has a $\{1,2\}$-inverse. In fact, if $A$ is singular then $A$ has infinitely many $\{1,2\}$-inverses. If $A$ is nonsingular, then its only $\{1,2\}$-inverse is $A^{-1}$. For $\nu=\mu$, a $v$-inverse of $A$, if exists, is unique and is called the group (generalized) inverse of $A$ and denoted by $A^{\#}$. A necessary and sufficient condition for $A^{\#}$ to exist is that $R(A)$ and $N(A)$ be complementary subspaces in $\bbR^n$, in which case $AA^{\#}$ is the projection matrix of $\bbR^n$ onto $R(A)$ along $N(A)$.
	\begin{definition}
		Let
		\begin{align}
		\Phi_{n,n}:=\bigg\{A=(a_{ij}) \in \bbR^{n \times n}\bigg| a_{ij}\geq 0 \quad i\neq j \quad \mbox{and}\quad A \quad \mbox{is irreducible}\bigg\}.
		\end{align}
	\end{definition}
	\begin{definition}
		Let $A \in \bbR^n$ be an essentially nonnegative matrix. Then $Q:=tI-A, t\geq \rho(A)$ is called an $n\times n$ $M$-matrix. Note that any $M$-matrix $Q$ belongs to the set 
		\begin{align}
		Z^{n,n}:=\bigg\{A=(a_{ij}) \in \bbR^{n\times n}\bigg| a_{ij}\leq 0 \quad \mbox{for} \quad i\neq j\}. 
		\end{align}
		If $A \in \Phi_{n,n}$ and $t=r(A)$, the matrix $Q$ given by $Q=\rho(A)I-A$ is called an $n \times n$ singular irreducible $M$-matrix.
	\end{definition}
	\begin{lemma}\cite[p. 7]{Deutsch1984a}\label{best1}  If $A$  is an $n \times n$ singular irreducible $M$-matrix, the following holds:
		\begin{itemize}
			\item there exists positive vectors $\psi$ and $\lambda$ such that $Q\psi=0$ and $\lambda^TQ=0$ to which we shall refer to as right and left Perron-Frobenius vectors of $Q$. These vectors are unique, up to positive scaling.
			\item $Q^{\#}$ exists as $0$ is a simple eigenvalue of $Q$.  
			\item $I-QQ^{\#}$ is the projection matrix of $\bbR^n$ onto $N(Q)$ along $R(Q)$.
			\item if $\psi$ and $\lambda$ are right and left Perron-Frobenius vectors of $Q$ normalized so that $\lambda^T \psi=1$, then
			\begin{align}
			I-QQ^{\#}=\psi \lambda^T.
			\end{align}
		\end{itemize}
		
	\end{lemma}
\end{definition}
\begin{lemma}\cite[Lemma 3.1]{Deutsch1984a}\label{best2}  For any $A \in \Phi_{n,n}$, we have
	\begin{align}
	\frac{\partial \rho(A)}{\partial A}=(I-QQ^{\#})^T,
	\end{align} where $Q=\rho(A)I-A$ and $Q^{\#}$ is the group inverse of $Q$.
\end{lemma}
From lemmas \ref{best1} and \ref{best2}, the following theorem holds
\begin{theorem}\label{Deutschthm} \label{deutthem}   For any $A \in \Phi_{n,n}$, we have 
	\begin{align}
	\frac{\partial \rho(A)}{\partial A}=\lambda \psi^T,
	\end{align} where $\frac{\partial \rho(A)}{\partial _{ij}}$ is the first-order partial derivatives of $\rho(\cdot)$ at $A$ with respect to $(i,j)$-th element is given by
	\begin{align}
	\lim_{t\to 0} \frac{\rho(A+tE_{ij})-\rho(A)}{t},
	\end{align} where $E_{ij}$ is the $n\times n$ matrix whose $(i,j)$-th entry is $1$ and whose remaining entries are $0$.
\end{theorem}

\begin{corollary} Assume that $\{X_n\}_{n=1}^{\infty}$ is a memoryless source which induces an i.i.d. sequence $\{\bQ_n\}_{n=1}^{\infty} \sim P_{\bQ}$ as Lemma~\ref{simlem}. Then, the following holds 
	\begin{align}
	\frac{\partial \log \rho(P_{\tilQ})}{\partial \tilQ}(\tilQ)=\frac{1}{\rho(P_{\tilQ})} \bbE[\bQ \exp(\mathrm{tr}(\tilQ \bQ))],
	\end{align}  which coincides with the result in~\cite[Eq. (112)]{GuoVerdu2005}.
\end{corollary}

\begin{IEEEproof} \blue{Recall the definitions of $\{\barQ_i\}_{i=1}^M$ in Subsection \ref{sub:nota}.} For this special case, $P_{\tilQ}$ which is defined in \eqref{defPtilQ}, becomes
	\begin{align}
	P_{\tilQ}=\left[\begin{array}{cccc} P(\barQ_0)e^{\mathrm{tr}(\tilQ \barQ_0)}& P(\barQ_1)e^{\mathrm{tr}(\tilQ \barQ_1)}& \cdots & P(\barQ_M)e^{\mathrm{tr}(\tilQ \barQ_M)}\\ P(\barQ_0)e^{\mathrm{tr}(\tilQ \barQ_0)}& P(\barQ_1)e^{\mathrm{tr}(\tilQ \barQ_1)}& \cdots & P(\barQ_M)e^{\mathrm{tr}(\tilQ \barQ_M)} \\ \vdots & \vdots & \vdots & \vdots \\ P(\barQ_0)e^{\mathrm{tr}(\tilQ \barQ_0)} & P(\barQ_1)e^{\mathrm{tr}(\tilQ \barQ_1)}& \cdots & P(\barQ_M)e^{\mathrm{tr}(\tilQ \barQ_M)}\end{array}\right].
	\end{align}
	This matrix has the  Perron-Frobenious eigenvalue 
	\begin{align}
	\rho(P_{\tilQ}) &= \mathrm{tr}(P_{\tilQ})\\
	&=\sum_{i=0}^M P(\barQ_i) e^{\mathrm{tr}(\tilQ \barQ_i)}\\
	&=\bbE_{\bQ}[e^{\mathrm{tr}(\tilQ \bQ)}]\\
	&=\calM(\tilQ),
	\end{align} which is the moment generating function for the random matrix $\bQ$.
	
	It is easy to see that the (normalized) right and left Perron vectors of $P_{\tilQ}$ are
	\begin{align}
	\psi&=(1,1,\cdots,1)^T,\\
	\lambda&=\frac{1}{\rho(P_{\tilQ})}(P(\barQ_0) e^{\mathrm{tr}(\tilQ \barQ_0)}, P(\barQ_1) e^{\mathrm{tr}(\tilQ \barQ_1)},\cdots, P(\barQ_M) e^{\mathrm{tr}(\tilQ Q_M)})^T.
	\end{align}
	Hence, from  Theorem \ref{deutthem}, we have
	\begin{align}
	\frac{\partial \rho(P_{\tilQ})}{\partial P_{\tilQ}}& =\frac{1}{\rho(P_{\tilQ})}\left[\begin{array}{cccc} P(\barQ_0)e^{\mathrm{tr}(\tilQ \barQ_0)}& P(\barQ_0)e^{\mathrm{tr}(\tilQ \barQ_0)}& \cdots &P(\barQ_0)e^{\mathrm{tr}(\tilQ \barQ_0)}\\ P(\barQ_1)e^{\mathrm{tr}(\tilQ \barQ_1)}& P(\barQ_1)e^{\mathrm{tr}(\tilQ \barQ_1)}& \cdots & P(\barQ_1)e^{\mathrm{tr}(\tilQ \barQ_1)} \\ \vdots & \vdots & \vdots & \vdots \\ P(\barQ_M)e^{\mathrm{tr}(\tilQ \barQ_M)} & P(\barQ_M)e^{\mathrm{tr}(\tilQ \barQ_M)}& \cdots & P(\barQ_M)e^{\mathrm{tr}(\tilQ \barQ_M)}\end{array}\right].
	\end{align} 
	
	Now, from the chain rule for derivatives, we have
	\begin{align}
	\frac{\partial \rho(P_{\tilQ})}{\partial \tilQ}(\tilQ)&=\frac{\partial \rho(P_{\tilQ})}{\partial P_{\tilQ}}o_{\mathrm{tr}}\frac{\partial P_{\tilQ}}{\partial \tilQ}\\
	&=\frac{1}{\rho(P_{\tilQ})}\left[\begin{array}{cccc} P(\barQ_0)e^{\mathrm{tr}(\tilQ \barQ_0)}& P(\barQ_0)e^{\mathrm{tr}(\tilQ \barQ_1)}& \cdots &P(\barQ_0)e^{\mathrm{tr}(\tilQ \barQ_M)}\\ P(\barQ_1)e^{\mathrm{tr}(\tilQ \barQ_1)}& P(\barQ_1)e^{\mathrm{tr}(\tilQ \barQ_2)}& \cdots & P(\barQ_1)e^{\mathrm{tr}(\tilQ \barQ_2)} \\ \vdots & \vdots & \vdots & \vdots \\ P(\barQ_M)e^{\mathrm{tr}(\tilQ \barQ_1)} & P(\barQ_M)e^{\mathrm{tr}(\tilQ \barQ_M)}& \cdots & P(\barQ_M)e^{\mathrm{tr}(\tilQ \barQ_M)}\end{array}\right]\nn\\
	& o_{\mathrm{tr}} \left[\begin{array}{cccc} \barQ_0 P(\barQ_0)e^{\mathrm{tr}(\tilQ \barQ_0)}& \barQ_1 P(\barQ_1)e^{\mathrm{tr}(\tilQ \barQ_1)}& \cdots & \barQ_M P(\barQ_M)e^{\mathrm{tr}(\tilQ \barQ_M)}\\ \barQ_0 P(\barQ_0)e^{\mathrm{tr}(\tilQ \barQ_0)}&\barQ_1 P(\barQ_1)e^{\mathrm{tr}(\tilQ \barQ_1)}& \cdots & \barQ_M P(\barQ_M)e^{\mathrm{tr}(\tilQ \barQ_M)} \\ \vdots & \vdots & \vdots & \vdots \\ \barQ_0 P(\barQ_0)e^{\mathrm{tr}(\tilQ \barQ_0)} & \barQ_1 P(\barQ_1)e^{\mathrm{tr}(\tilQ \barQ_1)}& \cdots & \barQ_M P(\barQ_M)e^{\mathrm{tr}(\tilQ \barQ_M)}\end{array}\right]\\
	&=\frac{1}{\rho(P_{\tilQ})}\bigg[\sum_{j=0}^M P(\barQ_j)e^{\mathrm{tr}(\tilQ \barQ_j)}\bigg] \bigg[\sum_{j=0}^M \barQ_j P(\barQ_j)e^{\mathrm{tr}(\tilQ \barQ_j)}\bigg] \\
	&=\frac{1}{\rho(P_{\tilQ})}\rho(P_{\tilQ})\bbE[\bQ e^{\mathrm{tr}(\tilQ \bQ)}]\\
	&=\bbE[\bQ e^{\mathrm{tr}(\tilQ \bQ)}].
	\end{align}
	In addition, by the chain rule, we also have
	\begin{align}
	\frac{\partial \log \rho(P_{\tilQ})}{\partial \tilQ}(\tilQ)&=\frac{1}{\rho(P_{\tilQ})}\frac{\partial  P_{\tilQ}}{\partial \tilQ}(\tilQ)\\
	&=\frac{1}{\rho(P_{\tilQ})}\bbE[\bQ e^{\mathrm{tr}(\tilQ \bQ)}].
	\end{align}
\end{IEEEproof}
Next, we use the above method to find the partial derivative $\frac{\partial \log \rho(P_{\tilQ})}{\partial \tilQ}$ for the more general Markov model considered in this paper.
\begin{lemma} \label{lem18} The following holds:
	\begin{align}
	\frac{\partial \log \rho(P_{\tilQ})}{\partial \tilQ}(\tilQ)=\frac{1}{\rho(P_{\tilQ})}\sum_{i=0}^M \lambda_i(\tilQ)\sum_{j=0}^M \psi_j(\tilQ)\barQ_j P(\barQ_j|\barQ_i)e^{\mathrm{tr}(\tilQ \barQ_j)},\label{colre}
	\end{align} 
	where $\lambda(\tilQ)$ and $\psi(\tilQ)$ are left and right eigenvectors associated with the Perron-Frobenius eigenvalue $\rho(P_{\tilQ})$ which are normalized such that $\lambda(\tilQ)^T \psi(\tilQ)=1$. Here, \blue{$\{\barQ_i\}_{i=1}^M$ are defined in Subsection \ref{sub:nota}.}
\end{lemma}
\begin{remark} It is easy to see that we can recover the above result for the i.i.d. case by using this lemma.
\end{remark}
\begin{IEEEproof} In this case, the matrix
	\begin{align}
	P_{\tilQ}=\left[\begin{array}{cccc} P(\barQ_0|\barQ_0)e^{\mathrm{tr}(\tilQ \barQ_0)}& P(\barQ_1|\barQ_0)e^{\mathrm{tr}(\tilQ \barQ_1)}& \cdots & P(\barQ_M|\barQ_0)e^{\mathrm{tr}(\tilQ \barQ_M)}\\ P(\barQ_0|\barQ_1)e^{\mathrm{tr}(\tilQ \barQ_0)}& P(\barQ_1|\barQ_1)e^{\mathrm{tr}(\tilQ \barQ_1)}& \cdots & P(\barQ_M|\barQ_1)e^{\mathrm{tr}(\tilQ \barQ_M)} \\ \vdots & \vdots & \vdots & \vdots \\ P(\barQ_0|\barQ_M)e^{\mathrm{tr}(\tilQ \barQ_0)} & P(\barQ_1|\barQ_M)e^{\mathrm{tr}(\tilQ \barQ_2)}& \cdots & P(\barQ_M|\barQ_M)e^{\mathrm{tr}(\tilQ \barQ_M)}\end{array}\right].
	\end{align}
	where
	\begin{align}
	P(\barQ_j|\barQ_i):=\bbP(\bQ_1=\barQ_j|\bQ_0=\barQ_i).
	\end{align}
	It follows from Theorem \ref{Deutschthm} that
	\begin{align}
	\frac{\partial \rho(P_{\tilQ})}{\partial P_{\tilQ}}=\lambda(\tilQ)\psi(\tilQ)^T
	\end{align} where $\lambda(\tilQ)$ and $\psi(\tilQ)$ are left and right eigenvectors associated with the eigenvalue $\rho(P_{\tilQ})$ which are normalized such that $\lambda(\tilQ)^T \psi(\tilQ)=1$. Then, we have
	\begin{align}
	\frac{\partial \rho(P_{\tilQ})}{\partial \tilQ}(\tilQ)&=\frac{\partial \rho(P_{\tilQ})}{\partial P_{\tilQ}}o_{\mathrm{tr}}\frac{\partial P_{\tilQ}}{\partial \tilQ} \label{eq259x} \\
	&=\lambda(\tilQ)\psi(\tilQ)^T\nn\\
	& o_{\mathrm{tr}}  \left[\begin{array}{cccc} \barQ_0 P(\barQ_0|\barQ_0)e^{\mathrm{tr}(\tilQ \barQ_0)}& \barQ_1 P(\barQ_1|\barQ_0)e^{\mathrm{tr}(\tilQ \barQ_1)}& \cdots & \barQ_M P(\barQ_M|\barQ_0)e^{\mathrm{tr}(\tilQ \barQ_M)}\\ \barQ_0 P(\barQ_0|\barQ_1)e^{\mathrm{tr}(\tilQ \barQ_0)}& \barQ_1 P(\barQ_1|\barQ_1)e^{\mathrm{tr}(\tilQ \barQ_1)}& \cdots & \barQ_M P(\barQ_M|\barQ_1)e^{\mathrm{tr}(\tilQ \barQ_M)} \\ \vdots & \vdots & \vdots & \vdots \\ \barQ_0 P(\barQ_0|\barQ_M)e^{\mathrm{tr}(\tilQ \barQ_0)} & \barQ_1 P(\barQ_1|\barQ_M)e^{\mathrm{tr}(\tilQ \barQ_2)}& \cdots & \barQ_M P(\barQ_M|\barQ_M)e^{\mathrm{tr}(\tilQ \barQ_M)}\end{array}\right]\\
	&=\sum_{i=0}^M \sum_{j=0}^M \lambda_i(\tilQ)\psi_j(\tilQ)\barQ_j P(\barQ_j|\barQ_i)e^{\mathrm{tr}(\tilQ \tilQ_j)}\\
	&=\sum_{i=0}^M \lambda_i(\tilQ)\sum_{j=0}^M \psi_j(\tilQ)\barQ_j P(\barQ_j|\barQ_i)e^{\mathrm{tr}(\tilQ \tilQ_j)}\label{req1}.
	\end{align}
	Now, by the chain rule, we also have
	\begin{align}
	\frac{\partial \log \rho(P_{\tilQ})}{\partial \tilQ}(\tilQ)&=\frac{1}{\rho(P_{\tilQ})}\frac{\partial  P_{\tilQ}}{\partial \tilQ}(\tilQ) \label{req2}.
	\end{align}
	Hence, we obtain~\eqref{colre} from \eqref{req1} and \eqref{req2}.
\end{IEEEproof}
\section{Proof of \eqref{jointmoment}} \label{proof:jointmoment}
To prove \eqref{jointmoment} we first recall that the Large Deviations Principle for probability measures also holds for any finite Borel measures on compact metric space (e.g.\cite{Young1990math}) or on Polish space \cite{Swart}. More specifically, Theorem \ref{Garliss} and Theorem \ref{varadhanthm} still holds for finite Borel measures on these spaces.

Now, since replicas $\bX^{(m)}$ are i.i.d.  and $\bX-(\bY,\bPhi)-\bX^{(m)}$ for all $m \in \{1,2,\cdots,\nu\}$, it holds that
\begin{align}
\bbE\bigg[X_k^{i_0} \tilX_k^{j_0} \langle X_k\rangle_q^{l_0}\bigg]&=\bbE\bigg[\bbE\bigg[X_k^{i_0} \tilX_k^{j_0} \langle X_k\rangle_q^{l_0}\bigg|\bY,\bPhi\bigg]\bigg]\\
&=\bbE\bigg[\bbE\bigg[X_k^{i_0} \tilX_k^{j_0}\bigg|\bY,\bPhi\bigg]\bbE\bigg[ \langle X_k\rangle_q^{l_0}\bigg|\bY,\bPhi\bigg]\bigg]\\
&=\bbE\bigg[\bbE\bigg[X_k^{i_0} \tilX_k^{j_0}\bigg|\bY,\bPhi\bigg]\bbE\bigg[ \bigg(\bbE_q[X_k|\bY,\bPhi]\bigg)^{l_0}\bigg|\bY,\bPhi\bigg]\bigg]\\
&=\bbE\bigg[\bbE\bigg[X_k^{i_0} \tilX_k^{j_0}\bigg|\bY,\bPhi\bigg] \bigg(\bbE_q[X_k|\bY,\bPhi]\bigg)^{l_0}\bigg]\\
&=\bbE\bigg[\bbE\bigg[X_k^{i_0}\bigg|\bY,\bPhi\bigg] \bbE\bigg[\tilX_k^{j_0}\bigg|\bY,\bPhi\bigg] \bigg(\bbE_q[X_k|\bY,\bPhi]\bigg)^{l_0}\bigg]\\
&=\bbE\bigg[\bbE\bigg[X_k^{i_0}\bigg|\bY,\bPhi\bigg] \bbE\bigg[\tilX_k^{j_0}\bigg|\bY,\bPhi\bigg]\prod_{a=1}^{l_0}\bbE_q[X_k^{(a)}|\bY,\bPhi]\bigg]\\
&=\bbE\bigg[\bbE\bigg[X_k^{i_0}\bigg|\bY,\bPhi\bigg] \bbE\bigg[\tilX_k^{j_0}\bigg|\bY,\bPhi\bigg]\prod_{a=1}^{l_0}\bbE_q[X_k^{(a)}|\bY,\bPhi]\bigg]\\
&=\bbE\bigg[\bbE\bigg[X_k^{i_0}\bigg|\bY,\bPhi\bigg] \bbE\bigg[\big(X_k^{(m)}\big)^{j_0}\bigg|\bY,\bPhi\bigg]\prod_{a=1}^{l_0}\bbE_q[X_k^{(a)}|\bY,\bPhi]\bigg]\\
&=\bbE \bigg[\bbE\bigg[X_k^{i_0} \big(X_k^{(m)}\big)^{j_0}\prod_{a=1}^{l_0}X_k^{(a)}\bigg|\bY,\bPhi\bigg] \bigg]\\
&=\bbE\bigg[X_k^{i_0} \big(X_k^{(m)}\big)^{j_0}\prod_{a=1}^{l_0}X_k^{(a)}\bigg] \label{naturalfact1}
\end{align} for all $k \in \{1,2,\cdots,n\}$ and all $m \in \{1,2,\cdots,\nu\}$. Since for $k\geq 2$, we know that
\begin{align}
\bbE\bigg[X_k^{i_0} \big(X_k^{(m)}\big)^{j_0}\prod_{a=1}^{l_0}X_k^{(a)}\bigg]&=\bbE\bigg[\bbE\bigg[X_k^{i_0} \big(X_k^{(m)}\big)^{j_0}\prod_{a=1}^{l_0}X_k^{(a)}\bigg|\sigma\big(X_{k-1},X_{k-1}^{(m)}, \{X_{k-1}^{(a)}\}_{a=1}^{l_0}\big) \bigg]\bigg] \label{ihave0}\\
&=\bbE\bigg[\bbE\bigg[X_k^{i_0}\bigg|X_{k-1}\bigg]\bbE\bigg[(X_k^{(m)})^{j_0} \bigg|X_{k-1}^{(m)}\bigg]\prod_{a=1}^{l_0}\bbE\bigg[X_k^{(a)}\bigg|X_{k-1}^{(a)}\bigg)  \bigg]\bigg] \label{ihave1}\\
&=\bbE\bigg[\bbE\bigg[\sfX_1^{i_0}\bigg|\sfX_0\bigg]\bbE\bigg[(\sfX_1^{(m)})^{j_0} \bigg|\sfX_0^{(m)}\bigg]\prod_{a=1}^{l_0}\bbE\bigg[\sfX_1^{(a)}\bigg|\sfX_0^{(a)}\bigg)  \bigg]\bigg] \label{ihave2}\\
&=\bbE\bigg[\bbE\bigg[\sfX_1^{i_0} \big(\sfX_1^{(m)}\big)^{j_0}\prod_{a=1}^{l_0}\sfX_1^{(a)}\bigg|\sigma\big(\sfX_0,\sfX_0^{(m)}, \{\sfX_0^{(a)}\}_{a=1}^{l_0}\big) \bigg]\bigg]\\
&=\bbE\bigg[\sfX_1^{i_0} \big(\sfX_1^{(m)}\big)^{j_0}\prod_{a=1}^{l_0}\sfX_1^{(a)} \bigg] \label{ihav3},
\end{align} where \eqref{ihave0} follows from the tower property \cite{Billingsley}, \eqref{ihave1} follows from the i.i.d. of replicas $\bX^{(a)}$ for all $a\in [\nu]$, and \eqref{ihave2} follows from the time-homogeneous property of the Markov chains $\{X_n^{(a)}\}_{n=1}^{\infty}$

It follows from \eqref{ihav3} that
\begin{align}
\bbE\bigg[X_k^{i_0} \big(X_k^{(m)}\big)^{j_0}\prod_{a=1}^{l_0}X_k^{(a)}\bigg]=\frac{1}{n}\bbE\bigg[ \sum_{j=1}^n X_j^{i_0} \big(X_j^{(m)}\big)^{j_0}\prod_{a=1}^{l_0}X_j^{(a)}\bigg], \quad \forall k \in \{1,2,\cdots,n\} \label{key1a}.
\end{align}
Now, let $\underline{\bX}:=[\bX^{(1)}, \bX^{(2)}, \cdots, \bX^{(\nu)}]]$ and set
\begin{align}
f(\bX^{(0)},\underline{\bX}_a):= \sum_{j=1}^n X_j^{i_0} \big(X_j^{(m)}\big)^{j_0}\prod_{a=1}^{l_0}X_j^{(a)}.
\end{align}
Note that
\begin{align}
\bbE\bigg[f(\bX_0,\underline{\bX}_a)\bigg|\bY,\bPhi, \bX^{(0)}\bigg]&=\bbE\bigg[ \sum_{j=1}^n X_j^{i_0} \big(X_j^{(m)}\big)^{j_0}\prod_{a=1}^{l_0}X_j^{(a)}\bigg|\bY,\bPhi, \bX^{(0)}\bigg]
\end{align} does not depend on $\nu$. Hence, by \cite[Lemma 1]{GuoVerdu2005}, we have
\begin{align}
\bbE\bigg[f(\bX^{(0)},\underline{\bX}_a)\bigg]=&\lim_{\nu \to 0} \frac{\partial }{\partial h}\log \bbE\bigg[Z^{(\nu)}(\bY,\bPhi,\bX^{(0)};h)\bigg]\bigg|_{h=0}
\end{align}
where
\begin{align}
Z^{(\nu)}(\bY,\bPhi,\bX^{(0)};h)=(2\pi \sigma^2)^{-\nu m/2}\bbE\bigg\{\exp \bigg[h\sum_{j=1}^n X_j^{i_0} \big(X_j^{(m)}\big)^{j_0}\prod_{a=1}^{l_0}X_j^{(a)}\bigg]\prod_{a=1}^{\nu} \exp\bigg[-\frac{1}{2\sigma^2}\|\bY-\bPhi \bX^{(a)}\|^2\bigg|\Phi \bigg]\bigg\}.
\end{align}
Hence, we have
\begin{align}
&\bbE\bigg[Z^{(\nu)}(\bY,\bPhi,\bX^{(0)};h)\bigg]\nn\\
&\qquad =\bbE\bigg[(2\pi)^{-\frac{m}{2}}(2\pi \sigma^2)^{-\frac{\nu m}{2}}\bbE\bigg\{\exp \bigg[h\sum_{j=1}^n X_j^{i_0} \big(X_j^{(m)}\big)^{j_0}\prod_{a=1}^{l_0}X_j^{(a)}\bigg]\nn\\
&\qquad \times \exp\bigg[-\frac{1}{2}\|\bY-\bPhi \bX^{(0)}\|^2\bigg]\prod_{a=1}^{\nu} \exp\bigg[-\frac{1}{2\sigma^2}\|\bY-\bPhi \bX_a\|^2\bigg|\Phi \bigg]\bigg\}\bigg] \label{loto1}.
\end{align}
Let $\tilde{\bU}=(U_1,U_2,\cdots,U_n)$, a vector of i.i.d. random variables each taking the same distribution as a row of the matrix $\bA$. Define
\begin{align}
V_a=\frac{1}{\sqrt{n}}\sum_{j=1}^n \sqrt{S_j} U_j X_j^{(a)}, \quad a=0,1,\cdots,\nu.  
\end{align} 
Now given $\bA, \underline{\bX}$, $m$ outputs $\{Y_k\}_{k=1}^m$ of the channel model in Section \ref{sec:setting} are independent since the sequence $\{S_k\}_{k=1}^{n}$ are i.i.d. Hence, \eqref{loto1} can be written as
\begin{align}
\bbE\bigg[Z^{(\nu)}(\bY,\bPhi,\bX^{(0)};h)\bigg]&=\bbE\bigg[\exp\bigg(m G_n^{(\nu)}(\bA,\underline{\bX})\bigg) \exp \bigg(h\sum_{j=1}^n X_j^{i_0} \big(X_j^{(m)}\big)^{j_0}\prod_{a=1}^{l_0}X_j^{(a)}\bigg)\bigg]\label{curebo1},
\end{align}
where $G_n^{(\nu)}(\bA,\underline{\bX})$ is as \cite[Eq. (98)]{GuoVerdu2005}, i.e.,
\begin{align}
G_n^{(\nu)}(\bA,\underline{\bX})&=-\frac{\nu}{2}\log (2 \pi \sigma^2)+\log \int \bbE\bigg\{\exp\bigg[-\frac{(y-\sqrt{\beta}V_0)^2}{2}\bigg] \nn\\
&\qquad \times \prod_{a=1}^{\nu} \exp\bigg[-\frac{(y-\sqrt{\beta}V_a)^2}{2} \bigg| \bA, \underline{\bX}\bigg\}\frac{dy}{\sqrt{2\pi}}\\
&=(2\pi \sigma^2)^{-\frac{\nu}{2}}\bigg(1+\frac{\nu}{\sigma^2}\bigg)^{-1/2}\bbE\bigg[\exp\bigg(-\frac{1}{2}\bV^T \bSigma \bV\bigg) \bigg|\bA, \underline{\bX}\bigg] \label{eq298},
\end{align} where $\bV=(V_1,V_2,\cdots,V_{\nu})^T$ and \eqref{eq298} follows from \cite[Eq. (A80)]{Guo}. Now, by Edgeworth expansion (e.g. \cite[Eq. (138)]{Tanaka2002a},\cite[Eq. (A75)]{Guo}), given $\bA,\underline{\bX}$ we have
\begin{align}
f_{\bV}(v;\bT_n)= \frac{1}{\sqrt{(2\pi)^{\nu+1}\det(\bT_n)}} \exp\bigg(-\frac{1}{2}v^T \bT_n^{-1}v\bigg)+O(n^{-1}),
\end{align} where
\begin{align}
\bT_n&=\frac{1}{n}\sum_{k=1}^n \bQ_k,\label{factu1}\\
\bQ_k&=\bigg\{S_k X_k^{(a)} X_k^{(b)}\bigg\}_{(a,b)}, \quad \forall k \in \{1,2,\cdots,n\}, \quad \forall a, b \in [\nu] \label{factu2}.
\end{align}
From \eqref{eq298}, \eqref{factu1} and \eqref{factu2}, given $\bA,\underline{\bX}$ we obtain
\begin{align}
G_n^{(\nu)}(\bA,\underline{\bX}) = G^{(\nu)}(\bT_n)+ O(n^{-1})\label{curebo2}
\end{align} where \cite[Eq. (100)]{GuoVerdu2005}
\begin{align}
G^{(\nu)}(\bT_n):=-\frac{1}{2}\log\det(I+\Sigma \bT_n)-\frac{1}{2}\log\bigg(1+\frac{\nu}{\sigma^2}\bigg)-\frac{\nu}{2}\log(2\pi \sigma^2)\label{def:GQ0},
\end{align} and $\Sigma$ is a $(\nu+1)\times (\nu+1)$ matrix
\begin{align}
\Sigma=\frac{\beta}{\sigma^2+\nu}\begin{bmatrix}\nu & -e^T\\-e & (1+\frac{\nu}{\sigma^2})I-\frac{1}{\sigma^2}e e^T\end{bmatrix},
\end{align} where $e$ is a $\nu \times 1$ column vector whose entries are all $1$.

From \eqref{curebo1} and \eqref{curebo2}, we have
\begin{align}
\bbE\bigg[Z^{(\nu)}(\bY,\bPhi,\bX^{(0)};h)\bigg]=\bbE_{p(\bA,\underline{\bX})}\bigg[\exp\bigg(m G^{(\nu)}(\bT_n)+O(n^{-1})\bigg) \exp \bigg(h\sum_{j=1}^n X_j^{i_0} \big(X_j^{(m)}\big)^{j_0}\prod_{a=1}^{l_0}X_j^{(a)}\bigg)\bigg].
\end{align}
Now, define a new measure
\begin{align}
\mu(\bA,\underline{\bX})(\rvF):=\int_{\rvF} \exp \bigg(h\sum_{j=1}^n X_j^{i_0} \big(X_j^{(m)}\big)^{j_0}\prod_{a=1}^{l_0}X_j^{(a)}\bigg) dp(\bA,\underline{\bX}) \quad \forall F \in \sigma(\bA,\underline{\bX}),
\end{align} where $p(\bA,\underline{\bX})$ is the joint probability between $\bA$ and $\underline{\bX}$ under the model setting in \ref{sec:setting}. It is obvious that $\mu(\bA,\underline{\bX})$ is a finite measure on Borel sets which are generated by $\bA,\underline{\bX}$ if 
\begin{align}
\bbE\bigg[ \exp \bigg(h\sum_{j=1}^n X_j^{i_0} \big(X_j^{(m)}\big)^{j_0}\prod_{a=1}^{l_0}X_j^{(a)}\bigg) \bigg]<\infty \label{eq307beuty},
\end{align}
By Cauchy Schwarz inequality, \eqref{eq307beuty} happens if $\bbE[ \exp (h X_k^{i_0} \big(X_k^{(m)}\big)^{j_0}\prod_{a=1}^{l_0}X_k^{(a)}) ]<\infty$ for all $k \in \{1,2,\cdots,n\}$, which is equivalent to 
\begin{align}
\bbE\bigg[\exp\bigg(hX_k^{i_0} \tilX_k^{j_0} \langle X_k\rangle_q^{l_0}\bigg)\bigg]< \infty, \quad \forall k=1,2,\cdots,n \label{eq308beauty}.
\end{align} by using the same proof techniques to obtain \eqref{naturalfact1}. This fact (i.e. \eqref{eq308beauty}), of course, holds if we assume that all the conditional joint moments among $\sfX_1, \sfX, \langle \sfX|\sfX_0 \rangle$ in the RHS and LHS of \eqref{jointmoment} are finite.

Under the measure $\mu(\bA,\underline{\bX})$, we have
\begin{align}
\phi_n(\tilQ)&=\bbE_\mu(\bA,\underline{\bX})\bigg[\exp(\mathrm{tr}(\tilQ \bT_n))\bigg]\\
&=\bbE_{p(\bA,\underline{\bX})} \bigg[\exp(\mathrm{tr}(\tilQ \bT_n)) \exp \bigg(h\sum_{j=1}^n X_j^{i_0} \big(X_j^{(m)}\big)^{j_0}\prod_{a=1}^{l_0}X_j^{(a)}\bigg)\bigg]\\
&=\bbE \bigg[\exp(\mathrm{tr}(\tilQ \sum_{j=1}^n \bQ_j)) \exp \bigg(h\sum_{j=1}^n X_j^{i_0} \big(X_j^{(m)}\big)^{j_0}\prod_{a=1}^{l_0}X_j^{(a)}\bigg)\bigg]\\
&=\bbE \bigg[\exp\bigg(\sum_{j=1}^n \mathrm{tr}(\tilQ \bQ_j) + h X_j^{i_0} \big(X_j^{(m)}\big)^{j_0}\prod_{a=1}^{l_0}X_j^{(a)}\bigg)\bigg]\\
&=\bbE \bigg[\exp\bigg(\sum_{j=1}^n K(\tilQ,\bQ_j)\bigg)\bigg] \label{cutefun}
\end{align} where
\begin{align}
K(\tilQ,\bQ_j):=\mathrm{tr}(\tilQ \bQ_j) + h X_j^{i_0} \big(X_j^{(m)}\big)^{j_0}\prod_{a=1}^{l_0}X_j^{(a)} \label{cupe1}.
\end{align} Note that the LHS of \eqref{cupe1} is a function of $\bQ_j$ which is defined in \eqref{factu2}. Moreover, it is know that $\{\bQ_n\}_{n=1}^{\infty}$ forms a irreducible Markov chain by Lemma \ref{simlem}.

\blue{Recall the definitions of $\{\barQ_i\}_{i=1}^M$ in Subsection \ref{sub:nota}.} Now, let
\begin{align}
P(\barQ_j|\barQ_i)=\bbP(\bQ_1=\barQ_j|\bQ_0=\barQ_i),
\end{align}
and
\begin{align}
\hat{\pi}(\barQ_i,\barQ_j)=P(\barQ_j|\barQ_i)e^{K(\tilQ,\barQ_j)}
\end{align}

Define $\rho\big(P_{\tilQ}^{\hat{\pi}}\big)$ be the Perron-Frobenius eigenvalue of the following irreducible non-negative matrix

\begin{align}
P_{\tilQ}^{\hat{\pi}}=\begin{bmatrix}\hat{\pi}(\barQ_0,\barQ_0)& \hat{\pi}(\barQ_0,\barQ_1)&\cdots &\hat{\pi}(\barQ_0,\barQ_M)\\\hat{\pi}(\barQ_1,\barQ_0)& \hat{\pi}(\barQ_1,\barQ_1)&\cdots &\hat{\pi}(\barQ_1,\barQ_M)\\ \vdots & \vdots & \ddots & \vdots \\ \hat{\pi}(\barQ_M,\barQ_0)& \hat{\pi}(\barQ_M,\barQ_1)&\cdots &\hat{\pi}(\barQ_M,\barQ_M)  \end{bmatrix}.
\end{align}
It follows from \eqref{cutefun} that
\begin{align}
\phi_n(\tilQ)&=\bbE \bigg[\exp\bigg(\sum_{j=1}^n K(\bQ_j)\bigg)\bigg]\\
&=\bbE\bigg[\sum_{\barQ_1, \barQ_2, \cdots, \barQ_n \in \calQ^n}P(\barQ_1,\barQ_2,\cdots, \barQ_n) \exp\bigg(\sum_{j=1}^n K(\barQ_j)\bigg)\bigg]\\
&=\bbE\bigg[\bbE\bigg[\sum_{\barQ_1, \barQ_2, \cdots, \barQ_n \in \calQ^n}P(\barQ_1,\barQ_2,\cdots, \barQ_n) \exp\bigg(\sum_{j=1}^n K(\barQ_j)\bigg)\bigg|\bQ_0\bigg]\bigg]\\
&=\bbE_{\bQ_0}\bigg[\sum_{\barQ_1, \barQ_2, \cdots, \barQ_n \in \calQ^n}P(\barQ_1|\bQ_0)P(\barQ_2|\barQ_1)\cdots P(\barQ_n|\barQ_{n-1}) \exp\bigg(\sum_{j=1}^n K(\barQ_j)\bigg)\bigg]\\
&=\bbE_{\bQ_0}\bigg[\sum_{\barQ_1, \barQ_2, \cdots, \barQ_n \in \calQ^n}\hat{\pi}(\bQ_0,\barQ_1)\hat{\pi}(\barQ_1,\barQ_2)\cdots \hat{\pi}(\barQ_{n-1},\barQ_n) \bigg]\\
&=\bbE_{\bQ_0}\bigg[\sum_{\barQ_n \in \calQ} \hat{\pi}_{\bQ_0,\barQ_n}^n \bigg]\\
\label{eqbestcu}
&=\bbE_{\bQ_0}\bigg[\rho\big(P_{\tilQ}^{\hat{\pi}}\big)\bigg]\\
&=\rho\big(P_{\tilQ}^{\hat{\pi}}\big),
\end{align} where \eqref{eqbestcu} follows from Lemma \ref{corimp}. 

Hence, by Theorem \ref{Garliss}, under the measure $\mu(\bA,\underline{\bX})$, $\bT_n$ satisfies the large deviations property, with rate function $I(Q)=\sup_{\tilQ}(\mbox{tr}(\tilQ Q) -\log \rho\big(P_{\tilQ}^{\hat{\pi}}\big)$. Specifically, for every initial state $\barQ_0 \in \calQ$, every closed set $\rvF \subset \calQ$ and every open set $\rvU \in \calQ$, the following holds:
\begin{align}
\limsup_{n}\frac{1}{n}\log\mu_{\bA,\underline{\bX}} \bigg(\bT_n\in \rvF|\barQ_0\bigg) \leq -\inf_{Q\in \rvF} I(Q),\\
\liminf_{n}\frac{1}{n}\log \mu_{\bA,\underline{\bX}} \bigg(\bT_n\in \rvU|\barQ_0\bigg) \leq -\inf_{Q \in \rvU} I(Q).
\end{align}
For any Borel set $F \in \calQ$, define a new measure
\begin{align}
P_n(F)=\mu_{\bA,\underline{\bX}}\big(\bT_n \in F\big).
\end{align}
It follows by Theorem \ref{varadhanthm} that for and bounded and continuous function $F: \calQ \to \bR$
\begin{align}
\lim_{n\to \infty} \frac{1}{n}\bbE_{\mu(\bA,\bX)}\big[e^{nF(\bT_n)}\big]&=\lim_{n\to \infty} \frac{1}{n}\log \int e^{nF(Q)} dP_n(Q) \label{eq34bext} \\
&=\sup_{Q}\bigg[F(Q)-I(Q)\bigg] \label{eq35bext}
\end{align} where $I(Q)=\sup_{\tilQ}(\mathrm{tr}(\tilQ Q) -\log \rho\big(P_{\tilQ}^{\hat{\pi}}\big))$. 

On the other hand, by the change of measure \cite{Royden}, we have
\begin{align}
\bbE_{\mu(\bA,\bX)}\big[e^{nF(\bT_n)}\big]=\bbE\bigg[e^{nF(\bT_n)}\exp \bigg(h\sum_{j=1}^n X_j^{i_0} \big(X_j^{(m)}\big)^{j_0}\prod_{a=1}^{l_0}X_j^{(a)}\bigg)  \bigg]\label{equ1}.
\end{align}
Hence, from \eqref{eq35bext} and \eqref{equ1} we obtain
\begin{align}
\bbE\bigg[e^{nF(\bT_n)}\exp \bigg(h\sum_{j=1}^n X_j^{i_0} \big(X_j^{(m)}\big)^{j_0}\prod_{a=1}^{l_0}X_j^{(a)}\bigg)=\sup_{Q}\bigg[F(Q)-I(Q)\bigg] 
\end{align} where
\begin{align}
I(Q)=\sup_{\tilQ}(\mathrm{tr}(\tilQ Q) -\log \rho\big(P_{\tilQ}^{\hat{\pi}}\big)).
\end{align}
Combining all the above results, we come up with the following theorem
\begin{theorem} \label{auxthem} Recall the definition of $G^{(\nu)}(Q)$ in Lemma \ref{lem:gv1}. In the large system limit, the following holds:
	\begin{align}
	\lim_{n \to \infty} \bbE\bigg[X_k^{i_0} \tilX_k^{j_0} \langle X_k\rangle_q^{l_0}\bigg]=  \frac{\partial}{\partial h}\lim_{\nu \to 0} \frac{\partial}{\partial \nu}\sup_Q \bigg[\beta^{-1}G^{(\nu)}(Q)-I^{(\nu,h)}(Q)\bigg]\bigg|_{h=0}\label{eq32ext},
	\end{align} where
	\begin{align}
	I^{(\nu,h)}(Q):=\sup_{\tilQ}\bigg[\mbox{tr}(\tilQ Q) - \log \rho\big(P_{\tilQ}^{\hat{\pi}}\big)\bigg] \label{eq33ext},
	\end{align} and $\log \rho\big(P_{\tilQ}^{\hat{\pi}}\big)$ is the Perron-Frobenius eigenvalue of the matrix 
	\begin{align}
	P_{\tilQ}^{\hat{\pi}}=\bigg\{\exp\big(\mathrm{tr}(\tilQ \barQ_j) + h X_j^{i_0} \big(X_j^{(m)}\big)^{j_0}\prod_{a=1}^{l_0}X_j^{(a)}\big) P(\barQ_j|\barQ_i) \bigg\}_{0\leq i,j\leq M}
	\end{align}
	and  $M=|\calQ|-1$ where $\calQ:=\{s x x^T \quad \mbox{for some} \enspace s \in \calS, \enspace x \in \calX^{\nu+1}\}$. \blue{Here, $\{\barQ_i\}_{i=1}^M$ ared defined in Subsection \ref{sub:nota}.}
\end{theorem}
Theorem \ref{auxthem} is similar to Theorem \ref{mainthm1}, except the matrix $P_{\tilQ}$ in Theorem \ref{mainthm1} is replace by a new matrix $P_{\tilQ}^{\hat{\pi}}$.

Using the same arguments and using the same replica assumptions, we can show a similar result  as Lemma \ref{rhoto1} that 
\begin{align}
\lim_{\nu \to 0}\rho\big(P_{\tilQ}^{\hat{\pi}}\big)=1.
\end{align}
Then, for a fixed $\tilQ \in \calQ$, the optimizer of the optimization problem in \eqref{eq32ext} of Theorem \ref{auxthem}, say $Q^*(h)$ can be expressed in the large system limit (cf. Proof of Lemma \ref{repre1})
\begin{align}
Q^*(h)=\lim_{\nu \to 0} \sum_{i=0}^M \lambda_i(\tilQ)\hatQ_i(h)
\end{align} where $(y_1(\tilQ),y_2(\tilQ),\cdots, y_M(\tilQ))^T$ is the positive (left) eigenvector (all elements are positive) of $P_{\tilQ}^{\hat{\pi}}$ such that $\sum_{i=1}^M \lambda_i(\tilQ)\to 1$, and 
\begin{align}
\hatQ_i(h)=\bbE\bigg[ \bQ_1\exp\big(\mathrm{tr}(\tilQ \bQ_1) + h X_1^{i_0} \big(X_1^{(m)}\big)^{j_0}\prod_{a=1}^{l_0}X_j^{(a)}\big)\bigg|\bQ_0=\barQ_i\bigg] \label{bunhiamon1ext}.
\end{align}
On the other hand, it holds from \eqref{bunhiamon1ext} that
\begin{align}
\frac{\partial }{\partial h}\hatQ_i(h)\bigg|_{h=0}=\bbE\bigg[  X_1^{i_0} \big(X_1^{(m)}\big)^{j_0}\prod_{a=1}^{l_0}X_1^{(a)} \bQ_1 \exp\big(\mathrm{tr}(\tilQ \bQ_1)\big)\bigg|\bQ_0=\barQ_i\bigg] \label{bunhiamon1ext2}.
\end{align}
Then, using the same arguments as ones to achieve \eqref{banav2}, it holds that for all $i \in [M]$,
\begin{align}
\frac{\partial}{\partial h}\lim_{\nu \to 0} \frac{\partial}{\partial \nu} \bigg[\beta^{-1}G^{(\nu)}(Q)-I^{(\nu,h)}(Q)\bigg]\bigg|_{h=0, Q=\hatQ_i} \to \bbE\bigg[\sfX_1^{i_0} \sfX^{j_0} \langle \sfX \rangle_q^{l_0}\bigg|\sfX_0=x_i^{(0)}\bigg]\label{bestest}
\end{align} if 
\begin{align}
\barQ_i=s_i \begin{bmatrix} x_i^{(0)}\\x_i^{(1)}\\ \vdots \\x_i^{(\nu)} \end{bmatrix}\begin{bmatrix} x_i^{(0)}& x_i^{(1)}& \cdots &x_i^{(\nu)} \end{bmatrix}
\end{align} for some $s_i \in \calS$ and $ (x_i^{(0)},x_i^{(1)}, \cdots, x_i^{(M)})^T \in \calX^{\nu+1}$. Interested readers can refer to \cite[Eqs. 165,166]{GuoVerdu2005} or \cite[Sec. (3.4.2)]{Guo} for some detailed calculations which lead to a similar equation as \eqref{bestest}.

Then using the concave property of the function $\frac{\partial}{\partial h}\lim_{\nu \to 0} \frac{\partial}{\partial \nu} \bigg[\beta^{-1}G^{(\nu)}(Q)-I^{(\nu,h)}(Q)\bigg]\bigg|_{h=0, Q}$  in $Q$ as in the proof of Theorem \ref{mainthm1}, we obtain \eqref{jointmoment}. 
\section{Extensions to Markov chains on a general Polish space in $\bbR$} \label{extend}
In this section, we sketch what we should change in our analysis when working with a Markov chains on a general Polish space in $\bbR$.
\begin{itemize}
	\item As the spectral method (Paulin), we define an associated linear operator $\bpi$ on $L_2$ to a the Markov kernel $\pi(x,y)$ such that
	\begin{align}
	\bpi(f)(x):=\int_{\calS} \pi(x,y) f(y) dy \label{eq558}. 
	\end{align}
	We call $f(\cdot)$ is an eigenvector of $\pi$ associated with an eigenvalue $\lambda$ if and only if $\bpi(f)(x)=\lambda f(x)$ for all $x \in \calS$. The existence of such $\lambda$ and $f$ is guaranteed (for example, let $f(y)=1/\|f\|, \forall y \in \calS$ and $\lambda=1$). Define $S_2$ be the set of eigenvalues of $\pi$. The Perron-Frobenius eigenvalue is defined as the supremum of all elements in this set\footnote{Since the linear operator is continuous (bounded), the set of eigenvalues is bounded.}. 
	\item Then, we show a similar fact as Corollary \ref{corimp}. More specifically, we show that for every positive function $h:\bbS \to \bbR_+$ and Markov chain $\{Z_n\}_{n=1}^{\infty}$ on an arbitrary space $\calV$ with stochastic kernel $Q(x,y)$, the following holds:
		\begin{align}
		\lim_{n\to \infty} \frac{1}{n}\log \bigg[\int_{\calV} Q^n(x,y) h(y) dy \bigg]=\log \rho(Q), \quad \forall x \in \calV \label{lava},
		\end{align} where $\rho(Q)$ is the Perron-Frobenius eigenvalue of $Q$.
  \item Using the same arguments as the proof of Theorem \ref{auxthm} with the fact \eqref{lava}, we conclude that $\bT_n=\frac{1}{n}\sum_{k=1}^n \bQ_k$ satisfies the large deviation bounds with rate $I(Q)=\sup_{\tilQ}(\tr(\tilQ Q)- \log \rho(P_{\tilQ})$, where $\rho(P_{\tilQ})$ is the Perron-Frobenius eigenvalue of the Markov chain $\bQ_0-\bQ_1 \cdots -\bQ_n$.
  \item  By Varadhan theorem on Polish space (Theorem \ref{varadhanthm}), we can show that Lemma \ref{markovvaradmod} still holds, i.e.,
   \begin{align}
  \lim_{n\to \infty} \frac{1}{n}\log \bbE\big[e^{nF(\bT_n)}\big]&=\lim_{n\to \infty} \frac{1}{n}\log \int e^{nF(Q)} dP_n(Q) \label{eq34bx} \\
  &=\sup_{Q}\bigg[F(Q)-I(Q)\bigg] \label{eq35bx}
   \end{align} for any bounded continuous function $F: \calQ \to \bbR$. The main difference is now $|\calQ|$ is unbounded or $M \to \infty$. 
   \item From \eqref{eq35bx}, by applying for a specific function $F$, we obtain Theorem \ref{mainthm1}.
   \item The rest is an optimization problem and the same arguments as previous section still work.
\end{itemize}
\subsection*{Acknowledgements}
The author is extremely grateful to Prof.\ Ramji Venkataramanan, the University of Cambridge, for many suggestions to improve the paper. The author also would like to thank the editor and reviewers for their suggestions to improve the paper.
\bibliographystyle{IEEEtran}	
\bibliography{isitbib}

\end{document}